\documentclass[fleqn,usenatbib]{mnras}

\usepackage[T1]{fontenc}
\usepackage{ae,aecompl}

\usepackage{graphicx}	% Including figure files
\usepackage{amsmath}	% Advanced maths commands
\usepackage{amssymb}	% Extra maths symbols
\usepackage[T1]{fontenc} 
\usepackage{academicons}
\usepackage{xcolor}
\usepackage{hyperref}
\usepackage{rotating}
\newcommand{\hb}{H$\beta$}

\newcommand{\orcid}[1]{\href{https://orcid.org/#1}{\textcolor[HTML]{A6CE39}{\aiOrcid}}}

\def\ltsima{$\buildrel<\over\sim$}
\def\lsim{\lower.5ex\hbox{\ltsima}~}
\def\gtsima{$\buildrel>\over\sim$}
\def\gsim{\lower.5ex\hbox{\gtsima}~}

\def\teff{\ifmmode T_{\rm eff} \else $T_{\mathrm{eff}}$\fi}

\def\hb{H$\beta$}

\def\cm2{cm$^{-2}$}

\def\ewo3{$EW_{\mathrm{[O\textsc{iii}]}}$}

\def\oiii{O{\sc iii}}

\def\nh{\ifmmode N_{\mathrm{HI}}\else $N_{\mathrm{HI}}$\fi}

\def\vexp{\ifmmode v_{\rm exp} \else v$_{\rm exp}$\fi}
\def\taua{\ifmmode \tau_{a}\else $\tau_{a}$\fi}

\newcommand{\jwst}{{\em JWST}}
\newcommand{\hst}{{\em HST}}
\usepackage{newtxtext,newtxmath}
\title[Highest-redshift galaxy candidates with JWST]{Revealing Galaxy Candidates out to $z \sim 16$ with JWST Observations of the Lensing Cluster SMACS0723}

\author[Atek et al.]{Hakim Atek$^{1}$\thanks{E-mail: hakim.atek@iap.fr},
Marko Shuntov$^{1}$,
Lukas J. Furtak$^{2}$,
Johan Richard$^{3}$,
Jean-Paul Kneib$^{4}$,
\newauthor Guillaume Mahler$^{5}$,
 Adi Zitrin$^{2}$,
H.J.~McCracken$^{1}$,
St\'ephane Charlot$^{1}$,
\newauthor Jacopo Chevallard$^{6}$,
Iryna Chemerynska$^{1}$
\\
$^{1}$Institut d'Astrophysique de Paris, CNRS, Sorbonne Universit\'e, 98bis Boulevard Arago, 75014, Paris, France\\
$^{2}$Physics Department, Ben-Gurion University of the Negev, P.O. Box 653, Be’er-Sheva 84105, Israel\\
$^{3}$Univ Lyon, Univ Lyon1, Ens de Lyon, CNRS, Centre de Recherche Astrophysique de Lyon UMR5574, F-69230, Saint-Genis Laval, France \\
$^{4}$Laboratoire d’Astrophysique, Ecole Polytechnique Fédérale de Lausanne, Observatoire de Sauverny, CH-1290 Versoix, Switzerland \\
$^{5}$Institute for Computational Cosmology, Durham University, South Road, Durham DH1 3LE, UK \\
$^{6}$Department of Physics, University of Oxford, Denys Wilkinson Building, Keble Road, Oxford OX1 3RH, UK  \\
}

\date{Accepted XXX. Received YYY; in original form ZZZ}

\pubyear{2022}
\begin{document}
\label{firstpage}
\pagerange{\pageref{firstpage}--\pageref{lastpage}}
\maketitle

\begin{abstract}
One of the main goals of the JWST is to study the first galaxies in the Universe. We present a systematic photometric analysis of very distant galaxies in the first JWST deep field towards the massive lensing cluster SMACS0723. As a result, we report the discovery of two galaxy candidates at $z\sim16$, only $250$ million years after the Big Bang. We also identify two candidates at $z\sim 12$ and 6 candidates at $z\sim 9-11$. Our search extended out to $z\lesssim21$ by combining color information across seven NIRCam and NIRISS filters. By modelling the Spectral Energy Distributions (SEDs) with \texttt{EAZY} and \texttt{BEAGLE}, we test the robustness of the photometric redshift estimates. While their intrinsic (un-lensed) luminosity is typical of the characteristic luminosity L$^*$ at $z>10$, our high-redshift galaxies typically show small sizes and their morphologies are consistent with disks in some cases. The highest-redshift candidates have extremely blue UV-continuum slopes $-3 < \beta <-2.4$, young ages $\sim 10-100$\,Myr, and stellar masses around $\log(M_{\star}/\mathrm{M}_{\odot})=8.8$ inferred from their SED modeling, which indicate a rapid build-up of their stellar mass. Our search clearly demonstrates the capabilities of JWST to uncover robust photometric candidates up to very high redshifts, and peer into the formation epoch of the first galaxies.
\end{abstract}


%%%%%%%%%%%%%%%%%%%%%%%%%%%%%%%%%%%%%%%%%%%%%%%%%%

%%%%%%%%%%%%%%%%% BODY OF PAPER %%%%%%%%%%%%%%%%%%

\section{Introduction}
Understanding the emergence of the first structures that hosted star formation has been one of the most important goals in modern astrophysics. One of the main routes to address this question is to identify the first generation of galaxies that we believe formed at $z\gtrsim15$. Great progress has been made in the last decade to uncover the most distant galaxies out to $z\sim11$, essentially thanks to the \textit{Hubble Space Telescope} (HST) and 8-10\,m class ground-based telescopes. The record-holder for the farthest galaxy known with a spectroscopic confirmation is GN-z11 \citep{oesch16,jiang21}, at a redshift of $z=10.96$. Remarkably, GN-z11 is a relatively bright and massive galaxy ($M_{\mathrm{UV}}=-22.1$\,AB), which challenges current galaxy formation models. The existence of a galaxy with a billion solar masses only $\sim400$\,Myr after the Big Bang indicates that galaxy formation was well underway very early in the history of the Universe.  
 
At slightly lower redshifts $z\sim7-9$, large samples of nearly 2,000 galaxies have been photometrically identified, thanks mostly to the infrared sensitivity of the \textit{Wide Field Camera Three} (WFC3) aboard HST \citep[e.g.][]{finkelstein15,bouwens21} or ground-based observations \citep[e.g.,][]{kauffmann22}. These large samples resulted in strong constraints on the shape of the ultra-violet (UV) luminosity function of galaxies at those epochs and its evolution with redshift. Moreover, {\em Spitzer Space Telescope} observations allowed to probe the rest-frame optical emission of these high-redshift galaxies, which enabled us to constrain their stellar mass \citep[e.g.][]{song16,bhatawdekar19,kikuchihara20,stefanon21}, albeit with important uncertainties due to source confusion and nebular emission contamination \citep{grazian15,furtak21}.
 
More recently, using deep {\em Spitzer} and ground-based observations, \citet{harikane22} reported the detection of two galaxy candidates at $z\sim12-13$. \textit{Atacama Large Millimeter/submillimeter Array} (ALMA) follow-up observations have revealed the tentative ($2\sigma$) detection of an [\ion{O}{iii}]$\lambda88\mu\mathrm{m}$ emission line, putting HD1 at a redshift of $z=13.27$. Again, these two candidates do not align with theoretical models, which predict a significantly lower density of such bright galaxies at $z>10$. While wide extra-galactic surveys have unveiled these surprisingly large numbers of luminous galaxies at $z\sim9-11$, deeper observations at longer wavelengths are needed to, (i) push the redshift frontier to uncover farther away galaxies and (ii) constrain the number density of faint galaxies at early times.

The advent of the {\em JWST} marks a new era in the detection and, more importantly, the characterization of early star-forming galaxies. The most important improvement is the significantly higher spatial resolution of JWST compared to {\em Spitzer}, which enables accurate photometric measurements in the near and mid-infrared range corresponding to the rest-frame optical for galaxies at $z>6$. This is crucial for deriving physical properties of galaxies through spectral energy distribution (SED) fitting, since it allows us to account for a larger dynamic range in the age of the underlying stellar populations. Also, for the first time, we will be able to obtain rest-frame optical spectra for these early-star forming galaxies, giving us access to several optical emission line, which are the gold standards for measuring galaxy properties such as star formation rate (SFR), gas-phase metallicity, dust content, etc.

%While pushing the observational frontiers to detect farther galaxies can inform us on the formation epoch of the first structures in the Universe, constraining their physical properties provides clues on the 
The first JWST observations consist of the Early Release Observations (ERO), which were obtained to showcase the observatory capabilities, and the Early Release Science (ERS) programs, aimed at testing the variety of instruments observing modes and to help the community understand and exploit \jwst\ data. One of the primary targets of the ERO observations is the galaxy cluster SMACS~J0723.3-7327 \citep[SMACS0723 hereafter;][]{ebeling01,repp18} which was recently observed with HST as part of the \textit{Reionization Lensing Cluster Survey} program \citep[RELICS;][]{coe19,salmon20}. One of the accepted ERS programs (1324, PI: Treu) also targets a lensing cluster, Abell~2744 (A2744), which is one of the \textit{Hubble Frontier Fields} \citep[HFF;][]{lotz17} clusters. These observations follow in the steps of numerous HST observing campaigns and efforts to use massive galaxy clusters as gravitational telescopes. The magnification provided by the strong gravitational lensing (SL) enables us to reach intrinsically fainter and lower-mass galaxies and thus to constrain the very faint and low-mass ends of the luminosity and stellar mass function \citep[e.g.][]{atek14b,atek18,livermore17,bouwens17b,ishigaki18,bhatawdekar19,kikuchihara20,furtak21}.

%Another ERS with an extragalacitc focus is the CEERS program, which covers the Extended Growth Stripe (EGS) with both MIRI and NIRCam imaging, and NIRSpec Spectroscopy  

In this work, we use the JWST ERO data of SMACS0723 to identify and characterize the most distant galaxies in the Universe. We use imaging data in 7 broad-band filters to select $z\gtrsim9$ galaxies based on the Lyman break technique coupled with SED fitting to estimate their photometric redshifts and physical properties. This paper is structured as follows. In section~\ref{sec:obs}, we present the observational data. In section~\ref{sec:sample}, we describe how the galaxy sample was selected. The SL model used in this work and our SED-fitting analysis are shown in sections~\ref{sec:SL} and~\ref{sec:SED-fitting} respectively. Finally, we compare our estimate with a compilation of literature results and discuss the implications on galaxy formation models in section~\ref{sec:discussion} before presenting our summary and conclusions in section~\ref{sec:conclusions}. Throughout the paper, magnitudes are in the AB system \citep{oke83} and we adopt a cosmology with H${_0} =70$ km s$^{-1}$ Mpc$^{-1}$, $\Omega_{\Lambda}=0.7$, and $\Omega_m=0.3$.

%--------------------------------------------------------------------
\section{Observations} \label{sec:obs}

\begin{table}
    \centering
    \caption{Limiting AB magnitudes (5$\sigma$) of the JWST/NIRCam and JWST/NIRISS imaging data used in this work. The depths were computed using random 0.3\arcsec circular apertures.}
    \begin{tabular}{lcccc}
    \hline
       SW filter & F090W & F115W & F150W  & F200W\\
       Depth     & 28.0  & 27.8  &  28.8  & 28.9\\\hline
       LW filter & F277W & F356W & F444W & \\
       Depth     & 29.2  &  29.3 & 29.1  & \\\hline
    \end{tabular}
    \label{tab:obs}
\end{table}

\begin{figure*}
    \centering
    \includegraphics[width=\textwidth]{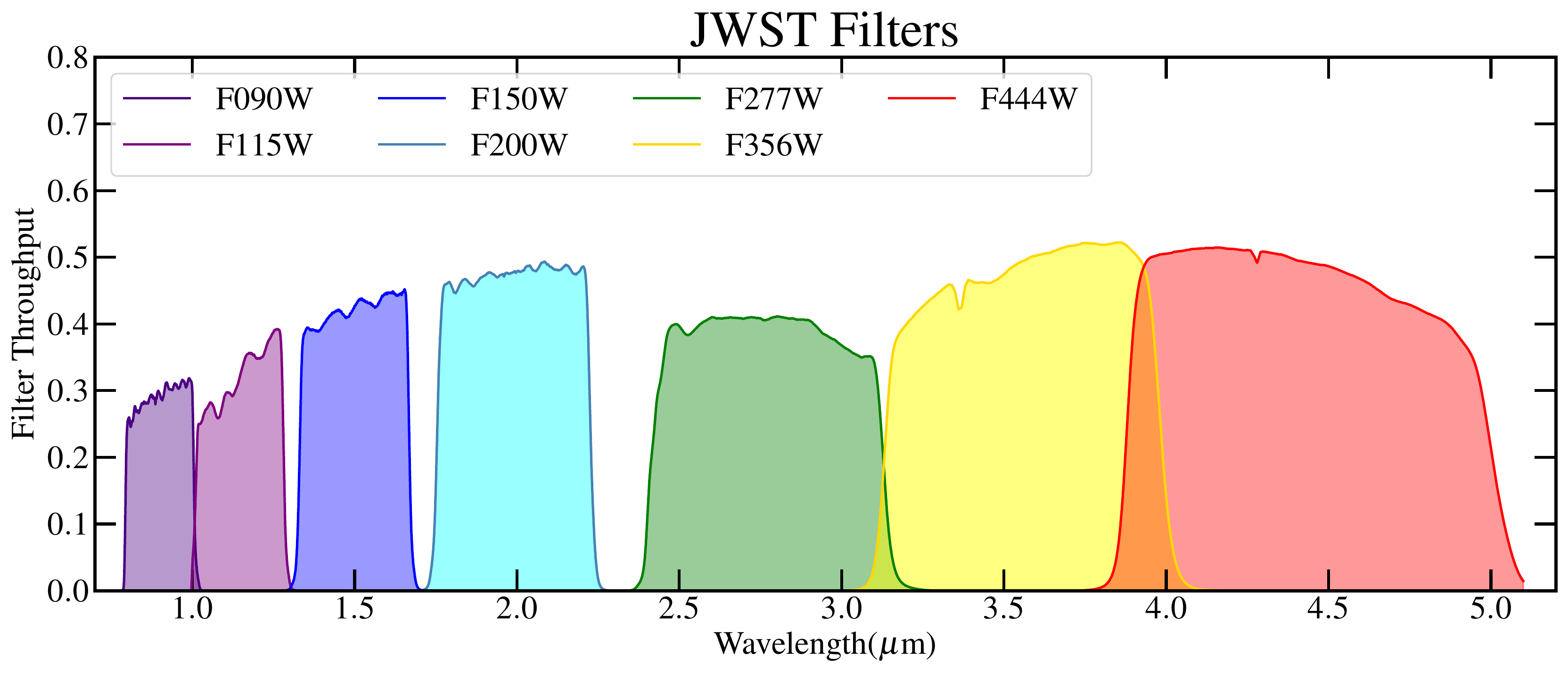}
    \caption{JWST/NIRCam and JWST/NIRISS through-put curves of the filter set of the JWST ERO observations of SMACS0723 used in this work.}
    \label{fig:filters}
\end{figure*}

The first JWST science observations, targeting the lensing cluster SMACS0723, were obtained as part of the Early Release Observations (ERO) program (ID~2736; PI: Pontoppidan), and have been released on Wednesday, July 13th. They consist of deep multi-wavelength imaging with NIRCam (Near-Infrared Camera) and MIRI (Mid-Infrared Instrument), NIRSpec (Near-Infrared Spectrograpgh) multi-object Spectroscopy, and NIRISS (Near-Infrared Imager and Siltless Spectrograpgh) wide-field slitless spectroscopy. We retrieved the data from the \texttt{Mikulski Archive for Space Telescopes} (\texttt{MAST}). Both the short wavelength (SW) and the long wavelength (LW) channels of NIRCam were used simultaneously to obtain uniform spectral coverage from $\lambda\sim1$\,\micron\ to $\lambda\sim5$\,\micron. Tab.~\ref{tab:obs} summarizes the set of filters used in this work, and Fig.~\ref{fig:filters} shows their transmission curves. Note that we complement the 6 available NIRCam filters with F115W-band imaging data from the \textit{Near-Infrared Imager and Slitless Spectrograph} \citep[NIRISS;][]{doyon12}. These however have a smaller field-of-view and only cover the core of the cluster.

For each of the filters, we reduced the raw {\em uncal} files using the most recent \jwst\ {\tt calwebb} pipeline {\tt v1.6.2} and the calibration context {\tt CRDS\_CTX = jwst\_0970.pmap}. In particular, the latest reference files contain updates from in-flight calibration, whereas ground calibrations were used in the early reductions of the ERO data. First, we applied the {\tt Detector1} stage to all files.  
%We also corrected residual artifacts in the NIRCam detectors using the CEERS script to remove vertical stripes. 
Second, we ran the images through the second reduction stage of the pipeline {\tt Image2}, which performs flat-field correction, WCS registration, and photometric calibration. Due to significant levels of background residuals in the calibrated images, and following recommendations from the ERS CEERS program, we independently ran background subtraction on individual {\em cal} files. In addition, the first NIRCam calibration observations have shown that in most cases the detectors/filters were more sensitive than predicted \citep{rigby22}. Therefore, we updated the photometric zeropoints of the {\em cal} files with recent measurements of \citet{boyer22} and G. Brammer\footnote{https://github.com/gbrammer/grizli/pull/107}. Finally, {\em cal} files were processed through the {\tt Image3} stage of the pipeline to create the final drizzled mosaic in each filter. All images were registered to GAIA DR2 astrometry and the relative registration was refined using {\tt TweakReg}. We adopted a pixel scale of 0.03\arcsec\ for both SW and LW channels. 
%As a sanity check, we also compared the final images with an independent reduction, which uses the {\tt grizli} software (Brammer et al. in prep.).%\footnote{\url{https://s3.amazonaws.com/grizli-v2/SMACS0723/Test/image_index.html}}.

In order to increase the detection efficiency in the extraction procedure described in section~\ref{sec:se++}, we created a deep LW image by stacking the three LW filters using their respective weight maps. We also create a stacked SW image of the non-detection filters blue-ward of the break.

One common limitation for the detection of faint sources behind massive cluster fields is the intra-cluster light (ICL). This emission in the central region of the cluster is emitted by tidally stripped stars and bright cluster members \citep[e.g.][]{atek14b,montes22}. In order to reduce the impact of this diffuse light, we apply a median filter to the detection image. We adopted a filter size of $\sim2$\arcsec\ $\times$ 2\arcsec, which significantly reduces the diffuse ICL and BCG light while minimizing residual artifacts. Note that the photometry measurement itself is still performed in the original image of each filter.

Finally, we also use spectroscopic observations of SMACS0723 obtained with the \textit{Near-Infrared Spectrograph} \citep[NIRSpec;][]{jakobsen22} as part of the same ERO program. We obtained level~3 data products from \texttt{MAST} which include 2D and 1D spectra of selected targets in the cluster field. 

\section{Sample} \label{sec:sample}

\subsection{Photometric Catalogs} \label{sec:se++}

\begin{figure}
    \centering
    \includegraphics[width=\columnwidth]{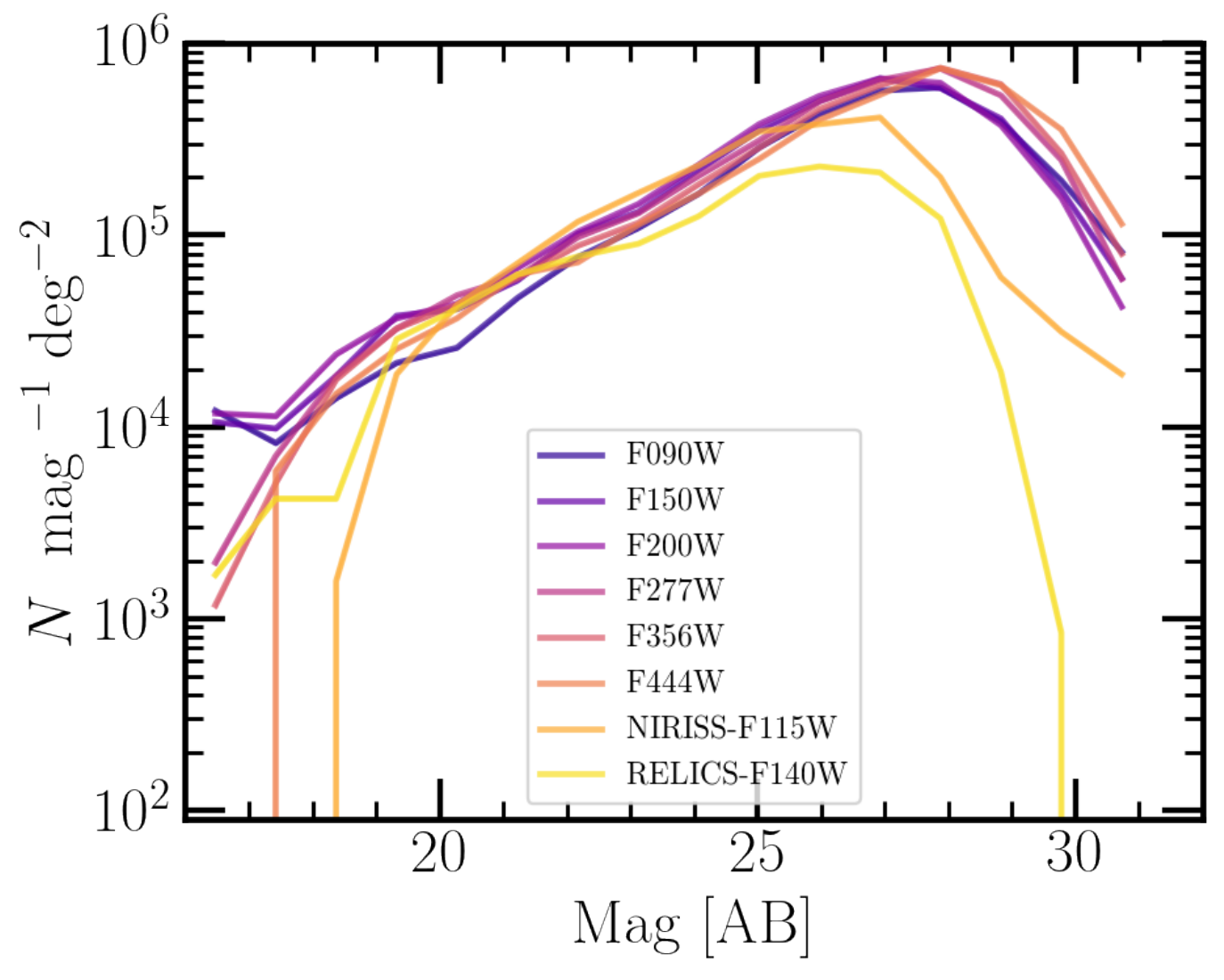}
    \caption{Magnitude number counts from the {\tt SE++} extraction in the seven NIRCam and NIRISS filters. Also over-plotted are the HST WFC3/IR F140W number counts from RELICS \citep{coe19,salmon20}. The plot illustrates the improvement in depth and the increase of the number of faint sources detected with JWST.}
    \label{fig:mag-number-counts}
\end{figure}

We carry out source detection and photometry measurements using {\tt SourceXtractor++} {\tt v0.17} \citep{bertin20,kummel20}. {\tt SourceXtractor++} ({\tt SE++}) is the successor of the classic and widely used {\sc SExtractor2}, rewritten in {\tt C++} in order to optimize computational efficiency and parallelization across multiple cores. The novelty and advantage of the code is its support for two-level detection thresholding, flexible multi-object and multi-frame model fitting as well as allowing for custom definition of models. Additionally, it can operate on multiple images based on their WCS information, bypassing the need to re-sample images on the same pixel scale. {\tt SE++} has been extensively tested in its performance on photometry and morphological parameter recovery from model fitting in the \textit{Euclid Morphology Challenge}, where it achieved the highest scores among all the tested codes \citep{HubertEMC2022, MerlinEMC2022}.
 
For the detection of sources, we use the ICL-subtracted LW stack (cf. section~\ref{sec:se++}) and its corresponding weight image as a weight map. The background map is computed on a grid with cell size of 128 pixels and smoothed with a box-filter of 5 pixels per side. Sources are detected via two-level thresholding which ensures the detection of faint and low surface brightness sources while minimizing false positives around bright sources. First, we detect all sources with at least 6 contiguous pixels and an integrated SNR threshold of 1.75. A second, higher threshold of 2.55 and 18 contiguous pixels is applied, whose main impact is to clear false positives around bright sources. Finally, a minimum area of 15 pixels is applied to clean spurious detections near bright sources and random noise peaks. Detected sources that are close enough such that they have connected pixels above the thresholds are grouped together and later fitted simultaneously.

Measurements are then carried out on all 6 NIRCam bands simultaneously. 
Additionally, we perform measurements on the NIRISS F115W-band separately because of its smaller size and shallower depth (cf. Tab.~\ref{tab:obs}), but using the same detection image. Errors in the photometry are estimated from an RMS map, obtained as the inverse of the square root of the weight image. The weight image contains the contribution of the sky, dark, flat and read noise to each pixel. As such, it considers the Poisson noise of the background, but not of the sources. Regardless, given the fact that of interest for this paper are the faintest sources, which are typically dominated by read and background noise, this approach is sufficient. The photometry for each source is obtained using \texttt{AUTO} apertures, as well as from model-fitting. The models and the priors on their parameters can be user-defined, and in our case we use a S\'ersic model \citep{sersic63} with the radius, S\'ersic index, angle, and axis ratio as free parameters. We find that this model performs well for the faint and small sources that we are interested in. The models are convolved with the point-spread-function (PSF) obtained using {\sc WebbPSF v1.0.0}. The PSF models are generated on a pixel scale of $0.03$ \arcsec/pix, on a stamp with a field-of-view of $\sim10$\arcsec\ (301 pixels).
% and rotated by an angle of $144.9$\,deg for NIRCam ($155.6$\,deg for NIRISS) to match the $x-y$ orientation of the final images.

For the purpose of this paper, we use the model magnitudes from the simultaneous fit of the NIRCam bands and the \texttt{AUTO} magnitudes for the separate NIRISS measurement. The resulting magnitude number counts are shown in Fig.~\ref{fig:mag-number-counts} for all bands used in this work. In order to check the photometric calibration of these early JWST observations, we conducted a comparison with well-calibrated HST images of the same cluster from RELICS. We measured fluxes in the HST/WFC3 F140W and in the JWST/NIRCam F150W images, using the same apertures and the HST mosaic as a detection image. The results are discussed in appendix~\ref{app:photometry-validation} and show a good agreement with a small residual offset. Fig.~\ref{fig:mag-number-counts} also shows consistent number counts and magnitude depths across the JWST filters. For comparison, we also show the number counts from the RELICS F140W-band.

\subsection{Dropout selection} \label{sec:dropout}

\begin{figure*}
     \centering
     \includegraphics[width=0.33\textwidth, keepaspectratio=true]{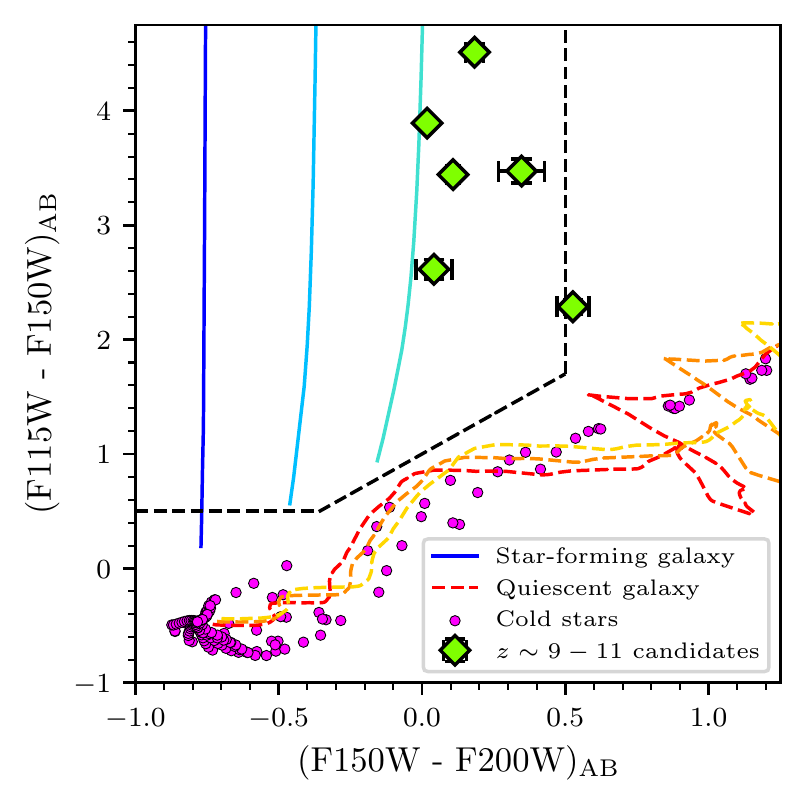}
     \includegraphics[width=0.33\textwidth, keepaspectratio=true]{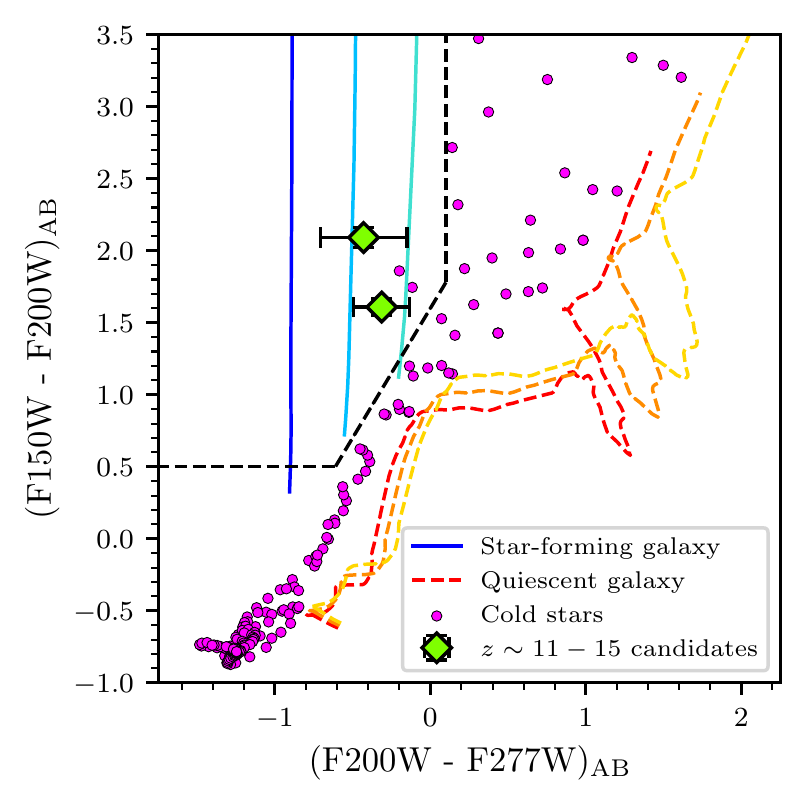}
     \includegraphics[width=0.33\textwidth, keepaspectratio=true]{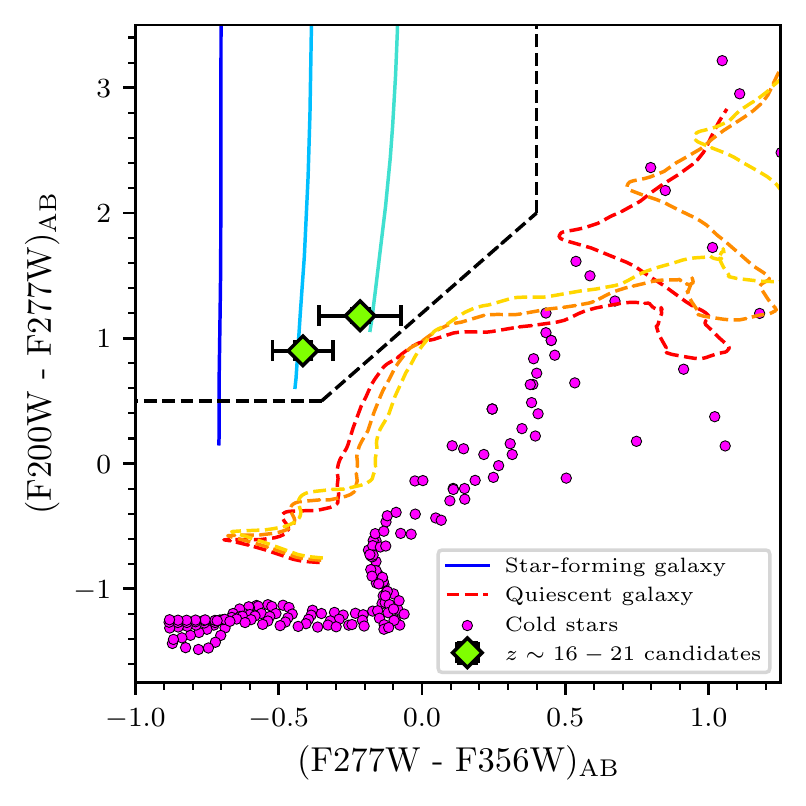}
     \caption{Color-color selection windows using the NIRCam broadband filters for high-redshift dropouts. {\em Left:} for $9<z<11$ {\em Middle:} for $11<z<15$; and {\em Right:} for $16<z<21$. The solid lines, from blue to turquoise, represent the color tracks of star-forming galaxy templates at high redshift generated with \texttt{BEAGLE} (cf. text for details). The dashed lines, from yellow to red, are quiescent galaxies, illustrating low-redshift galaxy contaminants, generated with \texttt{GRASIL} \citep{silva98}. For each star-forming template we applied three values of dust attenuation following the SMC dust law, $A_{V}$=[0,0.25,0.5], which are illustrated with the brighter blue or red colors in increasing order to illustrate the effect of dust attenuation on the color tracks. For the quiescent galaxies, we apply attenuation values in the range $A_{V}$=[1,2,3]. The purple circles represent the colors of a library of M-class stars and brown dwarfs \citep{chabrier00,allard01}, which also represent potential contaminants. The high-redshift dropouts selected in this work are represented as green diamonds.}
     \label{fig:color-color_selection}
\end{figure*}

\begin{figure*}
    \includegraphics[height=0.3\textwidth, keepaspectratio=true]{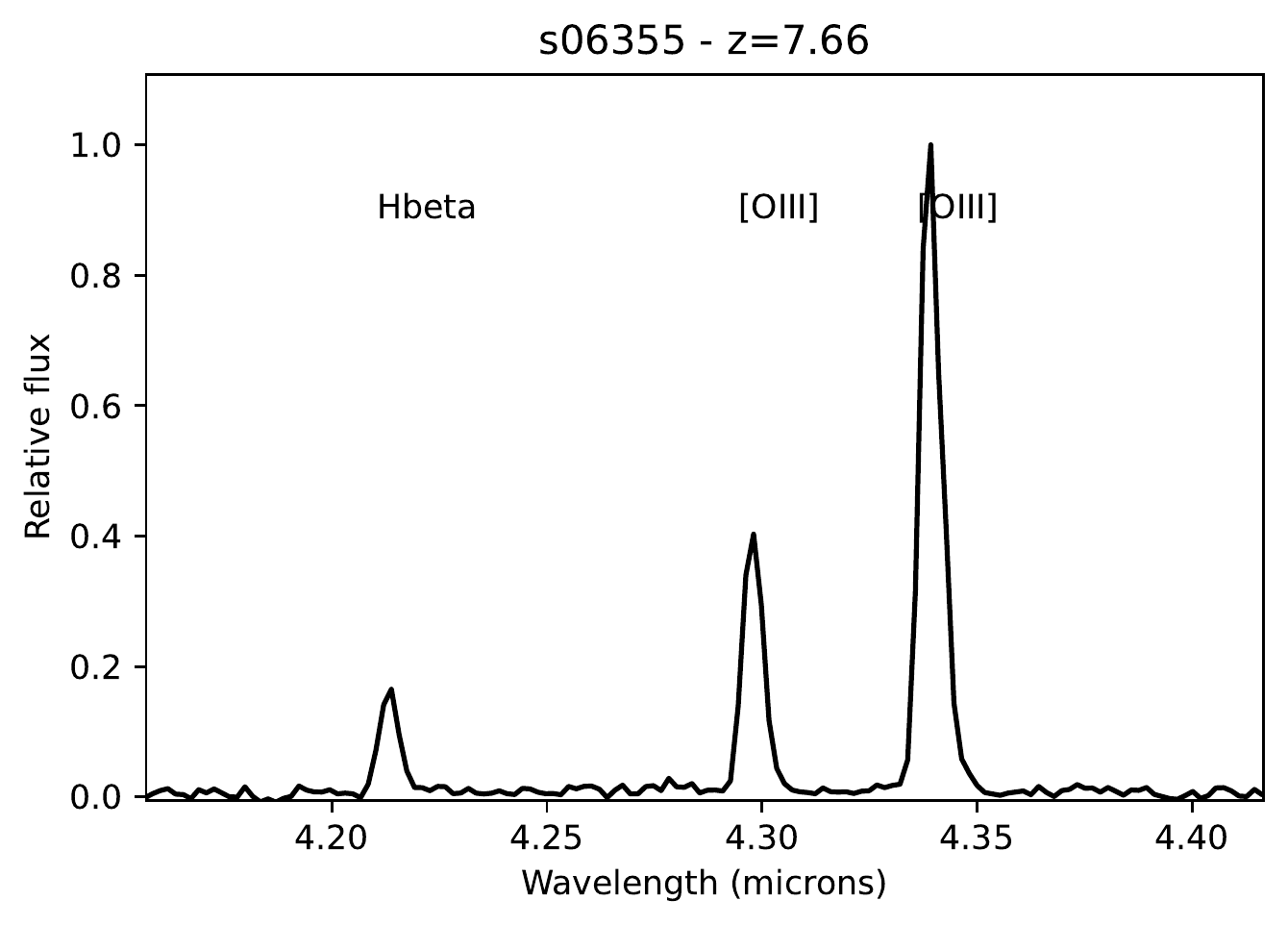}
     \includegraphics[height=0.3\textwidth, keepaspectratio=true]{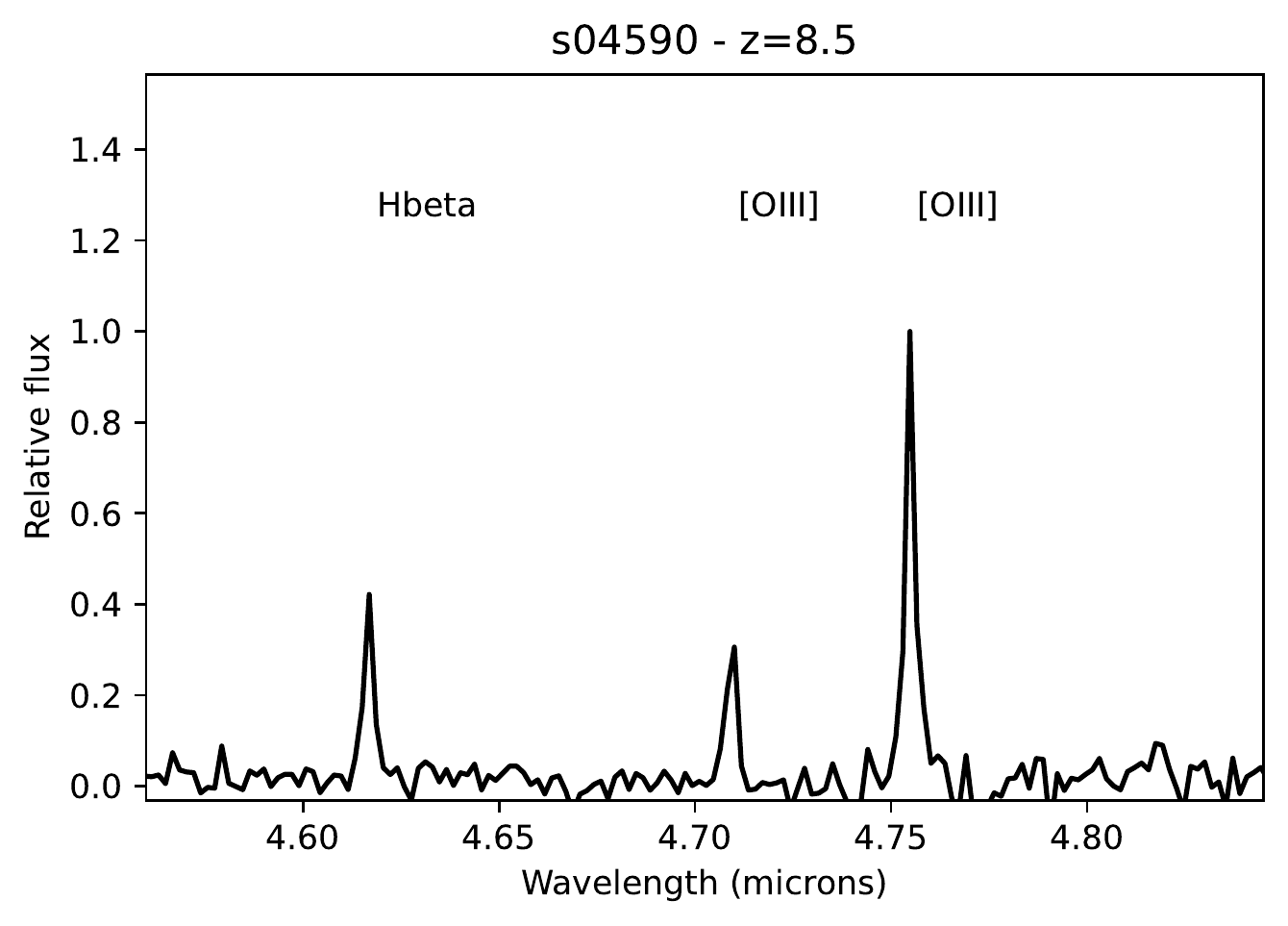}
    \caption{Examples of NIRSpec spectra using the G395M grism for 2 dropout galaxies at $z\sim7-9$, displaying very strong H$\beta$\ and [\ion{O}{iii}] emission lines.}
    \label{fig:nirspec}
\end{figure*}

We proceed to identify high-redshift galaxies using the Lyman-break, or dropout, selection technique \citep[e.g.][]{steidel96,atek15a,kawamata15,bouwens15}. In order to determine the best color-color selection criteria, we ran a set of simulations using different galaxy templates. We generated starburst galaxy templates using {\sc BEAGLE} \citep[][cf. section~\ref{sec:SED-fitting}]{chevallard16} at increasing redshifts from $z=6$ to $z=28$, including the intergalactic medium (IGM) attenuation from \citet{inoue14}. In addition, we explore a range of dust attenuation from $A_{V}=0$ to $A_{V}=0.5$ using the SMC extinction law \citep{pei92}, which has been found to well match the dust attenuation in high-redshift galaxies \citep[e.g.][]{capak15,reddy15,reddy18a}. We then computed synthetic photometry in the NIRCam broad-band filters and plotted the color tracks of these mock galaxies in Fig.~\ref{fig:color-color_selection}. In order to mitigate contamination from low-redshift interlopers that mimic the IGM absorption by their intrinsically red colors, we also explore the position of low-redshift quiescent galaxies, generated with \texttt{GRASIL} \citep{silva98} and available in the framework of the SWIRES template library \citep{polletta07}, on this color-color diagram. These again include several values of dust attenuation between $A_{V}=1$ and $A_{V}=3$. Finally, we include the last significant source of contamination, cold red stars and brown dwarfs, which can also mimic the Lyman break. We used the stellar templates from \citet{chabrier00} and \citet{allard01}. With this information in hand, we define the best color selection window that maximizes identification of the high-redshift sources while minimizing contamination for each redshift range. The resulting selection criteria are illustrated in Fig.~\ref{fig:color-color_selection}.
 
According to these simulations, we select galaxy candidates in the redshift ranges $z\sim9-11$ and $z\sim11-15$ respectively based on the following criteria:
 
\begin{equation*}
	\begin{array}{l}
		M_{115}-M_{150}>0.5\\
		M_{115}-M_{150}>1.0+1.4(M_{150}-M_{200})\\
		M_{150}-M_{200}<0.5
	\end{array}
\end{equation*}
 
\noindent and
 
 \begin{equation*}
	\begin{array}{l}
		M_{150}-M_{200}>0.5\\
		M_{150}-M_{200}>1.6+1.8(M_{200}-M_{277})\\
		M_{200}-M_{277}<0.2
	\end{array}
\end{equation*}
 
In addition to these color criteria, galaxy candidates must also satisfy a detection significance above 5$\sigma$ in the deep IR stacked image, and 4$\sigma$ in the individual detection filters. We also reject any candidate that shows a significant detection (at 2$\sigma$ level) in the stacked image that combines all filters blue-wards of the Lyman break. If a source is not detected in the dropout filter, we adopt the 1$\sigma$ lower limit measured in the same aperture to calculate the Lyman break.  
 
For $z\sim16-21$ candidates, we adopt the following criteria:

 \begin{equation*}
	\begin{array}{l}
		M_{200}-M_{277}>0.5\\
		M_{200}-M_{277}>1.2+2.0(M_{277}-M_{356})\\
		M_{277}-M_{356}<0.4
	\end{array}
\end{equation*}

Similarly to the $11<z<15$ selection, all candidates must satisfy the same detection and non-detection criteria detailed above. All candidates are visually inspected to check for spurious detection, artifacts, or bad PSF residuals. We also carefully check the candidates against other potential sources of contamination. For instance, very strong emission lines in intermediate-redshift ($z\sim3-5$) galaxies can contaminate the broadband flux and create a photometric dropout \citep[e.g.][]{atek11,debarros14}. In particular, strong [\ion{O}{iii}]$\lambda\lambda$4959,5007 + \hb\ lines are expected in high-redshift galaxies, \citep[e.g.][]{debarros19} which can lead to red colors. To address this issue, we perform SED fitting that takes nebular emission lines into account (cf. section~\ref{sec:SED-fitting}). Note also that most of the intermediate-redshift contaminants should be easily detected in the bluer filters, given the depth of the F090W and F150W filters and the observed hypothetical excess in the red filters. Finally, we discard all sources that have point-source morphologies in order to sieve out any stars that might scatter into our selection window. The colors of our selected galaxies are shown in Fig.~\ref{fig:color-color_selection} as green diamonds. Note that there are a few objects selected as high-redshift galaxies in the $z\sim11-15$ bin which do not completely satisfy all of the color-criteria. These objects were selected from their photometric redshift estimates with \texttt{EAZY} (cf. section~\ref{sec:SED-fitting}) because they have unambiguous high-redshift solutions and do not present morphologies consistent with stars.

In the present study, we have also analyzed the NIRSpec observations to investigate the spectroscopic properties of the high-redshift galaxies targeted by this program. Two examples at $z_{\mathrm{spec}}=7.66$ and $z_{\mathrm{spec}}=8.5$ are shown in Fig.~\ref{fig:nirspec}. These are the first rest-frame optical spectra observed for high-redshift ($z>6$) galaxies \citep[cf. also][]{schaerer22}. They show strong \hb\ and [\oiii] emission lines. The rest-frame equivalent width (EW) of these lines reaches up to $\sim400$\AA, which is still a lower limit because the continuum is hardly detected. This means that the flux contribution of emission lines to the broadband flux can reach up to $\sim30$\,\% in some galaxies.
 
Another consequence of the strong lensing is that some background sources will have multiple images. Using the most recent available parametric lensing model (cf. section~\ref{sec:SL}), we have not identified any possible multi-image systems. The magnification factors of the candidates range from $\mu\sim1$ to $\sim 4$.

\section{Lensing Model} \label{sec:SL}

\begin{figure}
    \centering
    \includegraphics[width=\columnwidth, keepaspectratio=true]{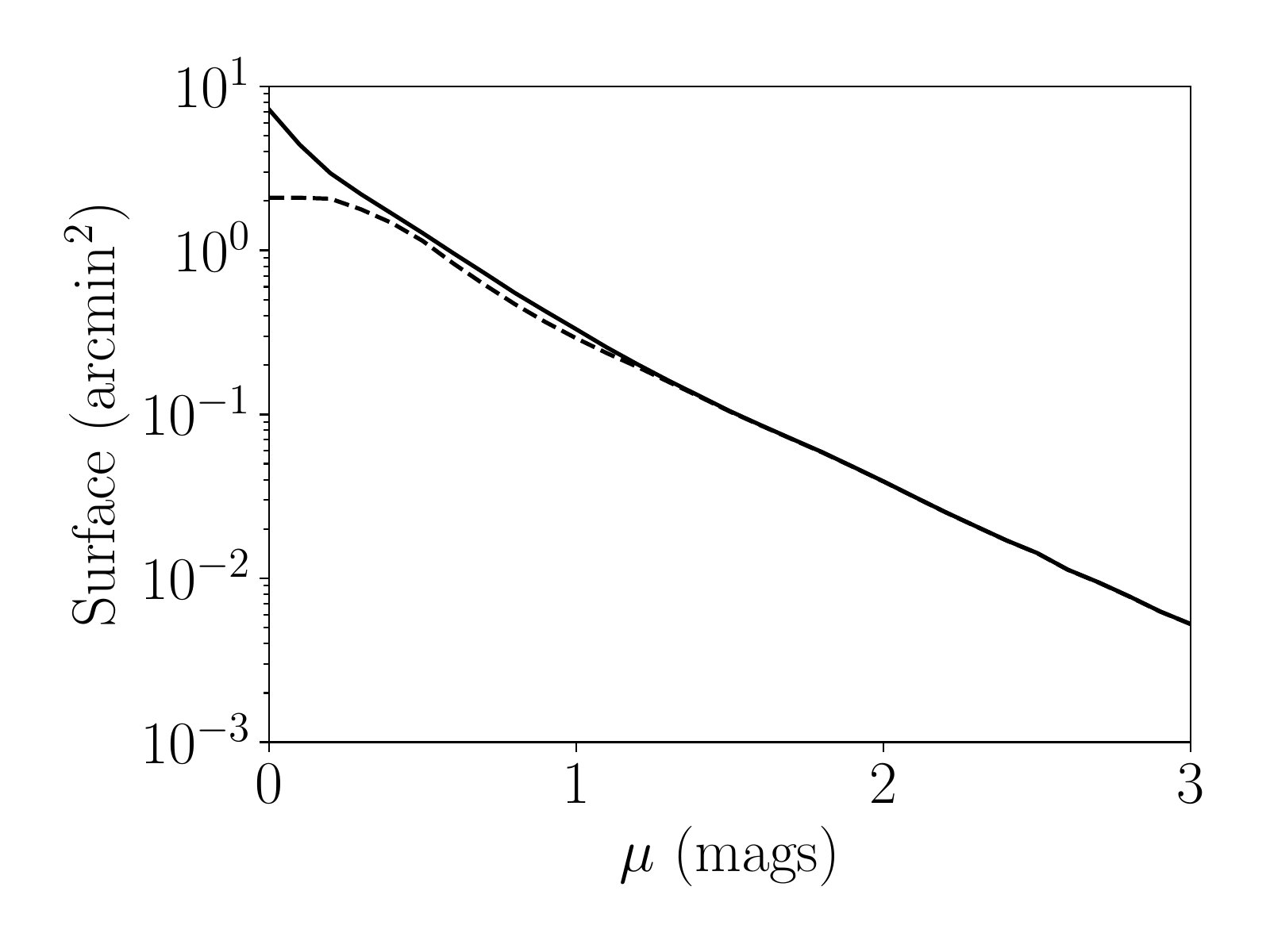}
    \caption{The cumulative surface area behind SMACS0723 at $z=9$ as a function of gravitational magnification $\mu$ (cf. section~\ref{sec:SL}), expressed in magnitudes. The solid line shows the area corresponding to the NIRCam field of view, whereas the dashed line is for the NIRISS pointing. The un-lensed total survey area is 7.23\,arcmin$^{2}$, and 2.1\,arcmin$^{2}$ for the NIRCam and NIRISS observations respectively. For reference, the observed area is $\sim 10.8$ arcmin$^{2}$ and $\sim 4.9$ arcmin$^{2}$ for NIRCam and NIRISS, respectively.}
    \label{fig:SL_surface-mag}
\end{figure}

In order to compute the gravitational magnifications of our objects, we use the most recent strong lensing mass model by \citet{mahler22}. The model was built using the  \texttt{lenstool} \citep{kneib96,jullo07,jullo09} software to optimize a parametric SL model for the mass distribution of the cluster. Following the approach used by \citet{richard14} for the HFF clusters \citep{lotz17} and \citet{fox} for the RELICS clusters, this model uses a combination of double Pseudo Isothermal Elliptical potentials \citep[dPIE;][]{eliasdottir07} for the total mass distribution at cluster- and galaxy-scales. The parameters of the galaxy-scale dPIE potentials are matched to the identification of cluster members on the red sequence from the RELICS HST catalogs (cf. \citealt{coe19} for details), fixing the center and elliptical parameters of the potential to its isophotes. The mass parameters of the dPIE profile (velocity dispersion $\sigma$, core radius r$_{\rm core}$, cut radius r$_{\rm cut}$) are scaled according to the luminosity L$^*$ of a reference galaxy using a constant mass-to-light ratio.

The SL constraints used in \citet{mahler22} are based on the released JWST/NIRCam images (cf. section~\ref{sec:obs}), and make use of spectroscopic redshifts measured with public data from the \textit{Multi Unit Spectroscopic Explorer} \citep[MUSE;][]{bacon10} on ESO's \textit{Very Large Telescope} (VLT) by \citet{golubchik} and with JWST/NIRSpec data in \citet{mahler22}. In summary, 48 secured images of 16 separate sources were identified, including 5 spectroscopic redshifts. The \citet{mahler22} SL model reproduces all systems with a residual lens plane RMS of $\sim0.7\arcsec$ between the predicted and observed locations of all images.

The cumulated surface area of magnification according to the model is shown in Fig.~\ref{fig:SL_surface-mag}. We make use of this model to compute the magnification factors and associated errors on each high redshift galaxy candidates. All magnifications assume $z=15$ as there is only very little variation of magnification with redshift at $z>7$. \texttt{lenstool} allows us to compute a statistical error on the magnification by sampling the posterior probability distribution of each parameter of the model.

\section{Spectral Energy Distribution Fitting} \label{sec:SED-fitting}
We continue our selection of high-redshift galaxy candidates by computing photometric redshifts by fitting galaxy spectral energy distributions (SEDs) to the multi-wavelength photometry.

The photometric redshifts are obtained with \texttt{EAZY} \citep{Brammer2008}, a Python-based SED fitting code. \texttt{EAZY} fits a non-negative linear combination of a set of basis templates to the observed flux densities for each galaxy. The fit is performed on the flux densities obtained from the model-fitting in {\tt SE++} (cf. section~\ref{sec:se++}), converted to units of $\mu$Jy, and corrected for Milky Way extinction internally in the code. Sources with valid flux density measurements available in at least three bands are considered for fitting. We use the template set \texttt{"tweak\_fsps\_QSF\_1\_v3.param"} that is derived from the Flexible Stellar Population Synthesis models (FSPS) \citep{Conroy2009, ConroyGunn2010}. The allowed redshift range is set to $0.01<z<22$, with a flat prior on the luminosity. Sources of interest are then selected by visually inspecting all sources with a best-fit $z_{\mathrm{phot}}>9$. In a second selection step, candidates are required to show a reasonably good SED fit.

In addition, we simultaneously infer physical parameters and photometric redshifts of the selected high-redshift candidates by fitting galaxy SEDs to the JWST photometry with the \texttt{BayEsian Analysis of GaLaxy sEds} tool \citep[\texttt{BEAGLE};][]{chevallard16}. This is done to refine our selection and obtain accurate uncertainty estimates on the physical parameters of the galaxies. Because the inclusion of nebular emission has been found to be crucial in fitting broad-band photometry of galaxies \citep{schaerer09,schaerer10,atek11,atek14c,mclure11,smit14,duncan14,reddy18b}, \texttt{BEAGLE} uses galaxy templates computed by \citet{gutkin16} which combine the stellar population templates by \citet{Bruzual_Charlot_2003} with the photoionization code \texttt{CLOUDY} \citep{ferland13}. It then accounts for IGM attenuation using the \citet{inoue14} models and applies a dust attenuation law (cf. section~\ref{sec:BEAGLE-setup} for details) to the galaxy emission. The \texttt{BEAGLE} tool is fully Bayesian, i.e. performs a Monte-Carlo Markov Chain (MCMC) analysis of the SED, which makes it ideally suited to determine both photometric redshifts and several physical parameters at the same time while robustly quantifying and combining their uncertainties. 

\subsection{\texttt{BEAGLE} set-up} \label{sec:BEAGLE-setup}
For our SED-fit with \texttt{BEAGLE}, we assume a constant star-formation history (SFH). While this is a simplification, it is a common assumption made in high-redshift galaxy analyses \citep[e.g.][]{eyles07,stark09,gonzales11,grazian15,kikuchihara20} and the results have been shown to not strongly differ from more flexible analytic SFH assumptions such as e.g. a delayed exponential SFH \citep{furtak21}. In order to avoid over-fitting due to the relatively low number of photometric bands available for this work, we limit the fit-parameters to four:

\begin{itemize}
    \item Photometric redshift $z_{\mathrm{phot}}$ both fixed to the \texttt{EAZY} result first and then with a uniform prior in the limits of $z_{\mathrm{phot}}\in[0, 25]$ for confirming the robustness of the photometric redshift estimate.
    \item Stellar mass $M_{\star}$ with a log-uniform prior in the limits of $\log(M_{\star}/\mathrm{M}_{\odot})\in[6, 10]$.
    \item Maximum stellar age $t_{\mathrm{age}}$ with a log-uniform prior in the limits of $\log(t_{\mathrm{age}}/\mathrm{yr})\in[7, t_{\mathrm{universe}}]$ where $t_{\mathrm{universe}}$ is the age of the Universe at the redshift of the galaxy.
    \item Effective \textit{V}-band dust attenuation optical depth $\hat{\tau}_V$ with a uniform prior in the limits of $\hat{\tau}_V\in[0, 0.5]$.
\end{itemize}

We furthermore fix the stellar metallicity to a constant value of $Z=0.1\,\mathrm{Z}_{\odot}$ since it has been shown that the broad-band photometry of high-redshift dropout galaxies is not sensitive to the metallicity in the very low metallicity regime expected for high-redshift galaxies \citep{furtak21}. Finally, we adopt an SMC dust extinction law \citep{pei92} which has been shown to match the observations of high-redshift galaxies best \citep{capak15,reddy15,reddy18a}, in particular in the low-metallicity range that we are probing \citep{shivaei20}.

The SED fit is run on all six broad-band filters without taking the gravitational magnification into account yet, in order to avoid including the uncertainties of magnification in the SED fit. Instead, we correct the stellar mass for magnification after the fit following our approach in \citet{furtak21}. Note though that the magnifications of our objects are not very high, $\mu\sim1-4$.

\subsection{Results} \label{sec:results}

\begin{table*}
\centering
\caption{Complete list of our high-redshift candidates identified behind SMACS0723 with their derived parameters.}
\hspace{-0.0cm}
\begin{tabular}{l|c|c|c|c|c|c|c|c|c}
ID & RA & Dec & $z_{\mathrm{phot}}$ & $M_{\mathrm{UV}}$ & $\beta$ & $\log(M_{\star}/\mathrm{M}_{\odot})$ & $\log(t_{\mathrm{age}}/\mathrm{yr})$ & $n$ & $r_{\rm e}$ [kpc]\\\hline

\multicolumn{10}{c}{$z\sim9-11$ candidates}\\\hline 

SMACS\_z10a & 7:23:26.252 & -73:26:56.940 & $ 9.78 ^{+0.02}_{- 0.02}$   & $-18.77 \pm 0.20$ & $-1.72\pm$0.04 & $ 9.11 ^{+0.07}_{- 0.07}$ & $ 8.68 ^{+0.2}_{- 0.2}$& $<1$ & $0.22$\\
SMACS\_z10b & 7:23:22.709 & -73:26:06.183 & $ 8.88 ^{+0.02}_{- 0.02}$   & $-20.78 \pm 0.13$ & $-1.36\pm$0.19 & $10.20 ^{+0.03}_{- 0.03}$ & $ 8.74 ^{+0.2}_{- 0.2}$& $<1$ & $0.65$\\
SMACS\_z10c & 7:23:20.169 & -73:26:04.233 & $ 9.77 ^{+0.02}_{- 0.02}$   & $-20.19 \pm 0.15$ & $-2.14\pm$0.12 & $ 9.53 ^{+0.02}_{- 0.02}$ & $ 8.68 ^{+0.2}_{- 0.2}$& $<1$ & $0.41$\\
SMACS\_z10d & 7:22:46.696 & -73:28:40.898 & $ 9.31 ^{+0.06}_{- 0.08}$   & $-19.76 \pm 0.18$ & $-2.22\pm$0.17 & $ 7.77 ^{+0.14}_{- 0.11}$ & $ 7.29 ^{+0.18}_{- 0.12}$& $<1$ & $0.55$\\
SMACS\_z10e  & 7:22:45.304 & -73:29:30.557 &$ 10.89 ^{+0.16}_{- 0.14}$  & $-18.91 \pm 0.26$ & $-2.03\pm$0.19 & $ 8.51 ^{+0.22}_{- 0.16}$ & $ 7.49 ^{+0.26}_{- 0.20}$& $<1$ & $0.33$\\
SMACS\_z11a  & 7:22:39.505 & -73:29:40.224 &$ 11.05 ^{+0.09}_{- 0.08}$  & $-18.55 \pm 0.38$ & $-1.97\pm$0.38 & $ 8.77 ^{+0.17}_{- 0.24}$ & $ 8.20 ^{+0.21}_{- 0.32}$& $<1$ & $0.39$\\\hline
 \multicolumn{10}{c}{$z\sim11-15$ candidates}\\\hline
%SMACS\_z11b & 7:22:53.848 & -73:28:23.348 & $ 11.07 ^{+0.15}_{- 0.18}$ & $-19.92 \pm 0.21$& $-2.35\pm$0.54 & $ 8.91 ^{+0.30}_{- 0.22}$ & $ 7.49 ^{+0.36}_{- 0.26}$ & $4.4\pm 0.5$ & $7.96$\\
SMACS\_z12a   & 7:22:47.380 & -73:30:01.785&$ 12.20 ^{+0.21}_{- 0.12}$ & $-19.75 \pm 0.23$& $-2.69\pm$0.16 & $ 8.14 ^{+0.21}_{- 0.17}$ & $ 7.61 ^{+0.26}_{- 0.20}$ & $<1$ & $0.47$\\
SMACS\_z12b & 7:22:52.261 & -73:27:55.497 & $ 12.26 ^{+0.17}_{- 0.16}$ & $-20.01 \pm 0.17$& $-2.82\pm$0.21 & $ 7.91 ^{+0.26}_{- 0.17}$ & $ 7.56 ^{+0.30}_{- 0.23}$ & $4.0\pm 1.0$ & $1.99$\\\hline
\multicolumn{10}{c}{$z>15$ candidates}\\\hline
SMACS\_z16a & 7:23:26.393 & -73:28:04.561 &  $ 15.92 ^{+0.17}_{- 0.15}$ & $-20.59 \pm 0.15$ & $-2.63\pm$0.13 & $ 8.79 ^{+0.32}_{- 0.33}$ & $ 7.65 ^{+0.36}_{- 0.39}$ & $1.1\pm 0.2$ & $0.40$\\
SMACS\_z16b   & 7:22:39.439 & -73:30:08.185 &$ 15.32 ^{+0.16}_{- 0.13}$ & $-20.96 \pm 0.14$ & $-2.40\pm$0.34 & $ 8.80 ^{+0.44}_{- 0.25}$ & $ 7.46 ^{+0.52}_{- 0.35}$ & $2.8\pm 0.6$ & $1.17$\\\hline

\end{tabular}
\label{tab:sample}
\end{table*}

\begin{figure*}
    \centering
    \includegraphics[width=0.99\textwidth]{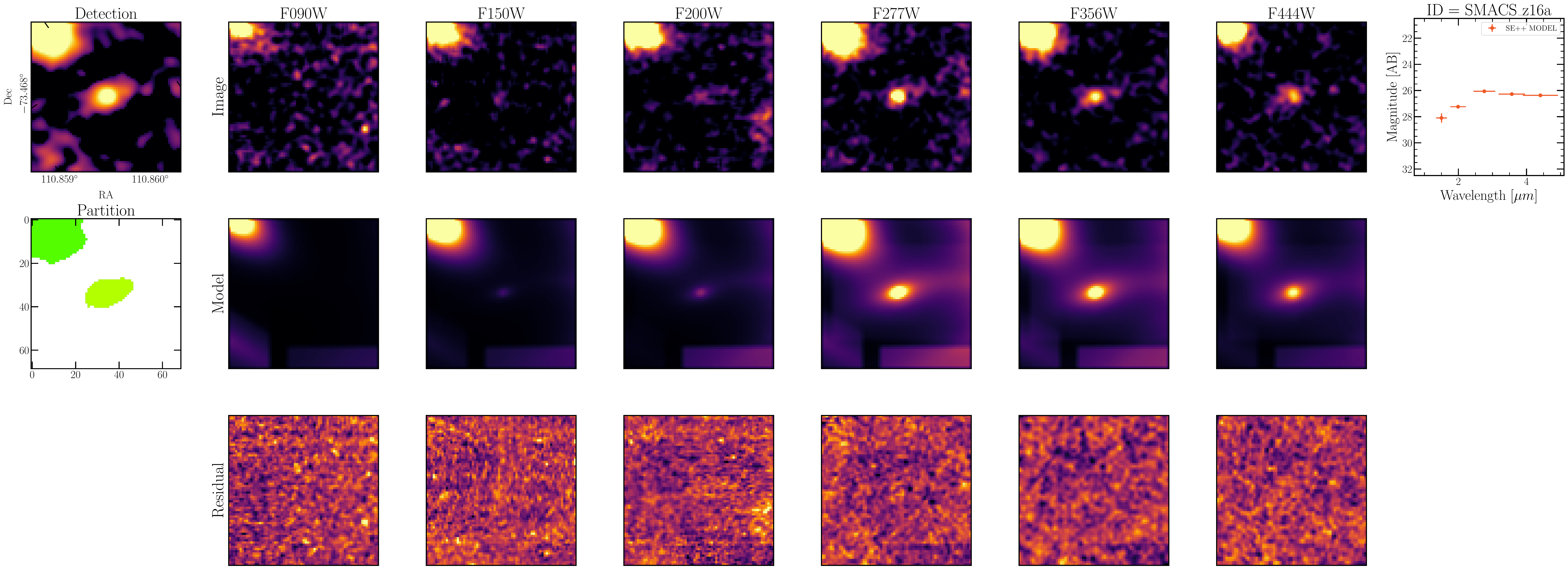}
    \caption{Cutout images of SMACS\_z16a in the JWST bands centered on the source. The left-most column shows the LW-stack detection image (cf. section~\ref{sec:obs}) and the source partition map. The following columns show the seven NIRCam and NIRISS science images in the top rows, best-fit model image in the middle rows and the residual in the bottom rows. The science and model images are scaled using a linear-stretch normalization, with minimum value equal to zero and maximum value 10 times the $2\sigma$-clipped standard deviation. The residuals are scaled linearly between $\pm7$ times the $2\sigma$-clipped standard deviation.
    }
    \label{fig:candidate-stamp-id-2455}
\end{figure*}

\begin{figure*}
    \centering
    \includegraphics[width=0.49\textwidth]{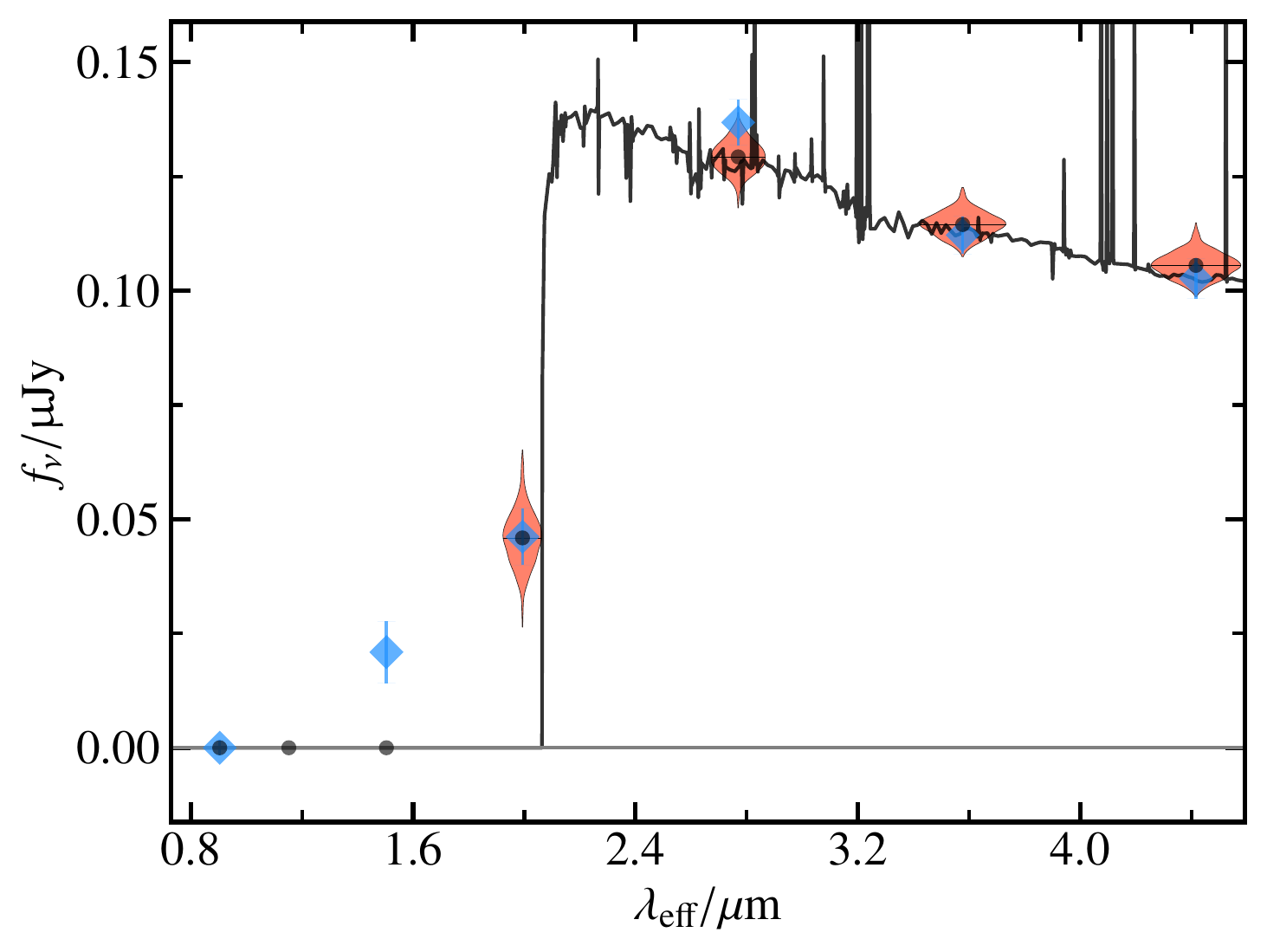}
    \includegraphics[width=0.49\textwidth]{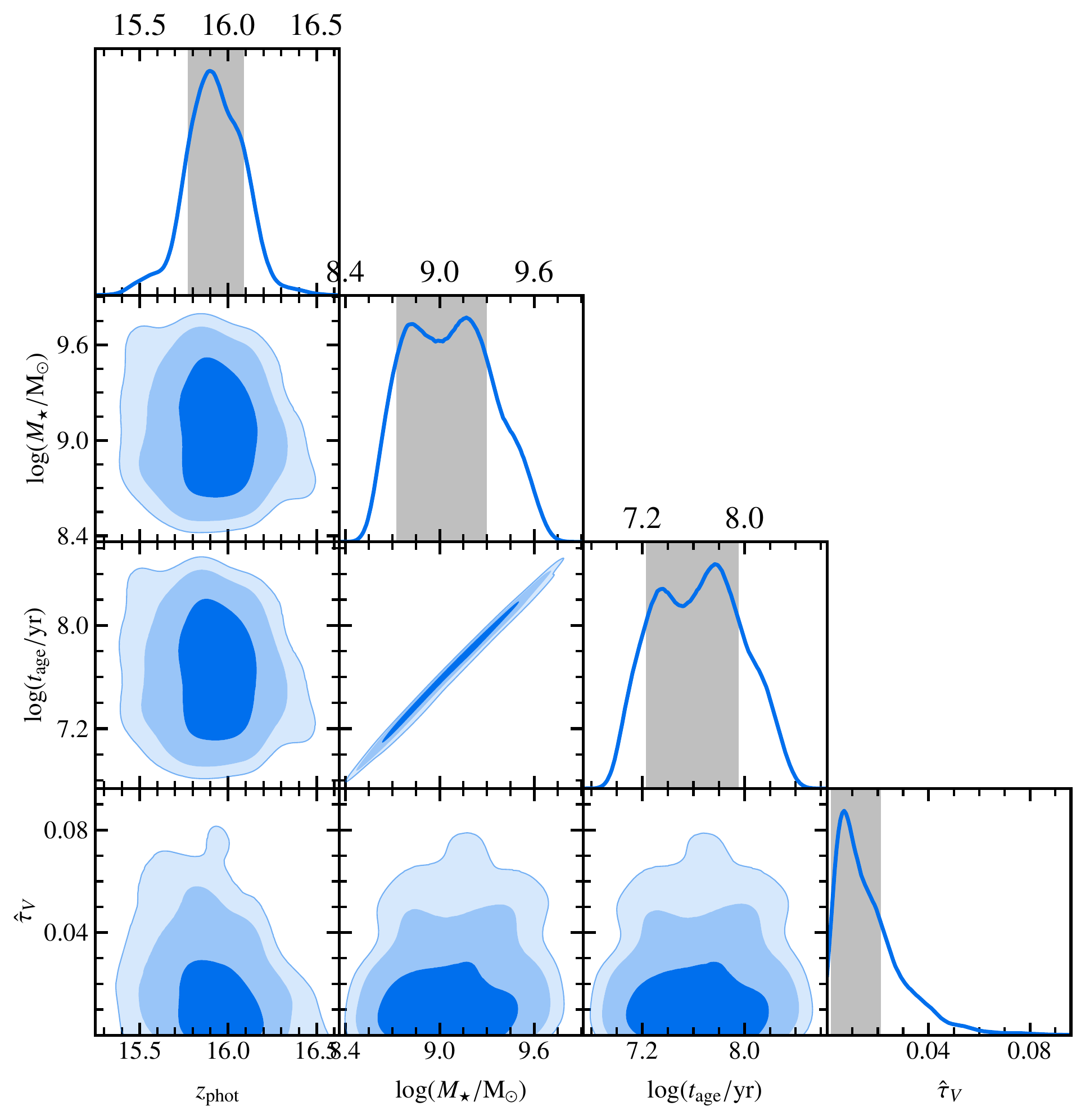}\\[1cm]
    \includegraphics[width=0.9\textwidth]{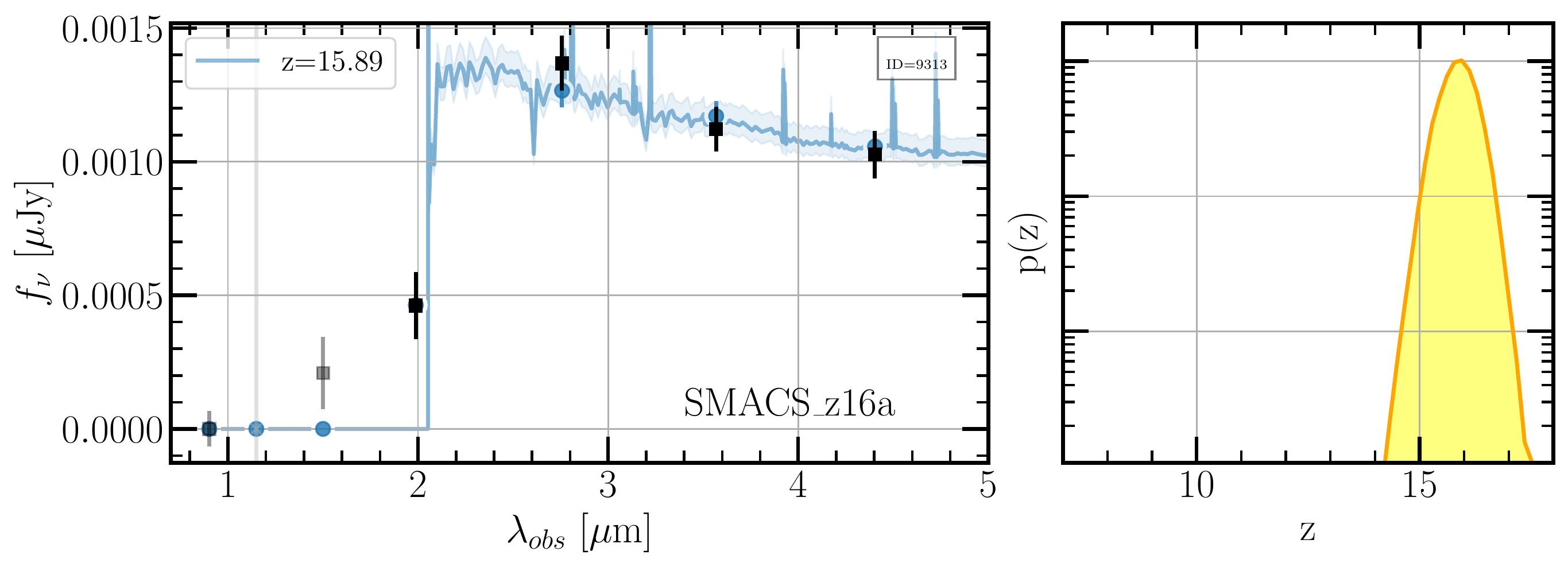}
    \caption{Best-fit solution for the SED and photometric redshift of SMACS\_z16a. {\em \bf Upper row:} Best-fit SED using the {\tt BEAGLE} code. {\em Left:} Best-fit SED (black solid curve) with the observed photometric data (blue points) and expected model photometric points (black points) and associated uncertainties (pink areas). {\em Right:} Triangle plot of the posterior probability distribution of the four fitted galaxy parameters: redshift, stellar mass, stellar age and attenuation. {\bf Bottom row:} Best-fit SED using the \texttt{EAZY} code. {\em Left panel:} The best-fit SED over-plotted over the observed flux densities (in dark squares). Model flux densities are shown in blue circles. The Lyman-break of the SED of this galaxy is estimated at $z=15.88$ and the redshift probability distribution function is shown in the {\em right} panel. Both codes agree on a high-redshift solution with a relatively narrow posterior distribution and which does not show a secondary peak at lower redshift.}
    \label{fig:SEF-fit-id-2455}
\end{figure*}

\begin{figure*}
    \centering
    \includegraphics[width=\textwidth]{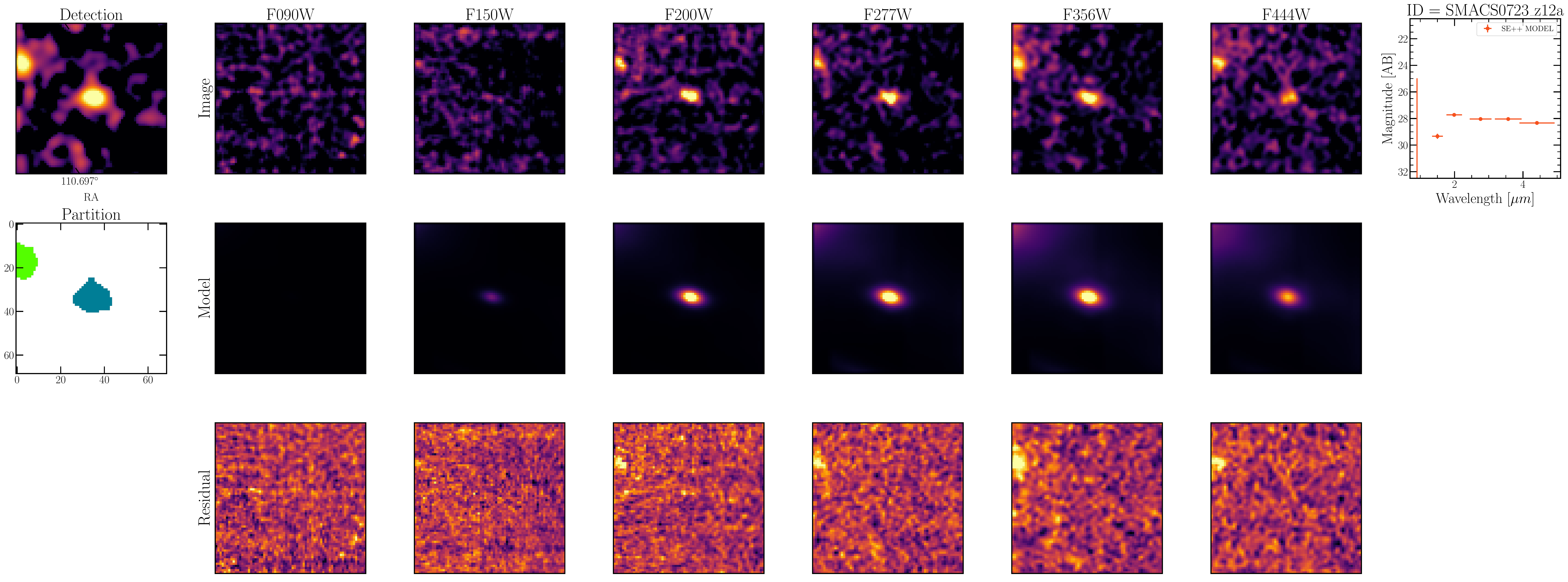}
    \caption{Same as Fig.~\ref{fig:candidate-stamp-id-2455}, this time for the $z\sim12$ candidate galaxy SMACS\_z12a.}
    \label{fig:candidate-stamp-id-93}
\end{figure*}

\begin{figure*}
    \centering
    \includegraphics[width=0.49\textwidth]{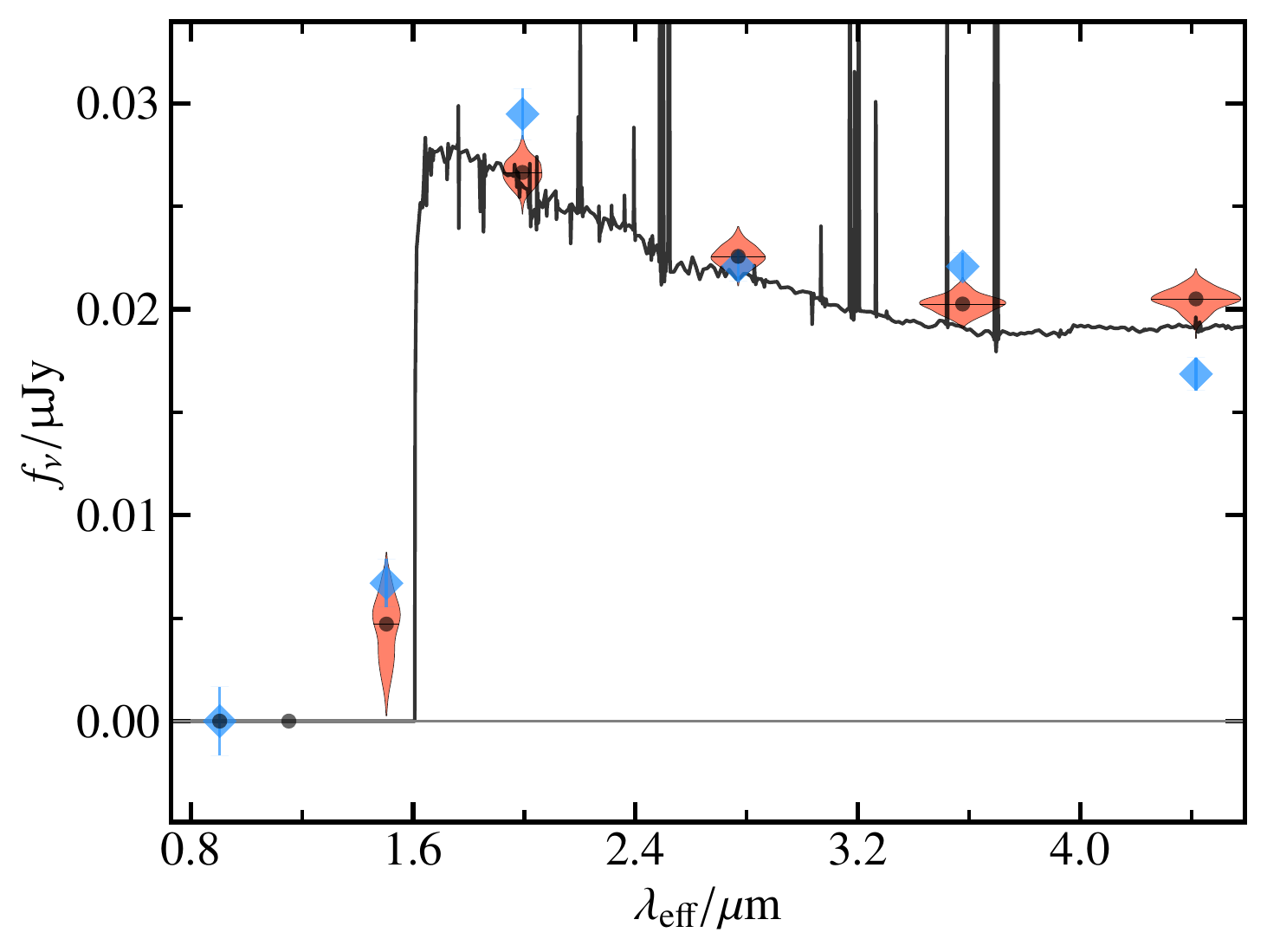}
    \includegraphics[width=0.49\textwidth]{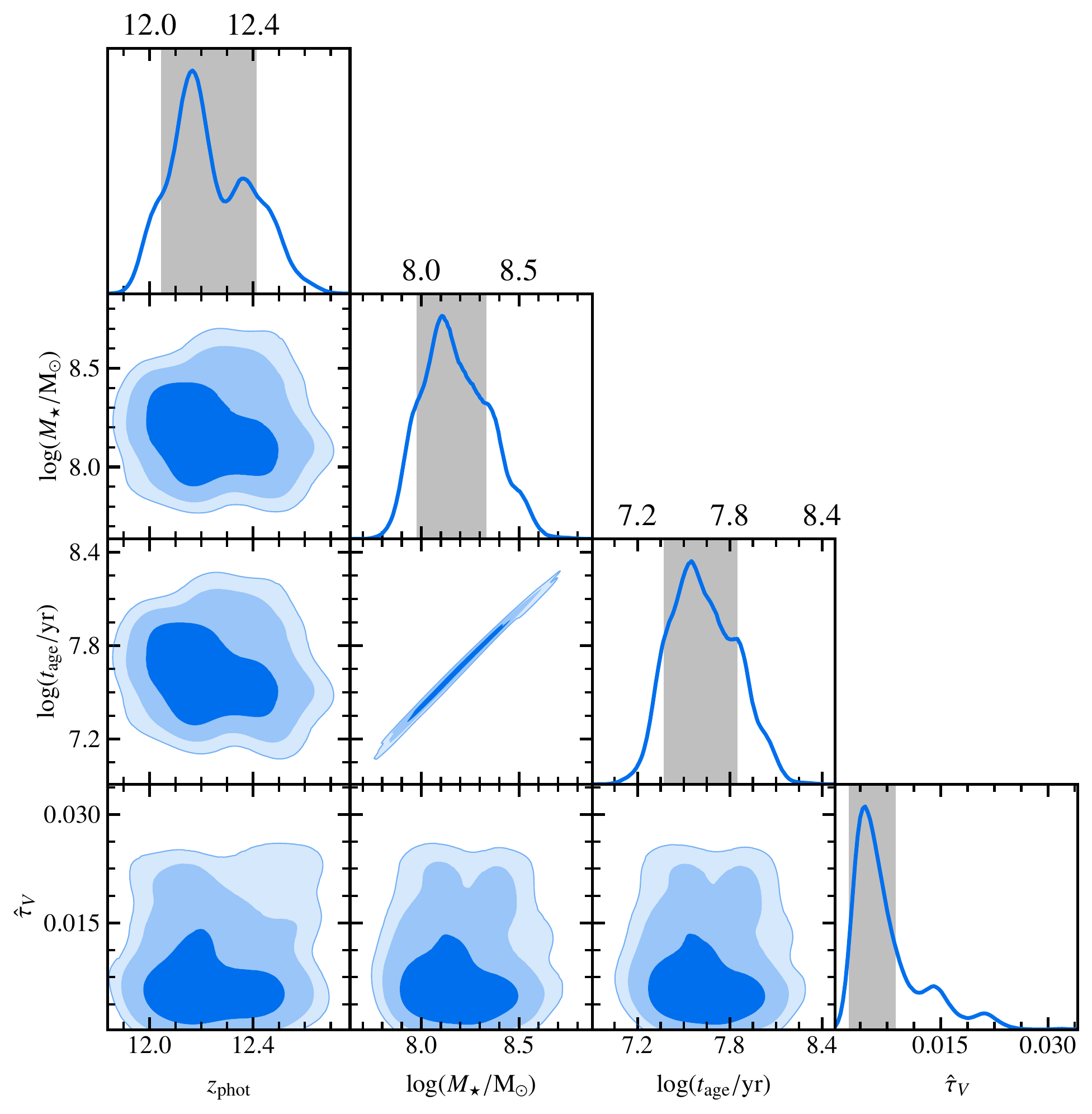}\\[1cm]
    \includegraphics[width=0.9\textwidth]{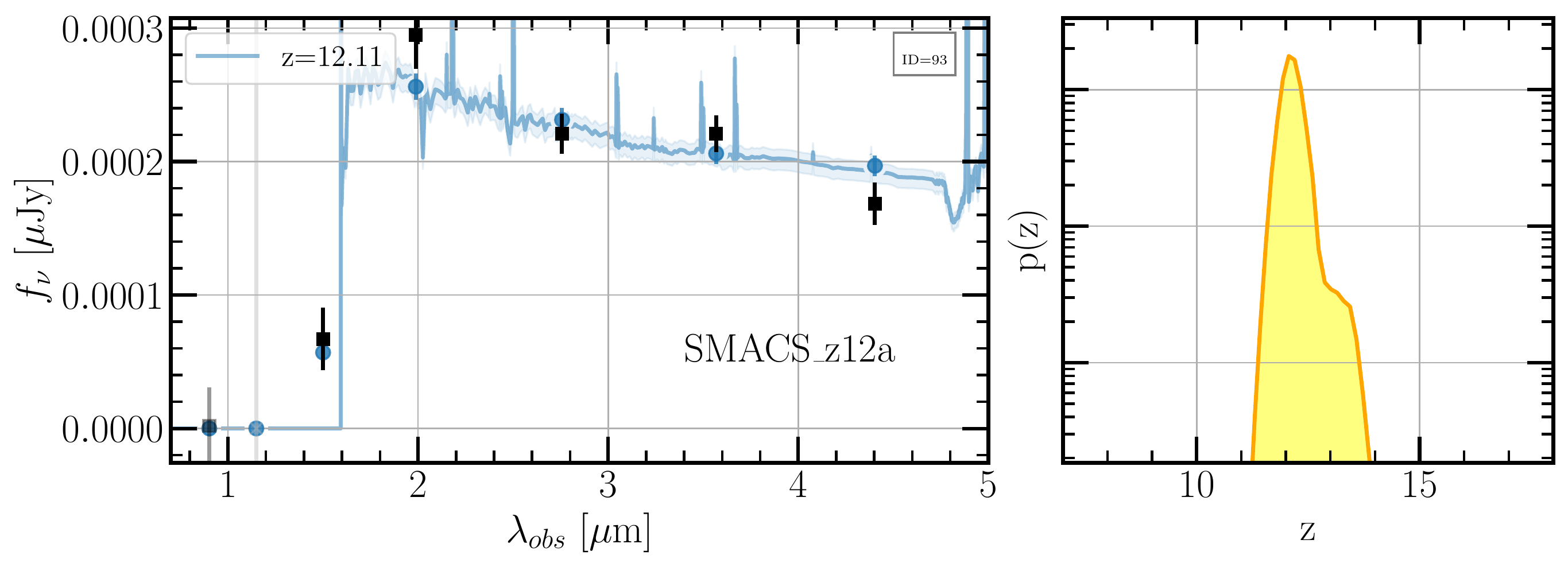}
    \caption{Same as Fig.~\ref{fig:SEF-fit-id-2455}, here showing the $z\sim12$ galaxy candidate SMACS\_z12a.}
    \label{fig:SEF-fit-id-93}
\end{figure*}

\begin{figure}
     \centering
     \includegraphics[width=0.95\linewidth]{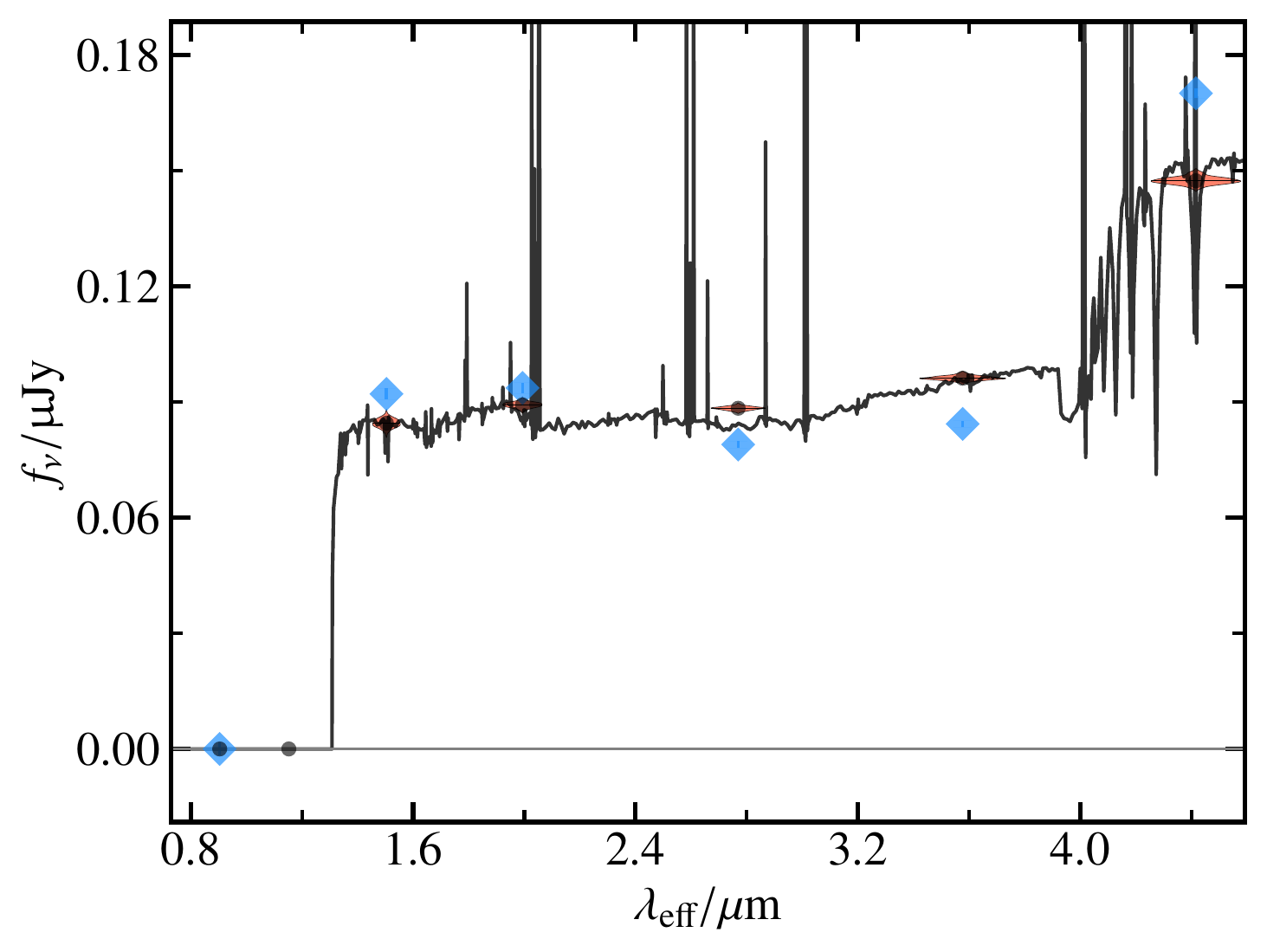}
     \caption{Example of a galaxy candidate, SMACS\_z10c, at $z\simeq9.76\pm0.01$ showing an excess in the F444W-band. The best-fit SED (same legend as Fig.~\ref{fig:SEF-fit-id-2455}) shows the presence of a Balmer break around 4.5\,\micron.}
     \label{fig:balmer}
\end{figure}

The final sample of high-redshift candidates consists of a total of 6 galaxies in the redshift range $9<z<11$, two galaxies at $11<z<15$, and two galaxies at $z>15$. All sources are new identifications, since \citet{salmon20} only reported galaxies below $z\sim8$ in this field. The complete results for our full sample are reported in Tab.~\ref{tab:sample}.

Our highest redshift galaxy, SMACS\_z16a, is consistently identified by the dropout selection and the two SED fitting procedures, as shown in Fig.~\ref{fig:SEF-fit-id-2455}. The photometric redshifts derived by both codes (cf. section~\ref{sec:SED-fitting}) are in excellent agreement: $z=15.92^{+0.17}_{-0.15}$ and $z=15.89\pm0.43$ for {\tt BEAGLE} and {\tt EAZY} respectively. This places the galaxy just $\sim250$\,Myr after the Big Bang. The galaxy exhibits an extremely blue UV continuum slope of $\beta =-2.63\pm 0.13$ which is consistent with a young and low-attenuation galaxy (cf. Fig. \ref{fig:SEF-fit-id-2455}). We note in general that no template in {\tt EAZY} is able to match the very blue UV continuum, whereas {\tt BEAGLE} allows for more flexibility in this range of galaxy templates. To obtain a more accurate estimate of the physical properties of our galaxies, we ran a second round of SED fitting with {\tt BEAGLE} using a prior in redshift from the PDF of {\tt EAZY} (cf. section~\ref{sec:BEAGLE-setup}). We derive a stellar mass of $\log(M_{\star}/\mathrm{M}_{\odot})=8.79^{+0.32}_{-0.33}$ and a stellar age of $\log(t_{\mathrm{age}}/\mathrm{yr})=7.65^{+0.36}_{-0.39}$ for SMACS\_z16a. This galaxy is also very compact, as can be seen in Fig.~\ref{fig:candidate-stamp-id-2455}, with an estimated size of $\sim0.4$\,kpc based on the model fitting of {\tt SE++} (cf. section~\ref{sec:se++}) and after correcting for the lensing distortion.  

In the next lower redshift range, one of the two most robust candidates, SMACS\_z12a, is presented in Fig.~\ref{fig:candidate-stamp-id-93}. Again, the source is well detected in all four filters and shows a clear continuum break. As before, there is a good agreement between the two photometric redshifts $z=12.20 ^{+0.21}_{-0.12}$ (\texttt{BEAGLE}) and $z=12.11 \pm 0.19$ (\texttt{EAZY}). The PDF from {\tt EAZY} is broader than the \texttt{BEAGLE} one, because of the challenge in matching the blue colors ($\beta \sim -2.60 \pm 0.32$) of this object (Fig.~\ref{fig:SEF-fit-id-93}). This source is also among the highest-redshift candidates identified photometrically in the literature \citep[e.g.][]{harikane22,naidu22,castellano22}, around 360\,Myr after the Big Bang. The derived parameters are very similar to SMACS\_z16a, with a stellar mass of $\log(M_{\star}/\mathrm{M}_{\odot})=8.14^{+0.21}_{-0.17}$, and an age of $\log(t_{\mathrm{age}}/\mathrm{yr})=7.61^{+0.21}_{-0.20}$.

In the lowest redshift range, $9<z<11$, some sources around $z\sim10$ show indication of a Balmer break as a clear excess in the F444W-band. An example is shown in Fig.~\ref{fig:balmer}, where flux excess is observed relative to a flat continuum in the bluer bands. This could indicate that an evolved stellar population was already in place at those early epochs, which is confirmed by the best-fit SED with a stellar age around $\sim400$\,Myr. An example for our $9<z<11$ objects is shown in appendix~\ref{app:z-10_example}. Of course, a stronger contribution from rest-frame optical emission lines is also possible, although this alternative does not yield the best-fit solution.

\section{Discussion} \label{sec:discussion}

\begin{figure*}
    \centering
    \includegraphics[width=0.77\textwidth, keepaspectratio=true]{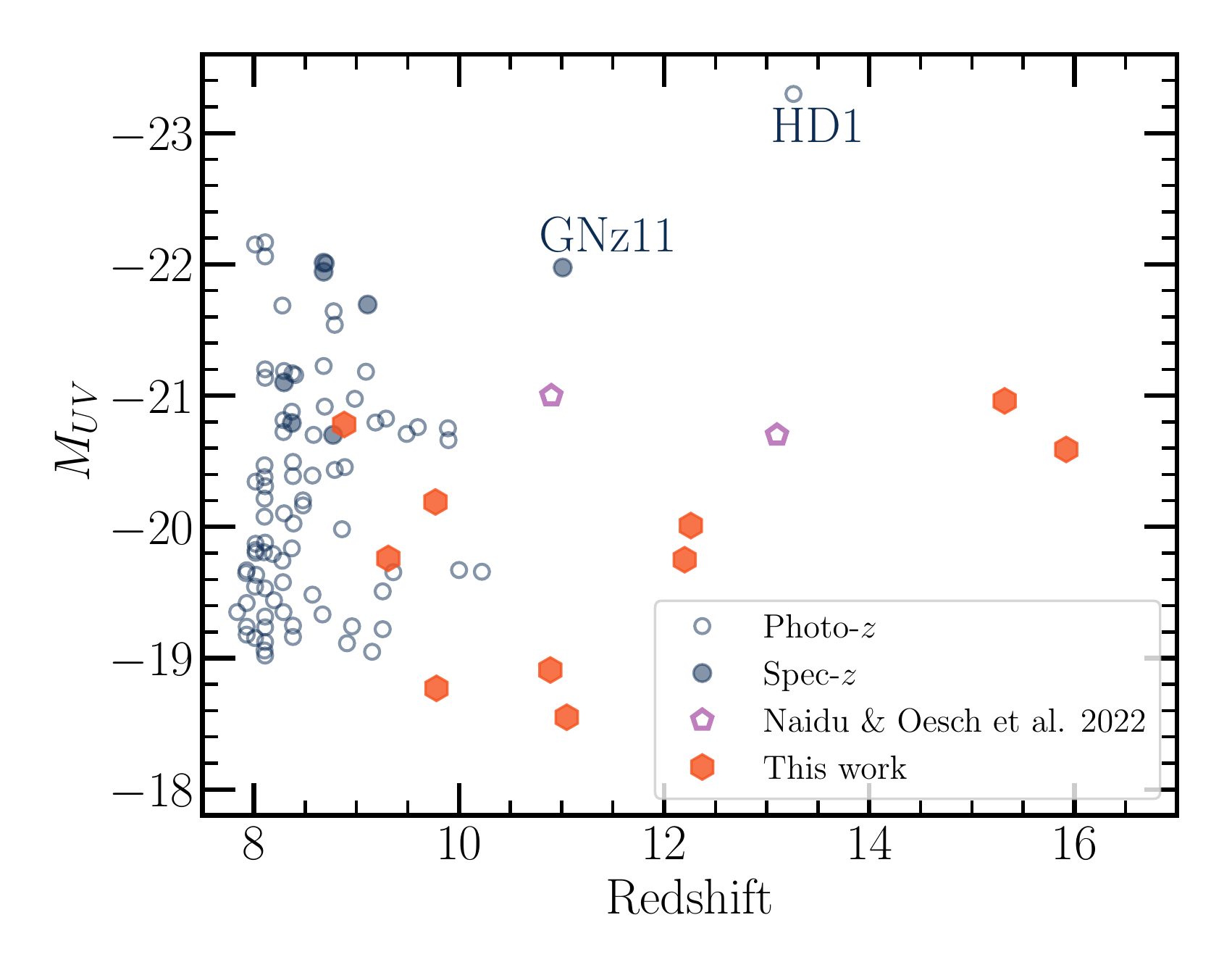}\\[-0.5cm]
    \caption{Absolute UV magnitude $M_{\mathrm{UV}}$ as a function of redshift for the high-redshift candidates in this work, along with a compilation of known galaxies at $z>8$. The candidates in this work are shown as orange filled hexagons, while empty hexagon marks the less secure candidate. The empty purple pentagons show the recently identified candidates from JWST in the GLASS parallel field by \citet{naidu22}. The empty blue circles show photometric-redshift galaxies, while the solid circles represent spectroscopically confirmed galaxies compiled from \citet{bouwens22}.}
    \label{fig:Muv_vs_z}
\end{figure*}

While observing through a lensing cluster offers invaluable flux amplification, it also results in a smaller survey area in the source plane. The field-of-view in the NIRCam image plane is about $9.7$\,arcmin$^2$, which translates into $7.23$\,arcmin$^2$ in the source plane after de-lensing. Although the area reduction is still reasonable (because one of the NIRCam modules is less affected by magnification), wide-area surveys are more prone to uncover rare and bright high-redshift galaxies \citep[e.g.][]{borsani16,oesch16,bowler20,harikane22}, while lensing-assisted surveys are sensitive to the fainter population. While some of our sources are relatively bright, they all have luminosities around, or below, the characteristic luminosity $M^{\star}$ at $z\sim9-10$ \citep{oesch18,bouwens21}. The highest redshift candidate in our sample, SMACS\_z16a, has an intrinsic UV magnitude of $M_{\mathrm{UV}}=-20.59\pm0.15$, which is fainter than previous very high redshift candidates Gn\_z11 and HD1, and roughly similar to the recently discovered candidate GLz13 \citep{naidu22}. Moreover, the yield of the galaxy hunt presented in this paper is surprisingly high, with the discovery of 2 candidates at $z\sim16$, 2 candidates at $z\sim12$, and 6 candidates at $z=10-11$. In Fig.~\ref{fig:Muv_vs_z}, we put these discoveries in the context of known high-redshift galaxies and compare our candidates to the published samples of galaxies at $z>8$. Not only do these results demonstrate the exceptional capabilities of JWST, but they also suggest a higher number density than predicted by an extrapolation of the $z\sim9-10$ luminosity function \citep[e.g.][]{bouwens21} and theoretical predictions \citep[e.g.,][]{ocvirk20,dayal22}. Some early theoretical interpretations suggest that these observations could be compatible with a total absence of dust in these galaxies, as their very blue UV slopes might also indicate \citep{ziparo22}. The evolution of the UV luminosity function observed through lensing clusters will be the scope of a future paper.

The inferred age of SMACS\_z16a indicates a recent formation history, which points to a rapid stellar mass assembly in this galaxy. In the context of early galaxy formation, one would expect the first galaxies to more resemble SMACS\_z16a in terms of physical parameters, than the bright and uncommon galaxies discussed earlier. As we are closing in on the formation epoch of the first galaxies, it will become increasingly challenging to find such evolved and relatively massive objects. Perhaps upcoming and future deep surveys through gravitational lenses will offer the best route to unveil these smallest structures. This first JWST cycle already includes several ERS, General Observer (GO), and Guaranteed Time (GTO) programs that target lensing clusters using different observing modes.

Another interesting finding is the small sizes of these galaxies. For instance, SMACS\_z16a has an effective radius of $r_{\rm e}=0.4$\,kpc, which is consistent with recent size measurement of high-redshift galaxies \citep{kawamata18,Adams2022}. More importantly, the S\'ersic index, derived from the {\tt SE++} modeling and source-plane reconstruction, hints to a disk-like morphology. We measured $n=1.1\pm0.2$ for SMACS\_z16a, which corresponds to a disk-like profile, although there is a real possibility that we are detecting only the brightest star-forming region in the galaxy. Such a morphology was also observed for GLz13 \citep{naidu22}, as well as in several other high-redshift candidates reported in \citet{Adams2022}. The existence of well settled disks at early epochs would provide strong constraints on theoretical models of galaxy formation.

While the nature of these ERO observations implies potential sources of uncertainties, we performed photometric quality verifications, and also combined multiple selection methods to build our sample of high-redshift sources. The dropout selection and the photometric redshifts from two independent codes significantly increase the robustness of the candidates. In the near future, deeper imaging in the short wavelength range will bring additional confirmation to these candidates, in addition to follow-up spectroscopy, e.g with NIRSpec, which will provide definitive redshifts. 

\section{Summary} \label{sec:conclusions}
In this paper, we presented our search for very high-redshift galaxies in the first JWST observations of the lensing cluster SMACS~0723. We report the discovery of the highest redshift candidate known to date at $z\sim16$. We also discover two candidates at $z\sim12$, and 11 candidates at $z\sim10-11$.  

We have used 6 NIRCam broadband filters, and one NIRISS filter, between $\sim0.8$\,\micron\ and 5\,\micron\, to identify Lyman break galaxies in the redshift ranges $9<z<11$, $11<z<15$, and $16<z<21$. We combined the dropout selection with SED-fitting to obtain accurate photometric redshifts using two different codes, {\tt BEAGLE} and {\tt EAZY}. We also simultaneously derive the physical properties of these galaxies. Our final sample is based on all candidates that satisfy the three selection methods and whose photometric redshifts from the two SED fits are in agreement. To determine the intrinsic luminosity of galaxies and account for potential multiple images, we use the one of the latest lensing models of SMACS~0723 produced from the same JWST data set in a companion paper \citep{mahler22}. The measured amplification factors range from 1 to $\sim4$, which gives us a lensing boost up to 1.5\,magnitudes in the best case.

With an intrinsic UV luminosity of $M_{\mathrm{UV}}=-20.59\pm0.15$, we find that SMACS\_z16a is fainter than typical $z>10$ galaxies in the literature and close to the characteristic luminosity $M^{\star}$ at $z\sim10$. The best-fit SED of SMACS\_z16a is compatible with a young age of $\log(t_{\mathrm{age}}/\mathrm{yr})=7.65^{+0.36}_{-0.39} $ and a stellar mass of $\log(M_{\star}/\mathrm{M}_{\odot})=8.79^{+0.32}_{-0.33}$. The bulk of its stellar mass build-up took place in the last 20\,Myr. This galaxy, like some other candidates, shows a very blue UV continuum-slope of $\beta=-2.63\pm 0.13$ consistent with young ages. In general, the galaxy candidates, from $\sim 10$ to $z \sim 16$, have ages in the range $\log(t_{\mathrm{age}}/\mathrm{yr})\sim =7.3 - 8.7$, and stellar masses of $\log(M_{\star}/\mathrm{M}_{\odot}) \sim  7.8 - 10.2$. Our source extraction and photometry procedure with {\tt SE++} also provide morphological measurements, since it relies on model-fitting. The measured sizes were reconstructed back to the lens plane using {\tt lenstool} to determine the intrinsic effective radius and the Sersi\'{c} index of the sources. Our measurements indicate very compact sources, with $r_{\rm e}$ below 1\,kpc for most of the candidates. Some of these sources also have S\'ersic indices compatible with disk-like a morphology. If true, it is surprising, for a galaxy like SMACS\_z16a, to evolve so quickly in the span of less than $\sim 250$\,Myr.

The discovery of one, possibly two, galaxy candidates only $\sim 250$\,Myr after the Big Bang, together with eight new candidates at $9<z<13$, announces great discoveries awaiting in the deep extra-galactic fields that will be observed by JWST. The number density, the physical properties and the morphologies of these galaxies will be extremely valuable in constraining hydrodynamical simulations of galaxy formation in a cosmological context. 

\section*{Acknowledgements}

We would like to thank Gabriel Brammer for the {\sc grizli} reduction notebooks. HA acknowledges support from CNES. LF and AZ acknowledge support by Grant No. 2020750 from the United States-Israel Binational Science Foundation (BSF) and Grant No. 2109066 from the United States National Science Foundation (NSF). AZ acknowledges support by the Ministry of Science \& Technology, Israel.

\noindent This work is based on observations obtained with the NASA/ESA/CSA \textit{JWST} and the NASA/ESA \textit{Hubble Space Telescope} (HST), retrieved from the \texttt{Mikulski Archive for Space Telescopes} (\texttt{MAST}) at the \textit{Space Telescope Science Institute} (STScI). STScI is operated by the Association of Universities for Research in Astronomy, Inc. under NASA contract NAS 5-26555. This work has made use of the \texttt{CANDIDE} Cluster at the \textit{Institut d'Astrophysique de Paris} (IAP), made possible by grants from the PNCG and the region of Île de France through the program DIM-ACAV+.

\noindent The JWST Early Release Observations (ERO) and associated materials were developed, executed, and compiled by the ERO production team: Hannah Braun, Claire Blome, Matthew Brown, Margaret Carruthers, Dan Coe, Joseph De- Pasquale, Nestor Espinoza, Macarena Garcia Marin, Karl Gordon, Alaina Henry, Leah Hustak, Andi James, Ann Jenkins, Anton Koekemoer, Stephanie LaMassa, David Law, Alexandra Lockwood, Amaya Moro-Martin, Susan Mullally, Alyssa Pagan, Dani Player, Klaus Pontoppidan, Charles Prof- fitt, Christine Pulliam, Leah Ramsay, Swara Ravindranath, Neill Reid, Massimo Robberto, Elena Sabbi, Leonardo Ubeda. The EROs were also made possible by the foundational efforts and support from the \jwst\ instruments, STScI planning and scheduling, and Data Management teams.
 
\section*{Data Availability}
The data underlying this article are publicly available on the \texttt{Mikulski Archive for Space Telescopes}\footnote{\url{https://archive.stsci.edu/}} (\texttt{MAST}), under program ID 2736.

\bibliographystyle{mnras}
\bibliography{references}

\begin{thebibliography}{}
\makeatletter
\relax
\def\mn@urlcharsother{\let\do\@makeother \do\$\do\&\do\#\do\^\do\_\do\%\do\~}
\def\mn@doi{\begingroup\mn@urlcharsother \@ifnextchar [ {\mn@doi@}
  {\mn@doi@[]}}
\def\mn@doi@[#1]#2{\def\@tempa{#1}\ifx\@tempa\@empty \href
  {http://dx.doi.org/#2} {doi:#2}\else \href {http://dx.doi.org/#2} {#1}\fi
  \endgroup}
\def\mn@eprint#1#2{\mn@eprint@#1:#2::\@nil}
\def\mn@eprint@arXiv#1{\href {http://arxiv.org/abs/#1} {{\tt arXiv:#1}}}
\def\mn@eprint@dblp#1{\href {http://dblp.uni-trier.de/rec/bibtex/#1.xml}
  {dblp:#1}}
\def\mn@eprint@#1:#2:#3:#4\@nil{\def\@tempa {#1}\def\@tempb {#2}\def\@tempc
  {#3}\ifx \@tempc \@empty \let \@tempc \@tempb \let \@tempb \@tempa \fi \ifx
  \@tempb \@empty \def\@tempb {arXiv}\fi \@ifundefined
  {mn@eprint@\@tempb}{\@tempb:\@tempc}{\expandafter \expandafter \csname
  mn@eprint@\@tempb\endcsname \expandafter{\@tempc}}}

\bibitem[\protect\citeauthoryear{{Adams} et~al.,}{{Adams}
  et~al.}{2022}]{Adams2022}
{Adams} N.~J.,  et~al., 2022, arXiv e-prints, \href
  {https://ui.adsabs.harvard.edu/abs/2022arXiv220711217A} {p. arXiv:2207.11217}

\bibitem[\protect\citeauthoryear{{Allard}, {Hauschildt}, {Alexander}, {Tamanai}
   \& {Schweitzer}}{{Allard} et~al.}{2001}]{allard01}
{Allard} F.,  {Hauschildt} P.~H.,  {Alexander} D.~R.,  {Tamanai} A.,
  {Schweitzer} A.,  2001, \mn@doi [\apj] {10.1086/321547}, \href
  {https://ui.adsabs.harvard.edu/abs/2001ApJ...556..357A} {556, 357}

\bibitem[\protect\citeauthoryear{{Atek} et~al.,}{{Atek} et~al.}{2011}]{atek11}
{Atek} H.,  et~al., 2011, \mn@doi [\apj] {10.1088/0004-637X/743/2/121}, \href
  {http://adsabs.harvard.edu/abs/2011ApJ...743..121A} {743, 121}

\bibitem[\protect\citeauthoryear{{Atek} et~al.,}{{Atek}
  et~al.}{2014a}]{atek14b}
{Atek} H.,  et~al., 2014a, \mn@doi [\apj] {10.1088/0004-637X/786/1/60}, \href
  {http://adsabs.harvard.edu/abs/2014ApJ...786...60A} {786, 60}

\bibitem[\protect\citeauthoryear{{Atek} et~al.,}{{Atek}
  et~al.}{2014b}]{atek14c}
{Atek} H.,  et~al., 2014b, \mn@doi [\apj] {10.1088/0004-637X/789/2/96}, \href
  {http://adsabs.harvard.edu/abs/2014ApJ...789...96A} {789, 96}

\bibitem[\protect\citeauthoryear{{Atek} et~al.,}{{Atek} et~al.}{2015}]{atek15a}
{Atek} H.,  et~al., 2015, \mn@doi [\apj] {10.1088/0004-637X/800/1/18}, \href
  {http://cdsads.u-strasbg.fr/abs/2015ApJ...800...18A} {800, 18}

\bibitem[\protect\citeauthoryear{{Atek}, {Richard}, {Kneib}  \&
  {Schaerer}}{{Atek} et~al.}{2018}]{atek18}
{Atek} H.,  {Richard} J.,  {Kneib} J.-P.,   {Schaerer} D.,  2018, \mn@doi
  [\mnras] {10.1093/mnras/sty1820}, \href
  {https://ui.adsabs.harvard.edu/abs/2018MNRAS.479.5184A} {479, 5184}

\bibitem[\protect\citeauthoryear{{Bacon} et~al.,}{{Bacon}
  et~al.}{2010}]{bacon10}
{Bacon} R.,  et~al., 2010, in {McLean} I.~S.,  {Ramsay} S.~K.,   {Takami} H.,
  eds,  Society of Photo-Optical Instrumentation Engineers (SPIE) Conference
  Series Vol. 7735, Ground-based and Airborne Instrumentation for Astronomy
  III. p. 773508, \mn@doi{10.1117/12.856027}

\bibitem[\protect\citeauthoryear{{Bertin}, {Schefer}, {Apostolakos},
  {{\'A}lvarez-Ayll{\'o}n}, {Dubath}  \& {K{\"u}mmel}}{{Bertin}
  et~al.}{2020}]{bertin20}
{Bertin} E.,  {Schefer} M.,  {Apostolakos} N.,  {{\'A}lvarez-Ayll{\'o}n} A.,
  {Dubath} P.,   {K{\"u}mmel} M.,  2020, in {Pizzo} R.,  {Deul} E.~R.,  {Mol}
  J.~D.,  {de Plaa} J.,   {Verkouter} H.,  eds,  Astronomical Society of the
  Pacific Conference Series Vol. 527, Astronomical Data Analysis Software and
  Systems XXIX. p.~461

\bibitem[\protect\citeauthoryear{{Bhatawdekar}, {Conselice},
  {Margalef-Bentabol}  \& {Duncan}}{{Bhatawdekar} et~al.}{2019}]{bhatawdekar19}
{Bhatawdekar} R.,  {Conselice} C.~J.,  {Margalef-Bentabol} B.,   {Duncan} K.,
  2019, \mn@doi [\mnras] {10.1093/mnras/stz866}, \href
  {https://ui.adsabs.harvard.edu/abs/2019MNRAS.486.3805B} {486, 3805}

\bibitem[\protect\citeauthoryear{{Bouwens} et~al.,}{{Bouwens}
  et~al.}{2015}]{bouwens15}
{Bouwens} R.~J.,  et~al., 2015, \mn@doi [\apj] {10.1088/0004-637X/803/1/34},
  \href {http://adsabs.harvard.edu/abs/2015ApJ...803...34B} {803, 34}

\bibitem[\protect\citeauthoryear{{Bouwens}, {Oesch}, {Illingworth}, {Ellis}  \&
  {Stefanon}}{{Bouwens} et~al.}{2017}]{bouwens17b}
{Bouwens} R.~J.,  {Oesch} P.~A.,  {Illingworth} G.~D.,  {Ellis} R.~S.,
  {Stefanon} M.,  2017, \mn@doi [\apj] {10.3847/1538-4357/aa70a4}, \href
  {http://adsabs.harvard.edu/abs/2017ApJ...843..129B} {843, 129}

\bibitem[\protect\citeauthoryear{{Bouwens} et~al.,}{{Bouwens}
  et~al.}{2021}]{bouwens21}
{Bouwens} R.~J.,  et~al., 2021, \mn@doi [\aj] {10.3847/1538-3881/abf83e}, \href
  {https://ui.adsabs.harvard.edu/abs/2021AJ....162...47B} {162, 47}

\bibitem[\protect\citeauthoryear{{Bouwens} et~al.,}{{Bouwens}
  et~al.}{2022}]{bouwens22}
{Bouwens} R.~J.,  et~al., 2022, \mn@doi [\apj] {10.3847/1538-4357/ac5a4a},
  \href {https://ui.adsabs.harvard.edu/abs/2022ApJ...931..160B} {931, 160}

\bibitem[\protect\citeauthoryear{{Bowler}, {Jarvis}, {Dunlop}, {McLure},
  {McLeod}, {Adams}, {Milvang-Jensen}  \& {McCracken}}{{Bowler}
  et~al.}{2020}]{bowler20}
{Bowler} R.~A.~A.,  {Jarvis} M.~J.,  {Dunlop} J.~S.,  {McLure} R.~J.,  {McLeod}
  D.~J.,  {Adams} N.~J.,  {Milvang-Jensen} B.,   {McCracken} H.~J.,  2020,
  \mn@doi [\mnras] {10.1093/mnras/staa313}, \href
  {https://ui.adsabs.harvard.edu/abs/2020MNRAS.493.2059B} {493, 2059}

\bibitem[\protect\citeauthoryear{{Boyer} et~al.,}{{Boyer}
  et~al.}{2022}]{boyer22}
{Boyer} M.~L.,  et~al., 2022, arXiv e-prints, \href
  {https://ui.adsabs.harvard.edu/abs/2022arXiv220903348B} {p. arXiv:2209.03348}

\bibitem[\protect\citeauthoryear{{Brammer}, {van Dokkum}  \& {Coppi}}{{Brammer}
  et~al.}{2008}]{Brammer2008}
{Brammer} G.~B.,  {van Dokkum} P.~G.,   {Coppi} P.,  2008, \mn@doi [\apj]
  {10.1086/591786}, \href
  {https://ui.adsabs.harvard.edu/abs/2008ApJ...686.1503B} {686, 1503}

\bibitem[\protect\citeauthoryear{Bruzual \& Charlot}{Bruzual \&
  Charlot}{2003}]{Bruzual_Charlot_2003}
Bruzual G.,  Charlot S.,  2003, \mn@doi [Monthly Notices of the Royal
  Astronomical Society] {10.1046/j.1365-8711.2003.06897.x}, 344, 1000

\bibitem[\protect\citeauthoryear{{Bruzual A.} \& {Charlot}}{{Bruzual A.} \&
  {Charlot}}{1993}]{bc93}
{Bruzual A.} G.,  {Charlot} S.,  1993, \mn@doi [\apj] {10.1086/172385}, \href
  {https://ui.adsabs.harvard.edu/abs/1993ApJ...405..538B} {405, 538}

\bibitem[\protect\citeauthoryear{{Capak} et~al.,}{{Capak}
  et~al.}{2015}]{capak15}
{Capak} P.~L.,  et~al., 2015, \mn@doi [\nat] {10.1038/nature14500}, \href
  {https://ui.adsabs.harvard.edu/abs/2015Natur.522..455C} {522, 455}

\bibitem[\protect\citeauthoryear{{Castellano} et~al.,}{{Castellano}
  et~al.}{2022}]{castellano22}
{Castellano} M.,  et~al., 2022, arXiv e-prints, \href
  {https://ui.adsabs.harvard.edu/abs/2022arXiv220709436C} {p. arXiv:2207.09436}

\bibitem[\protect\citeauthoryear{{Chabrier}, {Baraffe}, {Allard}  \&
  {Hauschildt}}{{Chabrier} et~al.}{2000}]{chabrier00}
{Chabrier} G.,  {Baraffe} I.,  {Allard} F.,   {Hauschildt} P.,  2000, \mn@doi
  [\apj] {10.1086/309513}, \href
  {http://adsabs.harvard.edu/abs/2000ApJ...542..464C} {542, 464}

\bibitem[\protect\citeauthoryear{{Chevallard} \& {Charlot}}{{Chevallard} \&
  {Charlot}}{2016}]{chevallard16}
{Chevallard} J.,  {Charlot} S.,  2016, \mn@doi [\mnras]
  {10.1093/mnras/stw1756}, \href
  {https://ui.adsabs.harvard.edu/abs/2016MNRAS.462.1415C} {462, 1415}

\bibitem[\protect\citeauthoryear{{Coe} et~al.,}{{Coe} et~al.}{2019}]{coe19}
{Coe} D.,  et~al., 2019, \mn@doi [\apj] {10.3847/1538-4357/ab412b}, \href
  {https://ui.adsabs.harvard.edu/abs/2019ApJ...884...85C} {884, 85}

\bibitem[\protect\citeauthoryear{{Conroy} \& {Gunn}}{{Conroy} \&
  {Gunn}}{2010}]{ConroyGunn2010}
{Conroy} C.,  {Gunn} J.~E.,  2010, \mn@doi [\apj]
  {10.1088/0004-637X/712/2/833}, \href
  {https://ui.adsabs.harvard.edu/abs/2010ApJ...712..833C} {712, 833}

\bibitem[\protect\citeauthoryear{{Conroy}, {Gunn}  \& {White}}{{Conroy}
  et~al.}{2009}]{Conroy2009}
{Conroy} C.,  {Gunn} J.~E.,   {White} M.,  2009, \mn@doi [\apj]
  {10.1088/0004-637X/699/1/486}, \href
  {https://ui.adsabs.harvard.edu/abs/2009ApJ...699..486C} {699, 486}

\bibitem[\protect\citeauthoryear{{Dayal} et~al.,}{{Dayal}
  et~al.}{2022}]{dayal22}
{Dayal} P.,  et~al., 2022, \mn@doi [\mnras] {10.1093/mnras/stac537}, \href
  {https://ui.adsabs.harvard.edu/abs/2022MNRAS.512..989D} {512, 989}

\bibitem[\protect\citeauthoryear{{De Barros}, {Oesch}, {Labb{\'e}}, {Stefanon},
  {Gonz{\'a}lez}, {Smit}, {Bouwens}  \& {Illingworth}}{{De Barros}
  et~al.}{2019}]{debarros19}
{De Barros} S.,  {Oesch} P.~A.,  {Labb{\'e}} I.,  {Stefanon} M.,
  {Gonz{\'a}lez} V.,  {Smit} R.,  {Bouwens} R.~J.,   {Illingworth} G.~D.,
  2019, \mn@doi [\mnras] {10.1093/mnras/stz940}, \href
  {https://ui.adsabs.harvard.edu/abs/2019MNRAS.489.2355D} {489, 2355}

\bibitem[\protect\citeauthoryear{{Doyon} et~al.,}{{Doyon}
  et~al.}{2012}]{doyon12}
{Doyon} R.,  et~al., 2012, in {Clampin} M.~C.,  {Fazio} G.~G.,  {MacEwen}
  H.~A.,   {Oschmann} Jacobus~M. J.,  eds,  Society of Photo-Optical
  Instrumentation Engineers (SPIE) Conference Series Vol. 8442, Space
  Telescopes and Instrumentation 2012: Optical, Infrared, and Millimeter Wave.
  p. 84422R, \mn@doi{10.1117/12.926578}

\bibitem[\protect\citeauthoryear{{Duncan} et~al.,}{{Duncan}
  et~al.}{2014}]{duncan14}
{Duncan} K.,  et~al., 2014, \mn@doi [\mnras] {10.1093/mnras/stu1622}, \href
  {https://ui.adsabs.harvard.edu/abs/2014MNRAS.444.2960D} {444, 2960}

\bibitem[\protect\citeauthoryear{{Ebeling}, {Edge}  \& {Henry}}{{Ebeling}
  et~al.}{2001}]{ebeling01}
{Ebeling} H.,  {Edge} A.~C.,   {Henry} J.~P.,  2001, \mn@doi [\apj]
  {10.1086/320958}, \href {http://adsabs.harvard.edu/abs/2001ApJ...553..668E}
  {553, 668}

\bibitem[\protect\citeauthoryear{{El{\'{\i}}asd{\'o}ttir}
  et~al.,}{{El{\'{\i}}asd{\'o}ttir} et~al.}{2007}]{eliasdottir07}
{El{\'{\i}}asd{\'o}ttir} {\'A}.,  et~al., 2007, preprint, \href
  {http://cdsads.u-strasbg.fr/abs/2007arXiv0710.5636E} {} (\mn@eprint {arXiv}
  {0710.5636})

\bibitem[\protect\citeauthoryear{{Euclid Collaboration} et~al.,}{{Euclid
  Collaboration} et~al.}{2022a}]{MerlinEMC2022}
{Euclid Collaboration} et~al., 2022a, arXiv e-prints, \href
  {https://ui.adsabs.harvard.edu/abs/2022arXiv220912906E} {p. arXiv:2209.12906}

\bibitem[\protect\citeauthoryear{{Euclid Collaboration} et~al.,}{{Euclid
  Collaboration} et~al.}{2022b}]{HubertEMC2022}
{Euclid Collaboration} et~al., 2022b, arXiv e-prints, \href
  {https://ui.adsabs.harvard.edu/abs/2022arXiv220912907E} {p. arXiv:2209.12907}

\bibitem[\protect\citeauthoryear{{Eyles}, {Bunker}, {Ellis}, {Lacy}, {Stanway},
  {Stark}  \& {Chiu}}{{Eyles} et~al.}{2007}]{eyles07}
{Eyles} L.~P.,  {Bunker} A.~J.,  {Ellis} R.~S.,  {Lacy} M.,  {Stanway} E.~R.,
  {Stark} D.~P.,   {Chiu} K.,  2007, \mn@doi [\mnras]
  {10.1111/j.1365-2966.2006.11197.x}, \href
  {https://ui.adsabs.harvard.edu/abs/2007MNRAS.374..910E} {374, 910}

\bibitem[\protect\citeauthoryear{{Ferland} et~al.,}{{Ferland}
  et~al.}{2013}]{ferland13}
{Ferland} G.~J.,  et~al., 2013, \rmxaa, \href
  {https://ui.adsabs.harvard.edu/abs/2013RMxAA..49..137F} {49, 137}

\bibitem[\protect\citeauthoryear{{Finkelstein} et~al.,}{{Finkelstein}
  et~al.}{2015}]{finkelstein15}
{Finkelstein} S.~L.,  et~al., 2015, \mn@doi [\apj]
  {10.1088/0004-637X/810/1/71}, \href
  {http://adsabs.harvard.edu/abs/2015ApJ...810...71F} {810, 71}

\bibitem[\protect\citeauthoryear{{Fox}, {Mahler}, {Sharon}  \& {Remolina
  Gonz{\'a}lez}}{{Fox} et~al.}{2022}]{fox}
{Fox} C.,  {Mahler} G.,  {Sharon} K.,   {Remolina Gonz{\'a}lez} J.~D.,  2022,
  \mn@doi [\apj] {10.3847/1538-4357/ac5024}, \href
  {https://ui.adsabs.harvard.edu/abs/2022ApJ...928...87F} {928, 87}

\bibitem[\protect\citeauthoryear{{Furtak}, {Atek}, {Lehnert}, {Chevallard}  \&
  {Charlot}}{{Furtak} et~al.}{2021}]{furtak21}
{Furtak} L.~J.,  {Atek} H.,  {Lehnert} M.~D.,  {Chevallard} J.,   {Charlot} S.,
   2021, \mn@doi [\mnras] {10.1093/mnras/staa3760}, \href
  {https://ui.adsabs.harvard.edu/abs/2021MNRAS.501.1568F} {501, 1568}

\bibitem[\protect\citeauthoryear{{Golubchik}, {Furtak}, {Meena}  \&
  {Zitrin}}{{Golubchik} et~al.}{2022}]{golubchik}
{Golubchik} M.,  {Furtak} L.~J.,  {Meena} A.~K.,   {Zitrin} A.,  2022, arXiv
  e-prints, \href {https://ui.adsabs.harvard.edu/abs/2022arXiv220705007G} {p.
  arXiv:2207.05007}

\bibitem[\protect\citeauthoryear{{Gonz{\'a}lez}, {Labb{\'e}}, {Bouwens},
  {Illingworth}, {Franx}  \& {Kriek}}{{Gonz{\'a}lez} et~al.}{2011}]{gonzales11}
{Gonz{\'a}lez} V.,  {Labb{\'e}} I.,  {Bouwens} R.~J.,  {Illingworth} G.,
  {Franx} M.,   {Kriek} M.,  2011, \mn@doi [\apjl]
  {10.1088/2041-8205/735/2/L34}, \href
  {https://ui.adsabs.harvard.edu/abs/2011ApJ...735L..34G} {735, L34}

\bibitem[\protect\citeauthoryear{{Grazian} et~al.,}{{Grazian}
  et~al.}{2015}]{grazian15}
{Grazian} A.,  et~al., 2015, \mn@doi [\aap] {10.1051/0004-6361/201424750},
  \href {https://ui.adsabs.harvard.edu/abs/2015A&A...575A..96G} {575, A96}

\bibitem[\protect\citeauthoryear{{Gutkin}, {Charlot}  \& {Bruzual}}{{Gutkin}
  et~al.}{2016}]{gutkin16}
{Gutkin} J.,  {Charlot} S.,   {Bruzual} G.,  2016, \mn@doi [\mnras]
  {10.1093/mnras/stw1716}, \href
  {https://ui.adsabs.harvard.edu/abs/2016MNRAS.462.1757G} {462, 1757}

\bibitem[\protect\citeauthoryear{{Harikane} et~al.,}{{Harikane}
  et~al.}{2022}]{harikane22}
{Harikane} Y.,  et~al., 2022, \mn@doi [\apj] {10.3847/1538-4357/ac53a9}, \href
  {https://ui.adsabs.harvard.edu/abs/2022ApJ...929....1H} {929, 1}

\bibitem[\protect\citeauthoryear{{Inoue}, {Shimizu}, {Iwata}  \&
  {Tanaka}}{{Inoue} et~al.}{2014}]{inoue14}
{Inoue} A.~K.,  {Shimizu} I.,  {Iwata} I.,   {Tanaka} M.,  2014, \mn@doi
  [\mnras] {10.1093/mnras/stu936}, \href
  {https://ui.adsabs.harvard.edu/abs/2014MNRAS.442.1805I} {442, 1805}

\bibitem[\protect\citeauthoryear{{Ishigaki}, {Kawamata}, {Ouchi}, {Oguri},
  {Shimasaku}  \& {Ono}}{{Ishigaki} et~al.}{2018}]{ishigaki18}
{Ishigaki} M.,  {Kawamata} R.,  {Ouchi} M.,  {Oguri} M.,  {Shimasaku} K.,
  {Ono} Y.,  2018, \mn@doi [\apj] {10.3847/1538-4357/aaa544}, \href
  {http://adsabs.harvard.edu/abs/2018ApJ...854...73I} {854, 73}

\bibitem[\protect\citeauthoryear{{Jakobsen} et~al.,}{{Jakobsen}
  et~al.}{2022}]{jakobsen22}
{Jakobsen} P.,  et~al., 2022, \mn@doi [\aap] {10.1051/0004-6361/202142663},
  \href {https://ui.adsabs.harvard.edu/abs/2022A&A...661A..80J} {661, A80}

\bibitem[\protect\citeauthoryear{{Jiang} et~al.,}{{Jiang}
  et~al.}{2021}]{jiang21}
{Jiang} L.,  et~al., 2021, \mn@doi [Nature Astronomy]
  {10.1038/s41550-020-01275-y}, \href
  {https://ui.adsabs.harvard.edu/abs/2021NatAs...5..256J} {5, 256}

\bibitem[\protect\citeauthoryear{{Jullo} \& {Kneib}}{{Jullo} \&
  {Kneib}}{2009}]{jullo09}
{Jullo} E.,  {Kneib} J.-P.,  2009, \mn@doi [\mnras]
  {10.1111/j.1365-2966.2009.14654.x}, \href
  {http://cdsads.u-strasbg.fr/abs/2009MNRAS.395.1319J} {395, 1319}

\bibitem[\protect\citeauthoryear{{Jullo}, {Kneib}, {Limousin},
  {El{\'{\i}}asd{\'o}ttir}, {Marshall}  \& {Verdugo}}{{Jullo}
  et~al.}{2007}]{jullo07}
{Jullo} E.,  {Kneib} J.-P.,  {Limousin} M.,  {El{\'{\i}}asd{\'o}ttir} {\'A}.,
  {Marshall} P.~J.,   {Verdugo} T.,  2007, \mn@doi [New Journal of Physics]
  {10.1088/1367-2630/9/12/447}, \href
  {http://adsabs.harvard.edu/abs/2007NJPh....9..447J} {9, 447}

\bibitem[\protect\citeauthoryear{{Kauffmann} et~al.,}{{Kauffmann}
  et~al.}{2022}]{kauffmann22}
{Kauffmann} O.~B.,  et~al., 2022, arXiv e-prints, \href
  {https://ui.adsabs.harvard.edu/abs/2022arXiv220711740K} {p. arXiv:2207.11740}

\bibitem[\protect\citeauthoryear{{Kawamata}, {Ishigaki}, {Shimasaku}, {Oguri}
  \& {Ouchi}}{{Kawamata} et~al.}{2015}]{kawamata15}
{Kawamata} R.,  {Ishigaki} M.,  {Shimasaku} K.,  {Oguri} M.,   {Ouchi} M.,
  2015, \mn@doi [\apj] {10.1088/0004-637X/804/2/103}, \href
  {http://cdsads.u-strasbg.fr/abs/2015ApJ...804..103K} {804, 103}

\bibitem[\protect\citeauthoryear{{Kawamata}, {Ishigaki}, {Shimasaku}, {Oguri},
  {Ouchi}  \& {Tanigawa}}{{Kawamata} et~al.}{2018}]{kawamata18}
{Kawamata} R.,  {Ishigaki} M.,  {Shimasaku} K.,  {Oguri} M.,  {Ouchi} M.,
  {Tanigawa} S.,  2018, \mn@doi [\apj] {10.3847/1538-4357/aaa6cf}, \href
  {https://ui.adsabs.harvard.edu/abs/2018ApJ...855....4K} {855, 4}

\bibitem[\protect\citeauthoryear{{Kikuchihara} et~al.,}{{Kikuchihara}
  et~al.}{2020}]{kikuchihara20}
{Kikuchihara} S.,  et~al., 2020, \mn@doi [\apj] {10.3847/1538-4357/ab7dbe},
  \href {https://ui.adsabs.harvard.edu/abs/2020ApJ...893...60K} {893, 60}

\bibitem[\protect\citeauthoryear{{Kinney}, {Calzetti}, {Bohlin}, {McQuade},
  {Storchi-Bergmann}  \& {Schmitt}}{{Kinney} et~al.}{1996}]{kc96}
{Kinney} A.~L.,  {Calzetti} D.,  {Bohlin} R.~C.,  {McQuade} K.,
  {Storchi-Bergmann} T.,   {Schmitt} H.~R.,  1996, \mn@doi [\apj]
  {10.1086/177583}, \href
  {https://ui.adsabs.harvard.edu/abs/1996ApJ...467...38K} {467, 38}

\bibitem[\protect\citeauthoryear{{Kneib}, {Ellis}, {Smail}, {Couch}  \&
  {Sharples}}{{Kneib} et~al.}{1996}]{kneib96}
{Kneib} J.~P.,  {Ellis} R.~S.,  {Smail} I.,  {Couch} W.~J.,   {Sharples} R.~M.,
   1996, \mn@doi [\apj] {10.1086/177995}, \href
  {https://ui.adsabs.harvard.edu/abs/1996ApJ...471..643K} {471, 643}

\bibitem[\protect\citeauthoryear{{K{\"u}mmel}, {Bertin}, {Schefer},
  {Apostolakos}, {{\'A}lvarez-Ayll{\'o}n}  \& {Dubath}}{{K{\"u}mmel}
  et~al.}{2020}]{kummel20}
{K{\"u}mmel} M.,  {Bertin} E.,  {Schefer} M.,  {Apostolakos} N.,
  {{\'A}lvarez-Ayll{\'o}n} A.,   {Dubath} P.,  2020, in {Pizzo} R.,  {Deul}
  E.~R.,  {Mol} J.~D.,  {de Plaa} J.,   {Verkouter} H.,  eds,  Astronomical
  Society of the Pacific Conference Series Vol. 527, Astronomical Data Analysis
  Software and Systems XXIX. p.~29

\bibitem[\protect\citeauthoryear{{Livermore}, {Finkelstein}  \&
  {Lotz}}{{Livermore} et~al.}{2017}]{livermore17}
{Livermore} R.~C.,  {Finkelstein} S.~L.,   {Lotz} J.~M.,  2017, \mn@doi [\apj]
  {10.3847/1538-4357/835/2/113}, \href
  {http://adsabs.harvard.edu/abs/2017ApJ...835..113L} {835, 113}

\bibitem[\protect\citeauthoryear{{Lotz} et~al.,}{{Lotz} et~al.}{2017}]{lotz17}
{Lotz} J.~M.,  et~al., 2017, \mn@doi [\apj] {10.3847/1538-4357/837/1/97}, \href
  {http://adsabs.harvard.edu/abs/2017ApJ...837...97L} {837, 97}

\bibitem[\protect\citeauthoryear{{Mahler} et~al.,}{{Mahler}
  et~al.}{2022}]{mahler22}
{Mahler} G.,  et~al., 2022, arXiv e-prints, \href
  {https://ui.adsabs.harvard.edu/abs/2022arXiv220707101M} {p. arXiv:2207.07101}

\bibitem[\protect\citeauthoryear{{McLure} et~al.,}{{McLure}
  et~al.}{2011}]{mclure11}
{McLure} R.~J.,  et~al., 2011, \mn@doi [\mnras]
  {10.1111/j.1365-2966.2011.19626.x}, \href
  {https://ui.adsabs.harvard.edu/abs/2011MNRAS.418.2074M} {418, 2074}

\bibitem[\protect\citeauthoryear{{Montes}}{{Montes}}{2022}]{montes22}
{Montes} M.,  2022, \mn@doi [Nature Astronomy] {10.1038/s41550-022-01616-z},
  \href {https://ui.adsabs.harvard.edu/abs/2022NatAs...6..308M} {6, 308}

\bibitem[\protect\citeauthoryear{{Naidu} et~al.,}{{Naidu}
  et~al.}{2022}]{naidu22}
{Naidu} R.~P.,  et~al., 2022, arXiv e-prints, \href
  {https://ui.adsabs.harvard.edu/abs/2022arXiv220709434N} {p. arXiv:2207.09434}

\bibitem[\protect\citeauthoryear{{Ocvirk} et~al.,}{{Ocvirk}
  et~al.}{2020}]{ocvirk20}
{Ocvirk} P.,  et~al., 2020, \mn@doi [\mnras] {10.1093/mnras/staa1266}, \href
  {https://ui.adsabs.harvard.edu/abs/2020MNRAS.496.4087O} {496, 4087}

\bibitem[\protect\citeauthoryear{{Oesch} et~al.,}{{Oesch}
  et~al.}{2016}]{oesch16}
{Oesch} P.~A.,  et~al., 2016, \mn@doi [\apj] {10.3847/0004-637X/819/2/129},
  \href {https://ui.adsabs.harvard.edu/abs/2016ApJ...819..129O} {819, 129}

\bibitem[\protect\citeauthoryear{{Oesch}, {Bouwens}, {Illingworth}, {Labb{\'e}}
   \& {Stefanon}}{{Oesch} et~al.}{2018}]{oesch18}
{Oesch} P.~A.,  {Bouwens} R.~J.,  {Illingworth} G.~D.,  {Labb{\'e}} I.,
  {Stefanon} M.,  2018, \mn@doi [\apj] {10.3847/1538-4357/aab03f}, \href
  {https://ui.adsabs.harvard.edu/abs/2018ApJ...855..105O} {855, 105}

\bibitem[\protect\citeauthoryear{{Oke} \& {Gunn}}{{Oke} \&
  {Gunn}}{1983}]{oke83}
{Oke} J.~B.,  {Gunn} J.~E.,  1983, \mn@doi [\apj] {10.1086/160817}, \href
  {http://adsabs.harvard.edu/abs/1983ApJ...266..713O} {266, 713}

\bibitem[\protect\citeauthoryear{{Pei}}{{Pei}}{1992}]{pei92}
{Pei} Y.~C.,  1992, \mn@doi [\apj] {10.1086/171637}, \href
  {https://ui.adsabs.harvard.edu/abs/1992ApJ...395..130P} {395, 130}

\bibitem[\protect\citeauthoryear{{Polletta} et~al.,}{{Polletta}
  et~al.}{2007}]{polletta07}
{Polletta} M.,  et~al., 2007, \mn@doi [\apj] {10.1086/518113}, \href
  {https://ui.adsabs.harvard.edu/abs/2007ApJ...663...81P} {663, 81}

\bibitem[\protect\citeauthoryear{{Reddy} et~al.,}{{Reddy}
  et~al.}{2015}]{reddy15}
{Reddy} N.~A.,  et~al., 2015, \mn@doi [\apj] {10.1088/0004-637X/806/2/259},
  \href {https://ui.adsabs.harvard.edu/abs/2015ApJ...806..259R} {806, 259}

\bibitem[\protect\citeauthoryear{{Reddy} et~al.,}{{Reddy}
  et~al.}{2018a}]{reddy18a}
{Reddy} N.~A.,  et~al., 2018a, \mn@doi [\apj] {10.3847/1538-4357/aaa3e7}, \href
  {https://ui.adsabs.harvard.edu/abs/2018ApJ...853...56R} {853, 56}

\bibitem[\protect\citeauthoryear{{Reddy} et~al.,}{{Reddy}
  et~al.}{2018b}]{reddy18b}
{Reddy} N.~A.,  et~al., 2018b, \mn@doi [\apj] {10.3847/1538-4357/aaed1e}, \href
  {https://ui.adsabs.harvard.edu/abs/2018ApJ...869...92R} {869, 92}

\bibitem[\protect\citeauthoryear{{Repp} \& {Ebeling}}{{Repp} \&
  {Ebeling}}{2018}]{repp18}
{Repp} A.,  {Ebeling} H.,  2018, \mn@doi [\mnras] {10.1093/mnras/sty1489},
  \href {https://ui.adsabs.harvard.edu/abs/2018MNRAS.479..844R} {479, 844}

\bibitem[\protect\citeauthoryear{{Richard} et~al.,}{{Richard}
  et~al.}{2014}]{richard14}
{Richard} J.,  et~al., 2014, preprint, \href
  {http://adsabs.harvard.edu/abs/2014arXiv1405.3303R} {} (\mn@eprint {arXiv}
  {1405.3303})

\bibitem[\protect\citeauthoryear{{Rigby} et~al.,}{{Rigby}
  et~al.}{2022}]{rigby22}
{Rigby} J.,  et~al., 2022, arXiv e-prints, \href
  {https://ui.adsabs.harvard.edu/abs/2022arXiv220705632R} {p. arXiv:2207.05632}

\bibitem[\protect\citeauthoryear{{Roberts-Borsani} et~al.,}{{Roberts-Borsani}
  et~al.}{2016}]{borsani16}
{Roberts-Borsani} G.~W.,  et~al., 2016, \mn@doi [\apj]
  {10.3847/0004-637X/823/2/143}, \href
  {https://ui.adsabs.harvard.edu/abs/2016ApJ...823..143R} {823, 143}

\bibitem[\protect\citeauthoryear{{Salmon} et~al.,}{{Salmon}
  et~al.}{2020}]{salmon20}
{Salmon} B.,  et~al., 2020, \mn@doi [\apj] {10.3847/1538-4357/ab5a8b}, \href
  {https://ui.adsabs.harvard.edu/abs/2020ApJ...889..189S} {889, 189}

\bibitem[\protect\citeauthoryear{{Schaerer} \& {de Barros}}{{Schaerer} \& {de
  Barros}}{2009}]{schaerer09}
{Schaerer} D.,  {de Barros} S.,  2009, \mn@doi [\aap]
  {10.1051/0004-6361/200911781}, \href
  {http://adsabs.harvard.edu/abs/2009A%26A...502..423S} {502, 423}

\bibitem[\protect\citeauthoryear{{Schaerer} \& {de Barros}}{{Schaerer} \& {de
  Barros}}{2010}]{schaerer10}
{Schaerer} D.,  {de Barros} S.,  2010, \mn@doi [\aap]
  {10.1051/0004-6361/200913946}, \href
  {https://ui.adsabs.harvard.edu/abs/2010A&A...515A..73S} {515, A73}

\bibitem[\protect\citeauthoryear{{Schaerer}, {Marques-Chaves}, {Oesch},
  {Naidu}, {Barrufet}, {Izotov}, {Guseva}  \& {Brammer}}{{Schaerer}
  et~al.}{2022}]{schaerer22}
{Schaerer} D.,  {Marques-Chaves} R.,  {Oesch} P.,  {Naidu} R.,  {Barrufet} L.,
  {Izotov} Y.~I.,  {Guseva} N.~G.,   {Brammer} G.,  2022, arXiv e-prints, \href
  {https://ui.adsabs.harvard.edu/abs/2022arXiv220710034S} {p. arXiv:2207.10034}

\bibitem[\protect\citeauthoryear{{S{\'e}rsic}}{{S{\'e}rsic}}{1963}]{sersic63}
{S{\'e}rsic} J.~L.,  1963, Boletin de la Asociacion Argentina de Astronomia La
  Plata Argentina, \href
  {https://ui.adsabs.harvard.edu/abs/1963BAAA....6...41S} {6, 41}

\bibitem[\protect\citeauthoryear{{Shivaei} et~al.,}{{Shivaei}
  et~al.}{2020}]{shivaei20}
{Shivaei} I.,  et~al., 2020, \mn@doi [\apj] {10.3847/1538-4357/aba35e}, \href
  {https://ui.adsabs.harvard.edu/abs/2020ApJ...899..117S} {899, 117}

\bibitem[\protect\citeauthoryear{{Silva}, {Granato}, {Bressan}  \&
  {Danese}}{{Silva} et~al.}{1998}]{silva98}
{Silva} L.,  {Granato} G.~L.,  {Bressan} A.,   {Danese} L.,  1998, \mn@doi
  [\apj] {10.1086/306476}, \href
  {https://ui.adsabs.harvard.edu/abs/1998ApJ...509..103S} {509, 103}

\bibitem[\protect\citeauthoryear{{Smit} et~al.,}{{Smit} et~al.}{2014}]{smit14}
{Smit} R.,  et~al., 2014, \mn@doi [\apj] {10.1088/0004-637X/784/1/58}, \href
  {http://adsabs.harvard.edu/abs/2014ApJ...784...58S} {784, 58}

\bibitem[\protect\citeauthoryear{{Song} et~al.,}{{Song} et~al.}{2016}]{song16}
{Song} M.,  et~al., 2016, \mn@doi [\apj] {10.3847/0004-637X/825/1/5}, \href
  {https://ui.adsabs.harvard.edu/abs/2016ApJ...825....5S} {825, 5}

\bibitem[\protect\citeauthoryear{{Stark}, {Ellis}, {Bunker}, {Bundy},
  {Targett}, {Benson}  \& {Lacy}}{{Stark} et~al.}{2009}]{stark09}
{Stark} D.~P.,  {Ellis} R.~S.,  {Bunker} A.,  {Bundy} K.,  {Targett} T.,
  {Benson} A.,   {Lacy} M.,  2009, \mn@doi [\apj]
  {10.1088/0004-637X/697/2/1493}, \href
  {https://ui.adsabs.harvard.edu/abs/2009ApJ...697.1493S} {697, 1493}

\bibitem[\protect\citeauthoryear{{Stefanon}, {Bouwens}, {Labb{\'e}},
  {Illingworth}, {Gonzalez}  \& {Oesch}}{{Stefanon} et~al.}{2021}]{stefanon21}
{Stefanon} M.,  {Bouwens} R.~J.,  {Labb{\'e}} I.,  {Illingworth} G.~D.,
  {Gonzalez} V.,   {Oesch} P.~A.,  2021, \mn@doi [\apj]
  {10.3847/1538-4357/ac1bb6}, \href
  {https://ui.adsabs.harvard.edu/abs/2021ApJ...922...29S} {922, 29}

\bibitem[\protect\citeauthoryear{Steidel, Giavalisco, Pettini, Dickinson  \&
  Adelberger}{Steidel et~al.}{1996}]{steidel96}
Steidel C.~C.,  Giavalisco M.,  Pettini M.,  Dickinson M.,   Adelberger K.~L.,
  1996, \mn@doi [The Astrophysical Journal] {10.1088/1538-4357/462/1/L17}, 462,
  L17

\bibitem[\protect\citeauthoryear{{Ziparo}, {Ferrara}, {Sommovigo}  \&
  {Kohandel}}{{Ziparo} et~al.}{2022}]{ziparo22}
{Ziparo} F.,  {Ferrara} A.,  {Sommovigo} L.,   {Kohandel} M.,  2022, arXiv
  e-prints, \href {https://ui.adsabs.harvard.edu/abs/2022arXiv220906840Z} {p.
  arXiv:2209.06840}

\bibitem[\protect\citeauthoryear{{de Barros}, {Schaerer}  \& {Stark}}{{de
  Barros} et~al.}{2014}]{debarros14}
{de Barros} S.,  {Schaerer} D.,   {Stark} D.~P.,  2014, \mn@doi [\aap]
  {10.1051/0004-6361/201220026}, \href
  {http://adsabs.harvard.edu/abs/2014A%26A...563A..81D} {563, A81}

\makeatother
\end{thebibliography}

\appendix

\section{Photometry validation} \label{app:photometry-validation}

\begin{figure}
    \centering
    \includegraphics[width=0.9\columnwidth, keepaspectratio=true]{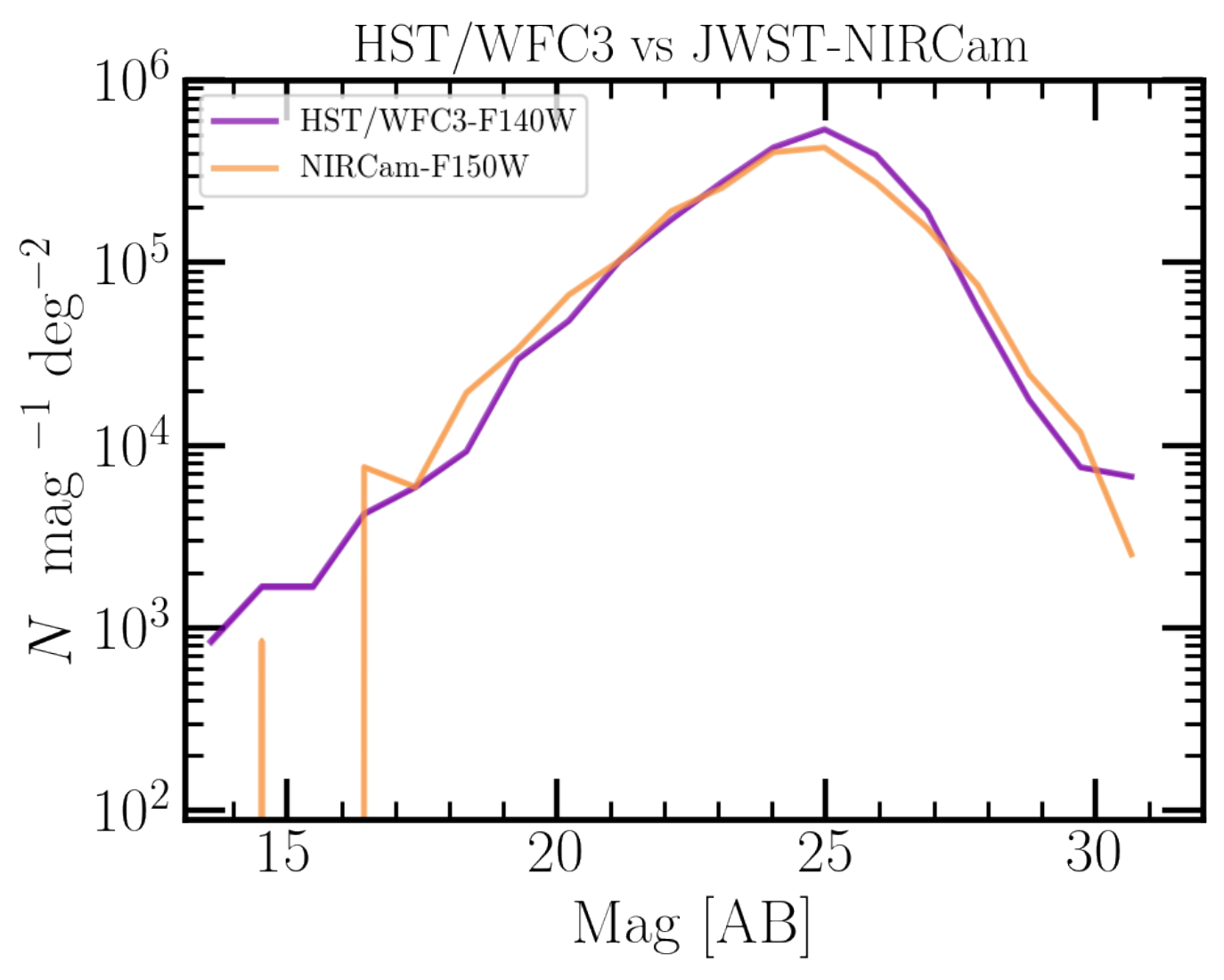}
    \caption{Magnitude number counts in the HST WFC3/IR F140W and JWST NIRCam F150W bands. The magnitudes are obtained for objects detected in the shallower F140W-band and measured simultaneously in F140W and F150W. The number counts show a good agreement and no photometric offset is present.}
    \label{fig:mag-numbercounts-comparison}
\end{figure}

\begin{figure}
    \centering
    \includegraphics[width=\columnwidth, keepaspectratio=true]{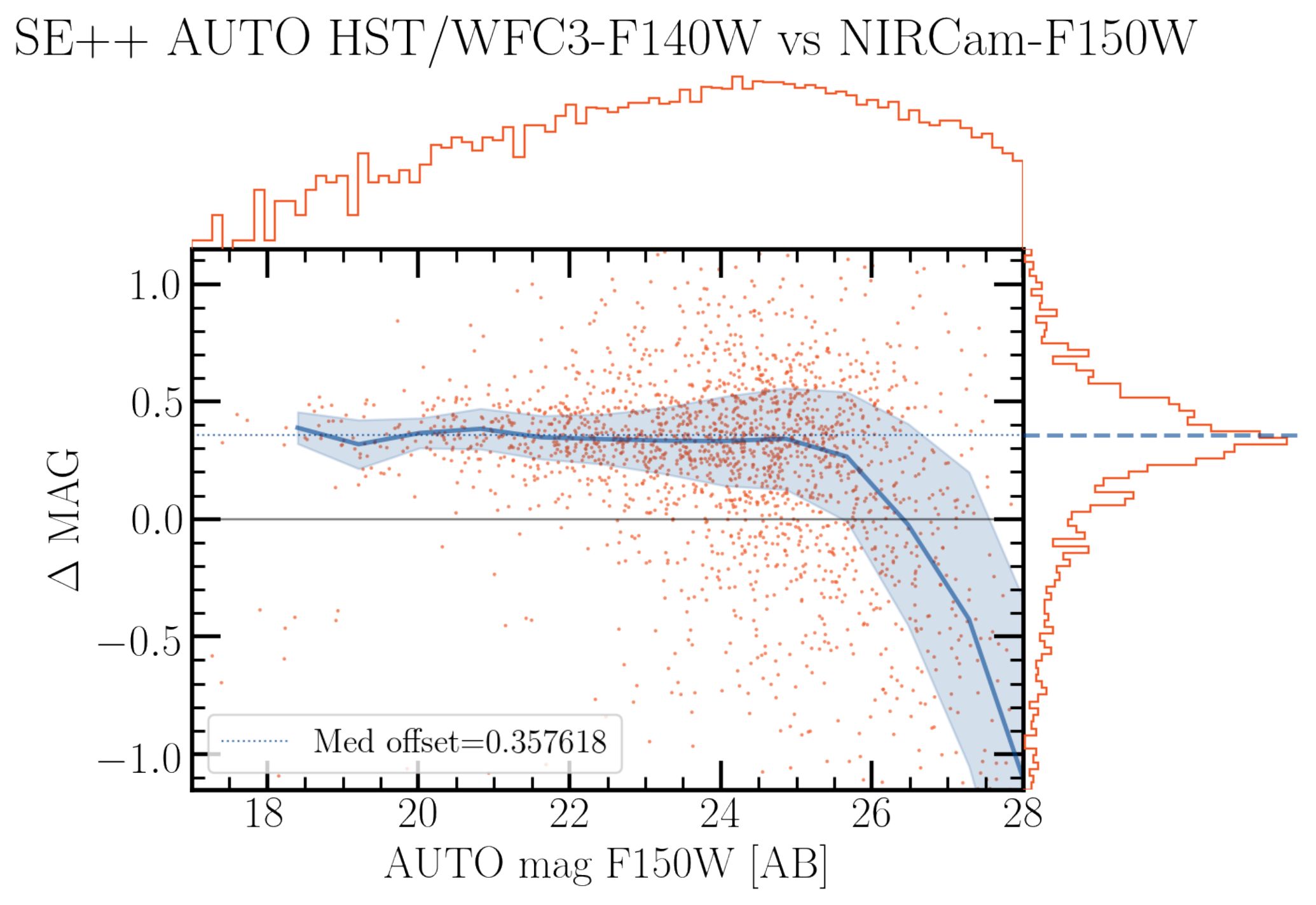}
    \caption{Magnitude comparison between \hst-WFC3/IR F140W and \jwst-NIRCam F150W bands. The blue line marks the running median offset and the blue envelope the $\sigma$-clipped standard deviation. The magnitudes of the two bands show a relatively tight dispersion and an offset of about 0.24 mag, that can be explained  by the difference between the filter curves and pivot wavelengths.}
    \label{fig:mag-comparison}
\end{figure}

We perform photometry validation by comparing our photometry with the well known and calibrated HST WFC3/IR images from the RELICS survey. We run {\tt SE++} on a stacked detection image and do photometric measurements simultaneously on the WFC3/IR F140W and NIRCam F150W bands. 

Fig.~\ref{fig:mag-numbercounts-comparison} shows the \texttt{AUTO} magnitude number counts in NIRCam F150W and WFC3/IR F140W as measured with {\tt SE++}. The number counts are in excellent agreement, which is not surprising given the same detection image and with minimal or no offsets. This confirms the accurate zero-point calibration of our JWST data reduction, explained in section~\ref{sec:obs}. The comparison of magnitudes is shown in Fig~\ref{fig:mag-number-counts}, where on the $y-$axis we show the magnitude difference between the F140W and F150W bands. There is a relatively tight scatter and a median offset of about $0.36$ towards fainter F140W magnitudes. Most of this offset can be attributed to the difference between the F140W and F150W filter curves and pivot wavelengths. Using {\tt Pysynphot}, we performed synthetic photometry in the two filters, assuming a variety of galaxy templates from BC95 \citep{bc93} and KC96 \citep{kc96} libraries, and a redshift range of $z=0-5$. We found a median offset of $\sim 0.3$\,magnitudes.

\section{Example of a $z \sim 10$ candidate} \label{app:z-10_example}

\begin{figure*}
    \centering
    \includegraphics[width=0.9\textwidth]{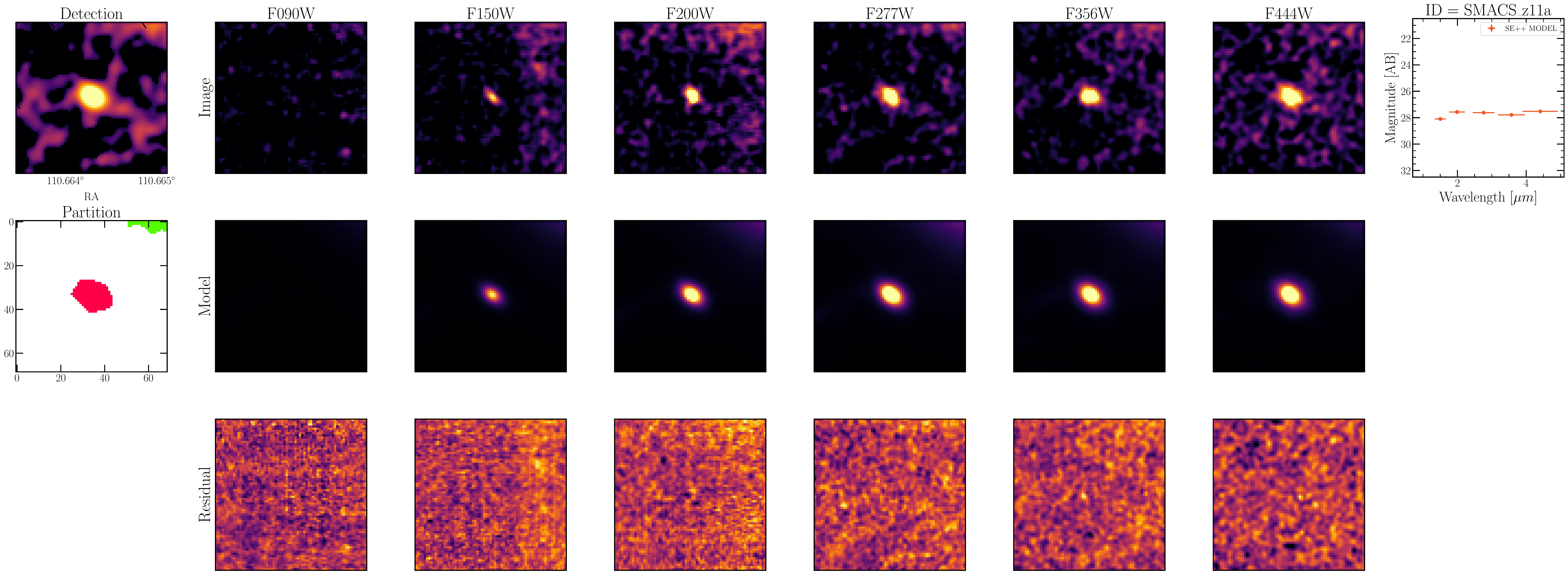}
    \caption{Same as Fig.~\ref{fig:candidate-stamp-id-2455}, this time for the $z\sim12$ candidate galaxy SMACS\_z11a.}
    \label{fig:candidate-stamp-id-323}
\end{figure*}

\begin{figure*}
    \centering
    \includegraphics[width=0.49\textwidth]{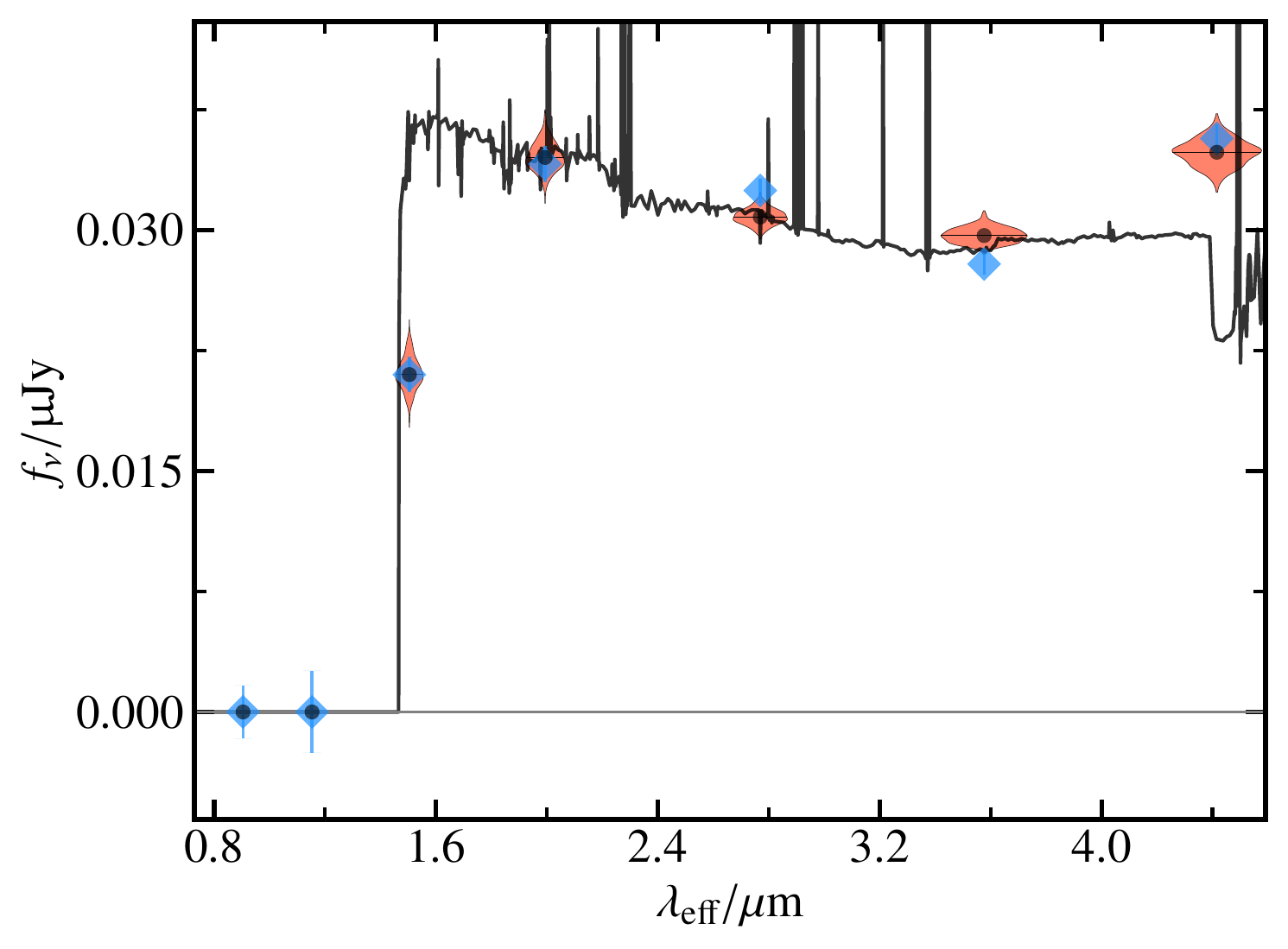}
    \includegraphics[width=0.49\textwidth]{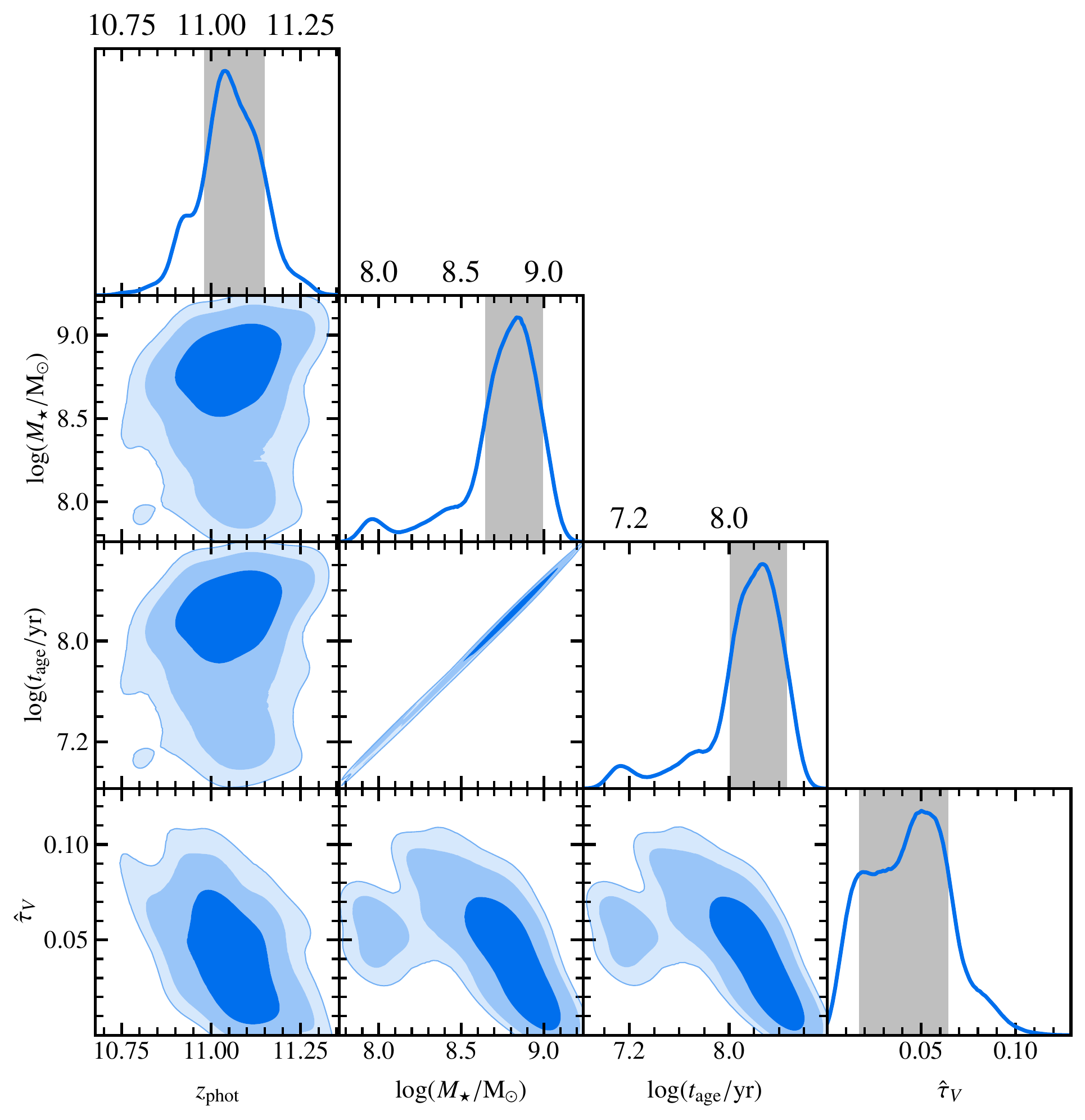}\\[1cm]
    \includegraphics[width=0.9\textwidth]{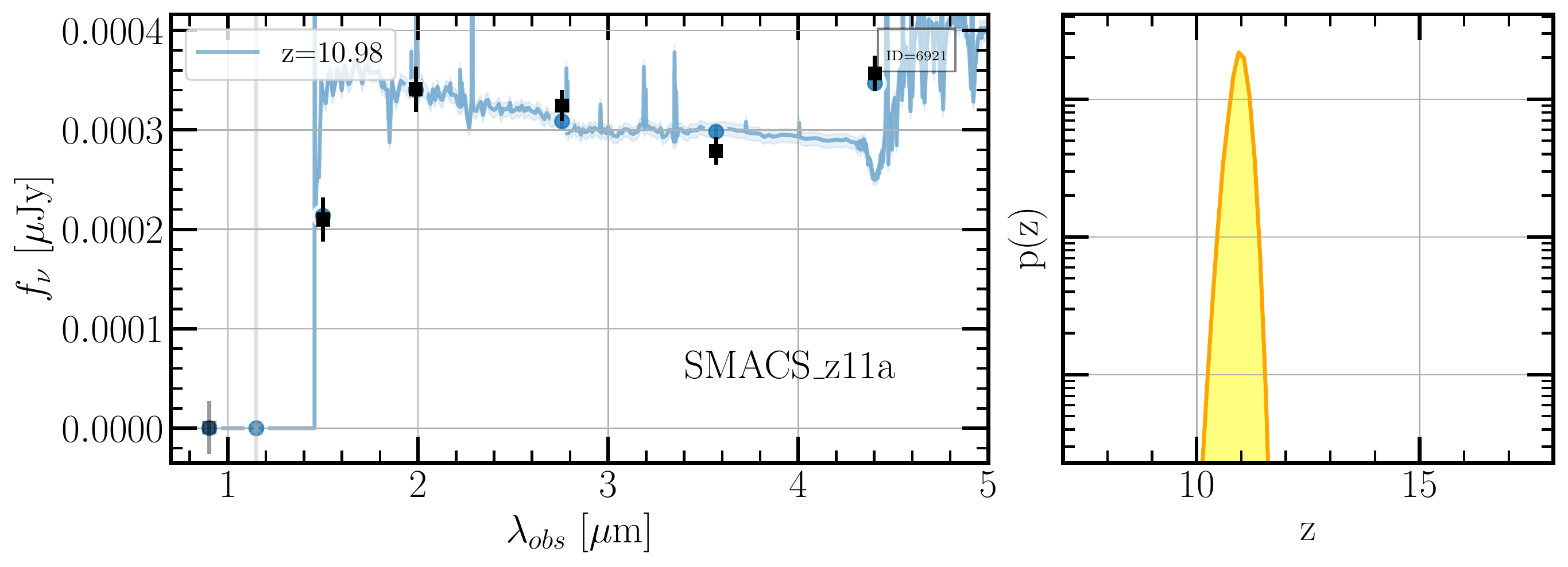}
    \caption{Same as Fig.~\ref{fig:SEF-fit-id-2455}, here showing the galaxy candidate SMACS\_z11a.}
    \label{fig:SEF-fit-id-323}
\end{figure*}

In addition to the $z\sim16$ and $z\sim12$ examples shown in the main text (cf. section~\ref{sec:results}), we display here an example of a $z\sim10$ galaxy, the candidate SMACS\_z11a, in Figs~\ref{fig:candidate-stamp-id-323} and~\ref{fig:SEF-fit-id-323}.

\section{Full candidate list}

\begin{figure*}
    \centering
    \includegraphics[width=0.9\textwidth]{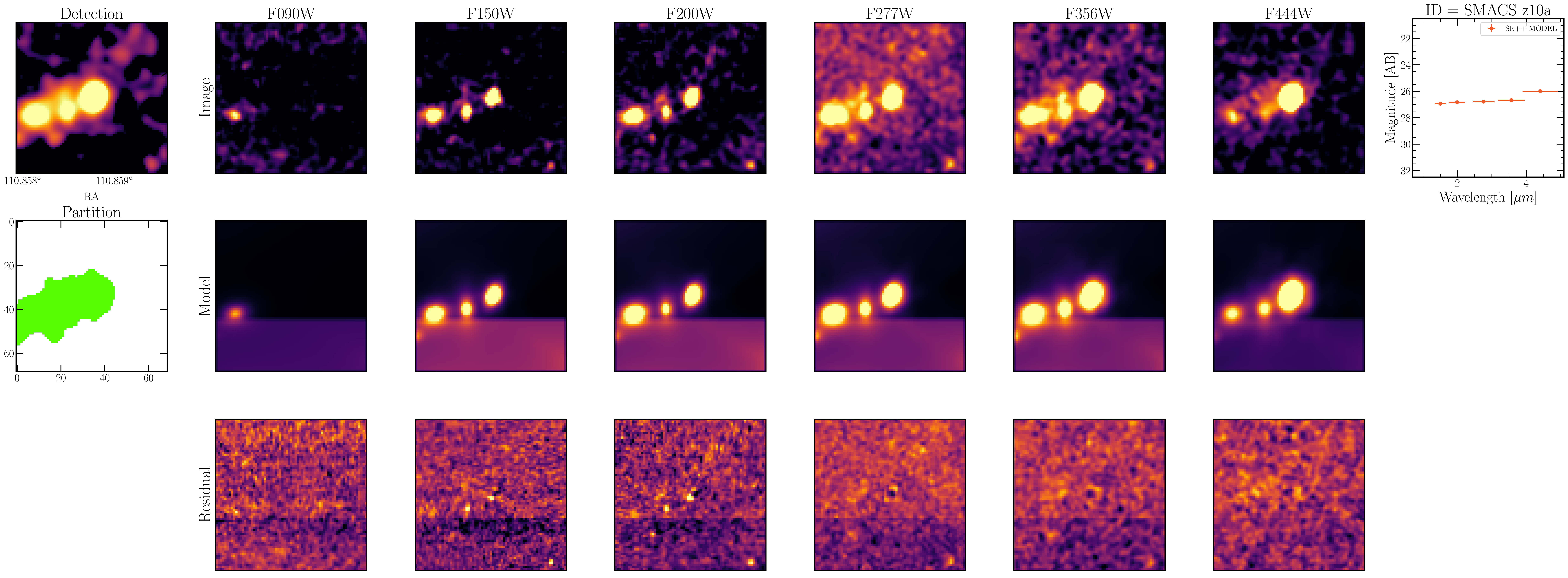}
    \includegraphics[width=0.49\textwidth]{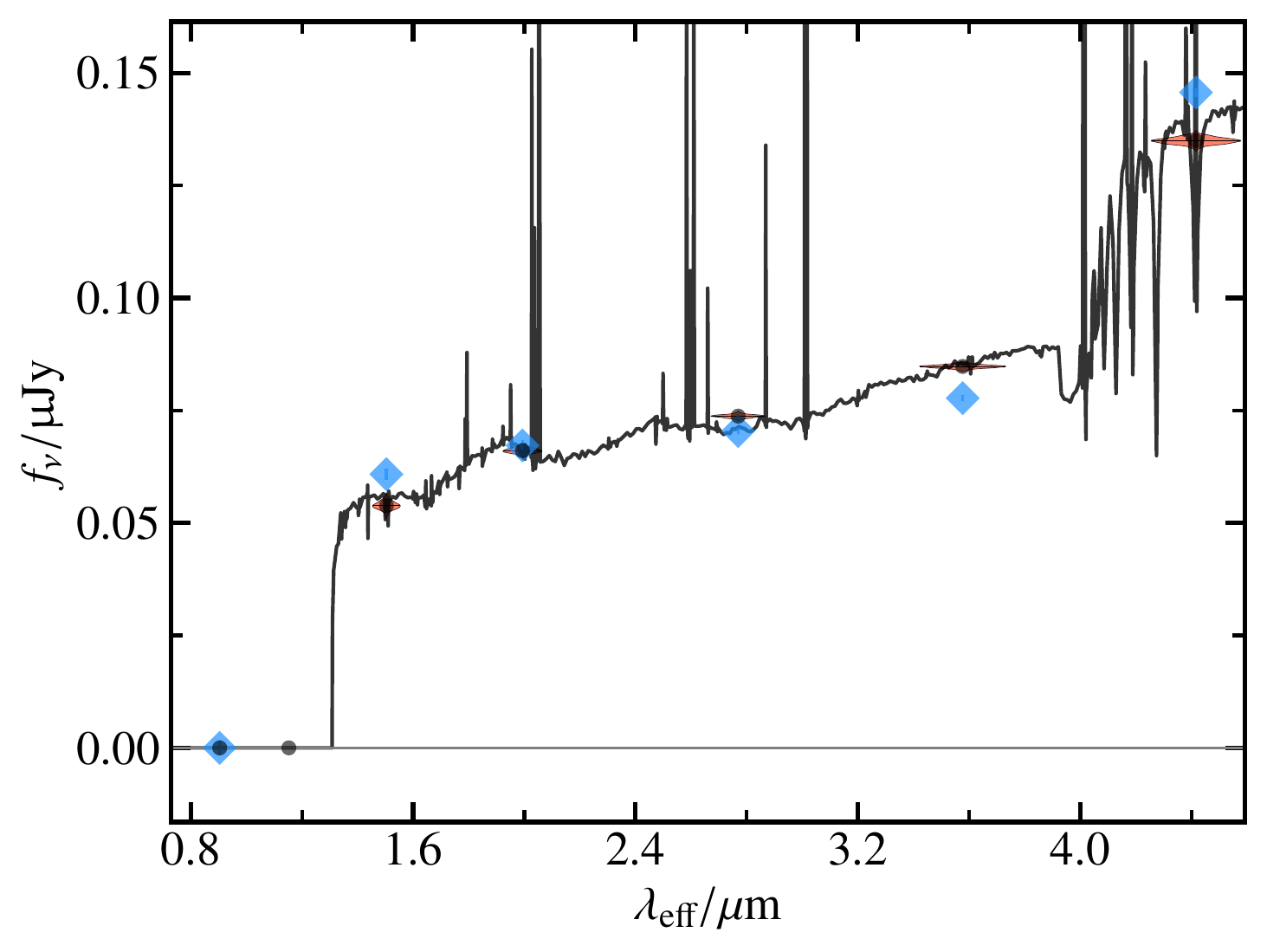}
    \includegraphics[width=0.49\textwidth]{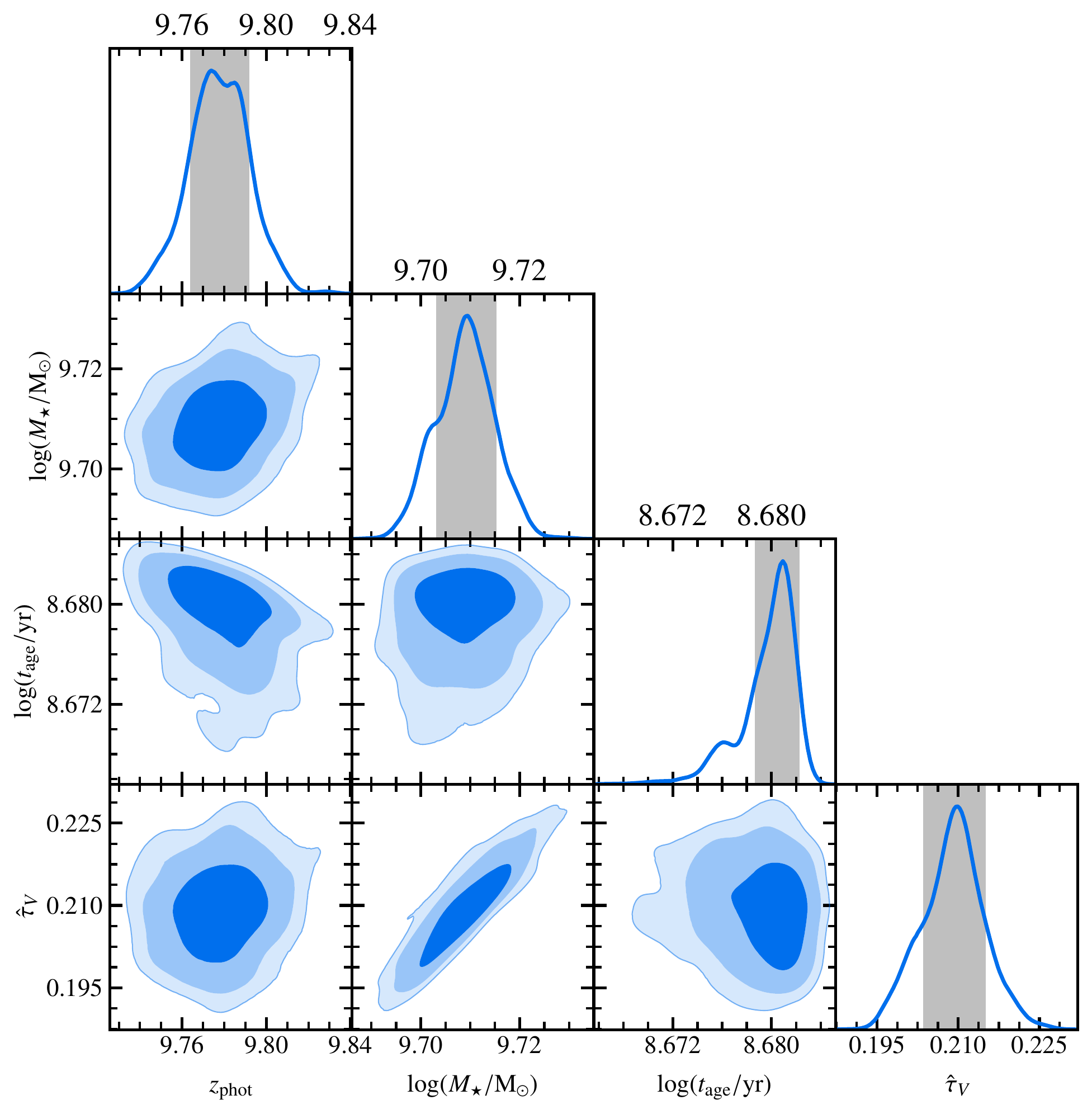}\\[1cm]
    \includegraphics[width=0.9\textwidth]{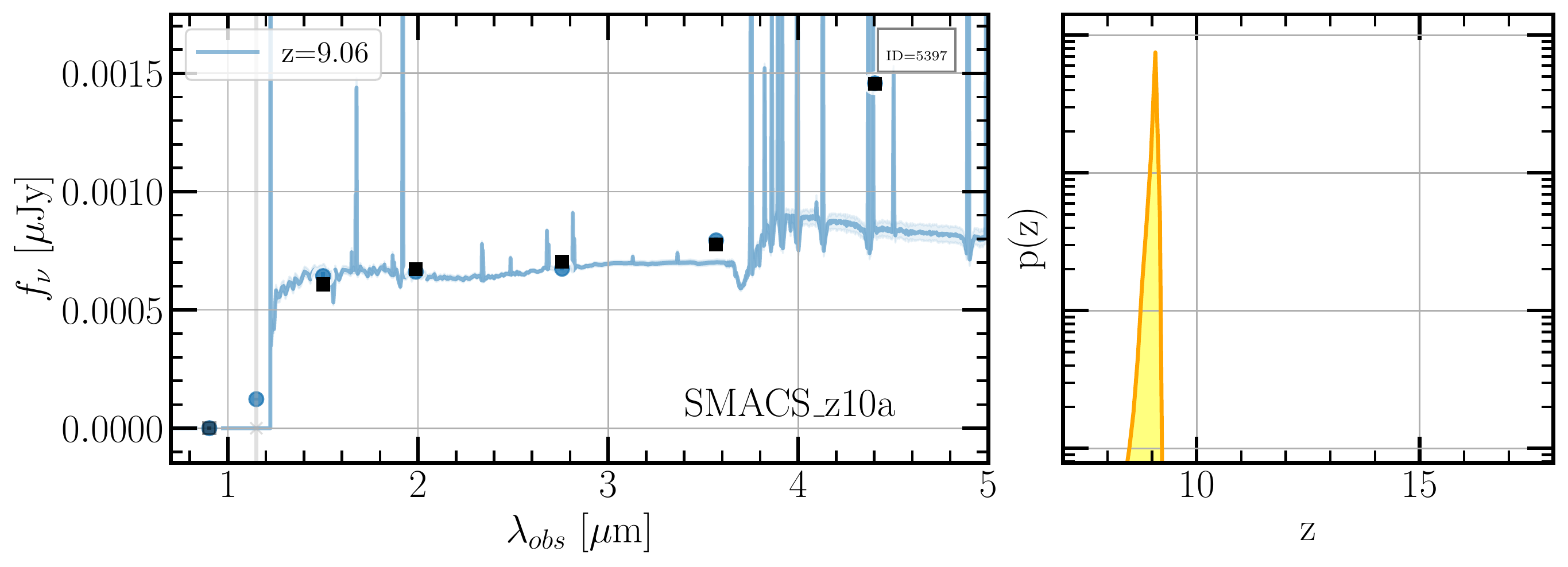}
    \caption{Photometric data and best-fit SED results for candidate SMACS\_z10a.}
    \label{fig:id-4482}
\end{figure*}

\begin{figure*}
    \centering
    \includegraphics[width=0.9\textwidth]{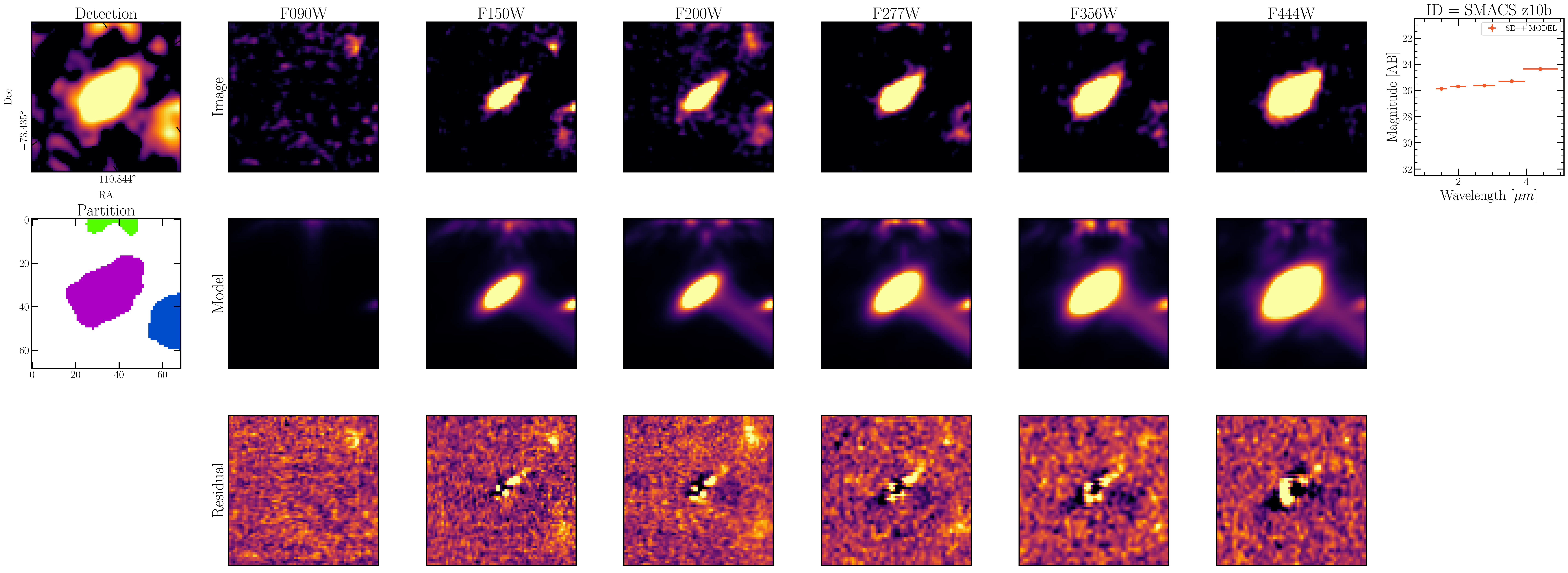}
    \includegraphics[width=0.49\textwidth]{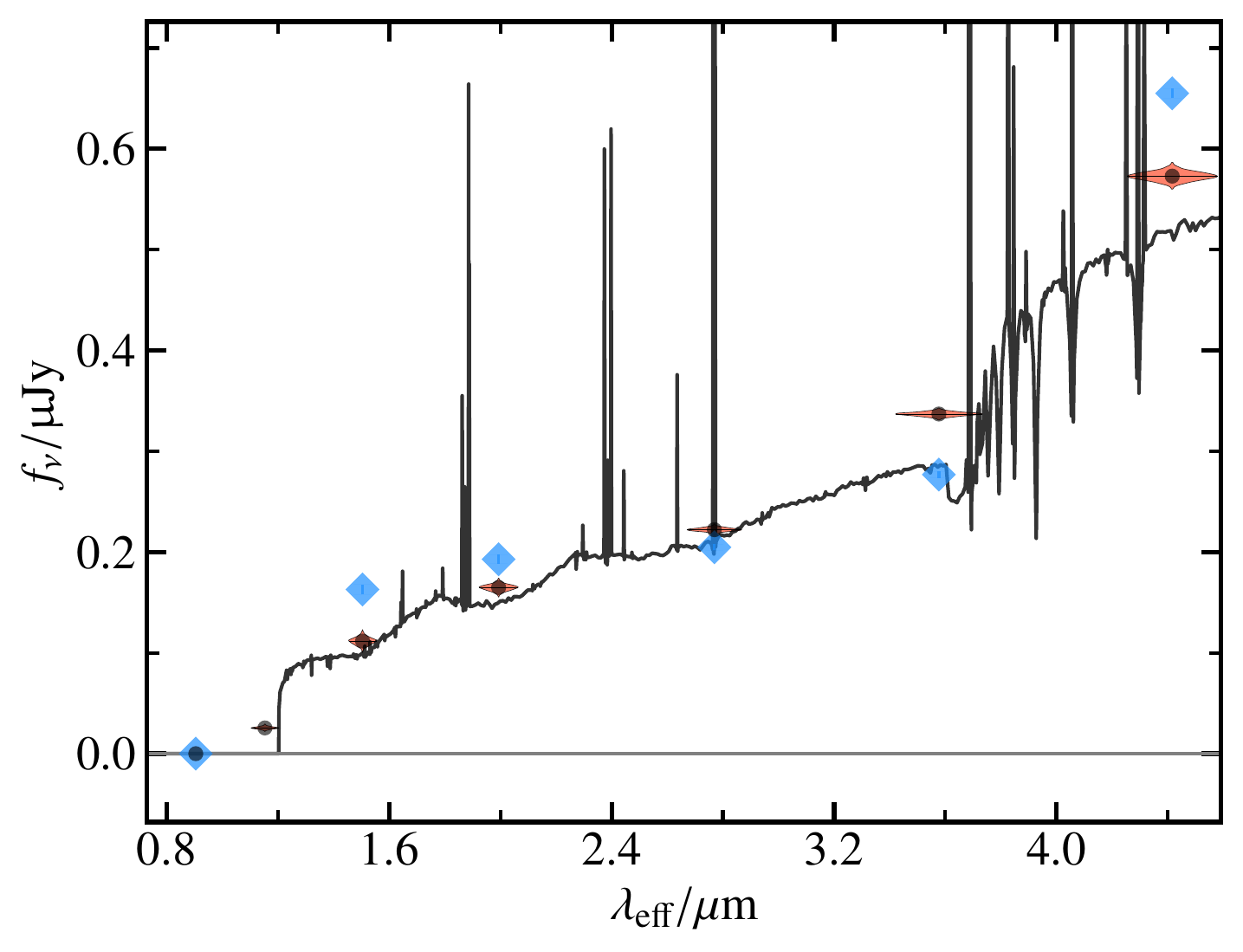}
    \includegraphics[width=0.49\textwidth]{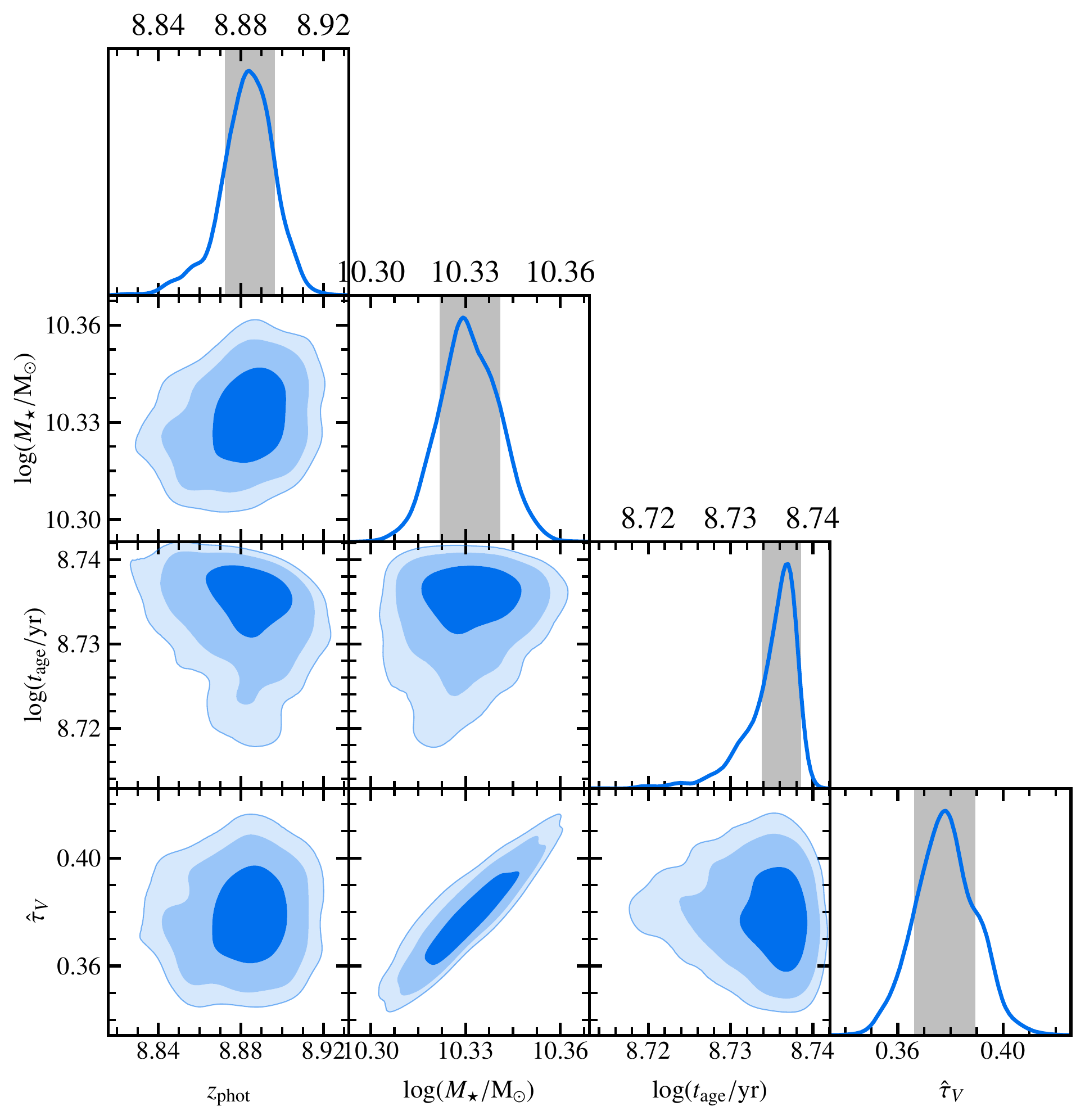}\\[1cm]
    \includegraphics[width=0.9\textwidth]{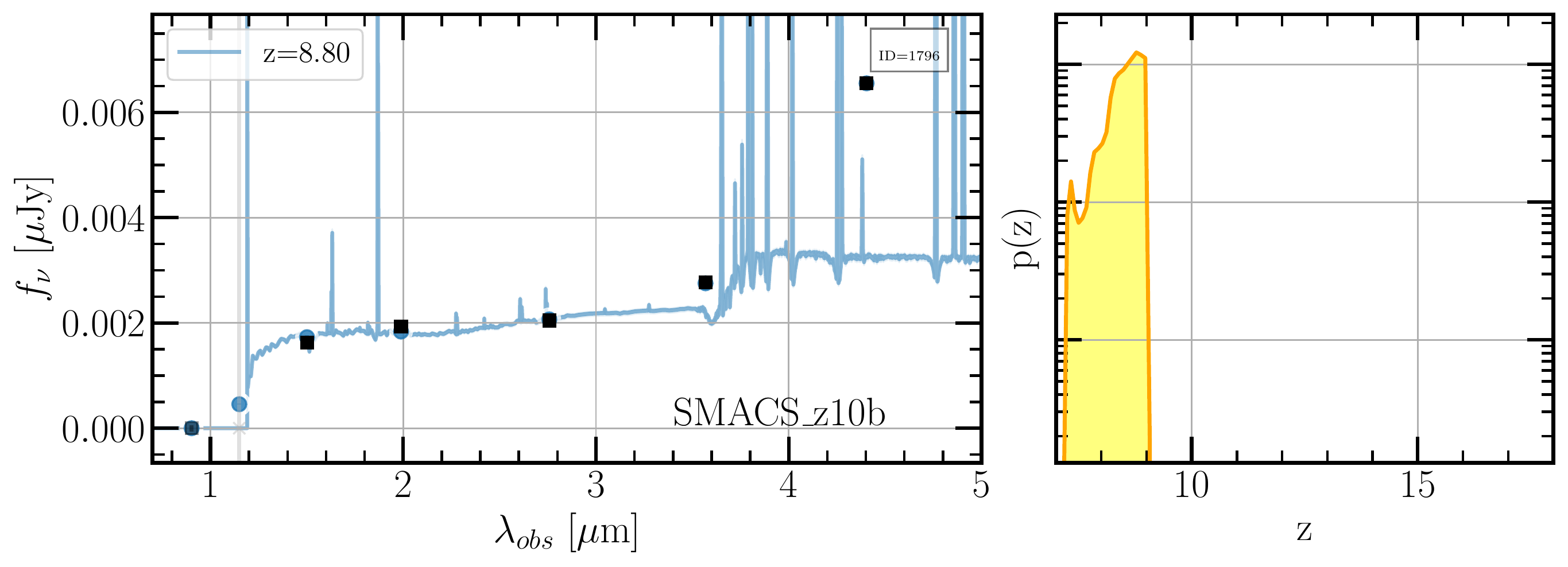}
    \caption{Photometric data and best-fit SED results for candidate SMACS\_z10b.}
    \label{fig:id-5388}
\end{figure*}

\begin{figure*}
    \centering
    \includegraphics[width=0.9\textwidth]{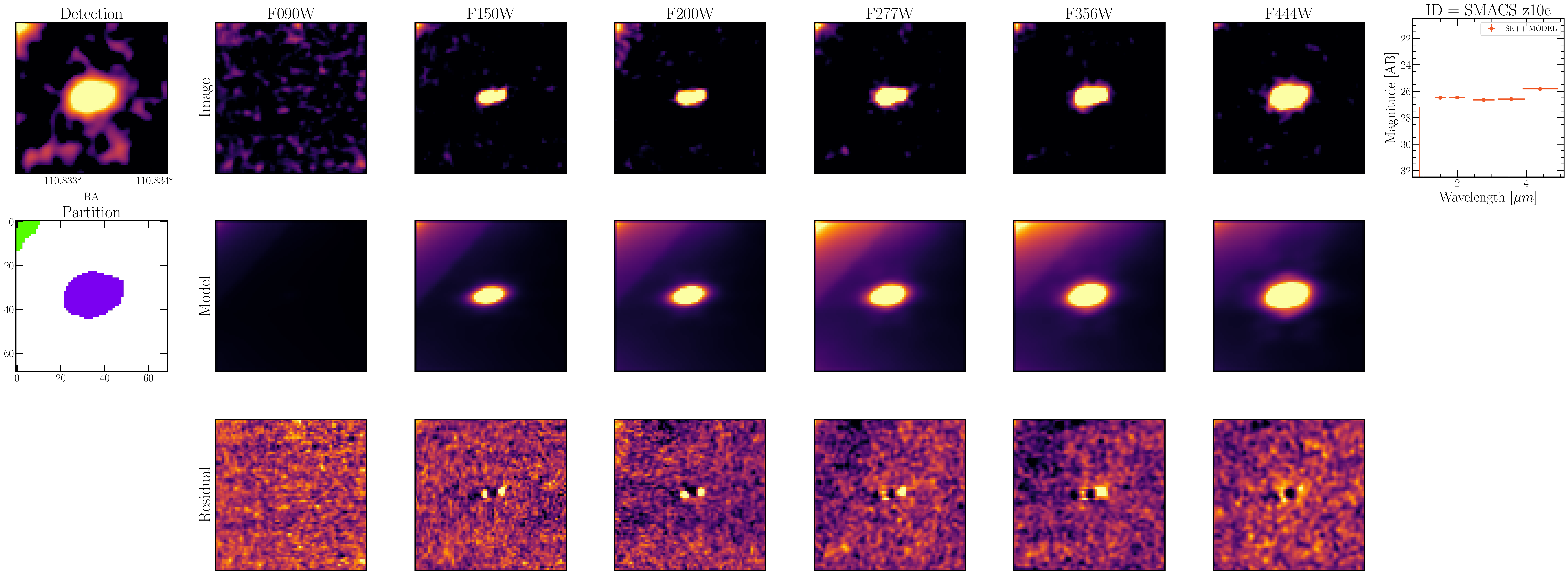}
    \includegraphics[width=0.49\textwidth]{5414_BEAGLE_marginal_SED_phot.pdf}
    \includegraphics[width=0.49\textwidth]{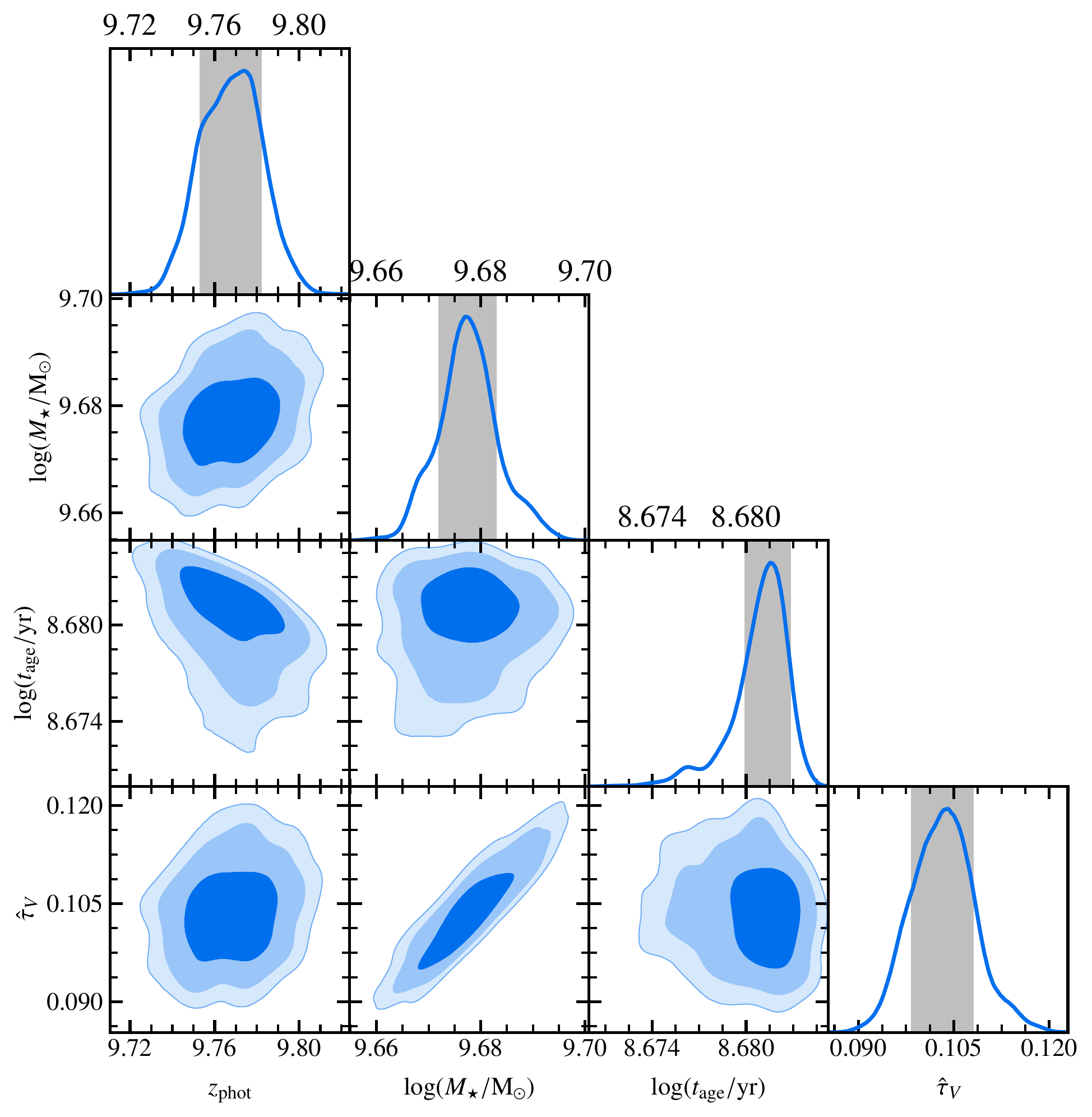}\\[1cm]
    \includegraphics[width=0.9\textwidth]{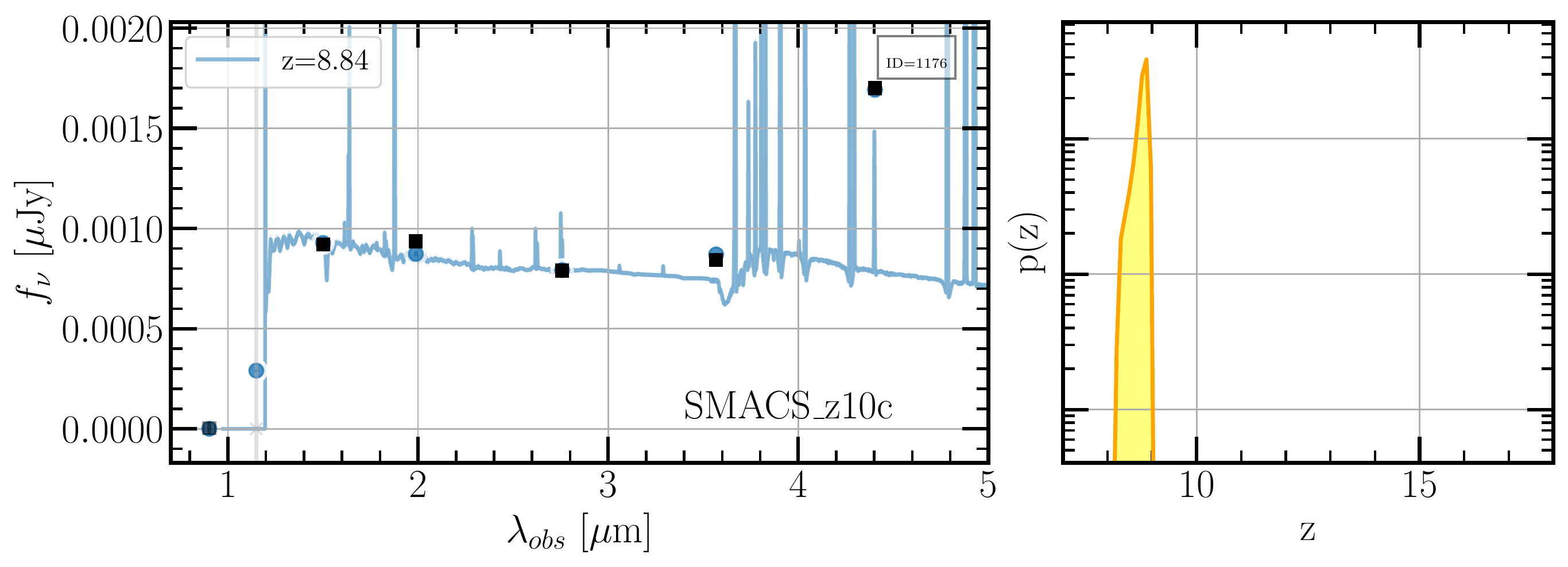}
    \caption{Photometric data and best-fit SED results for candidate SMACS\_z10c.}
    \label{fig:id-5414}
\end{figure*}

\begin{figure*}
    \centering
    \includegraphics[width=0.9\textwidth]{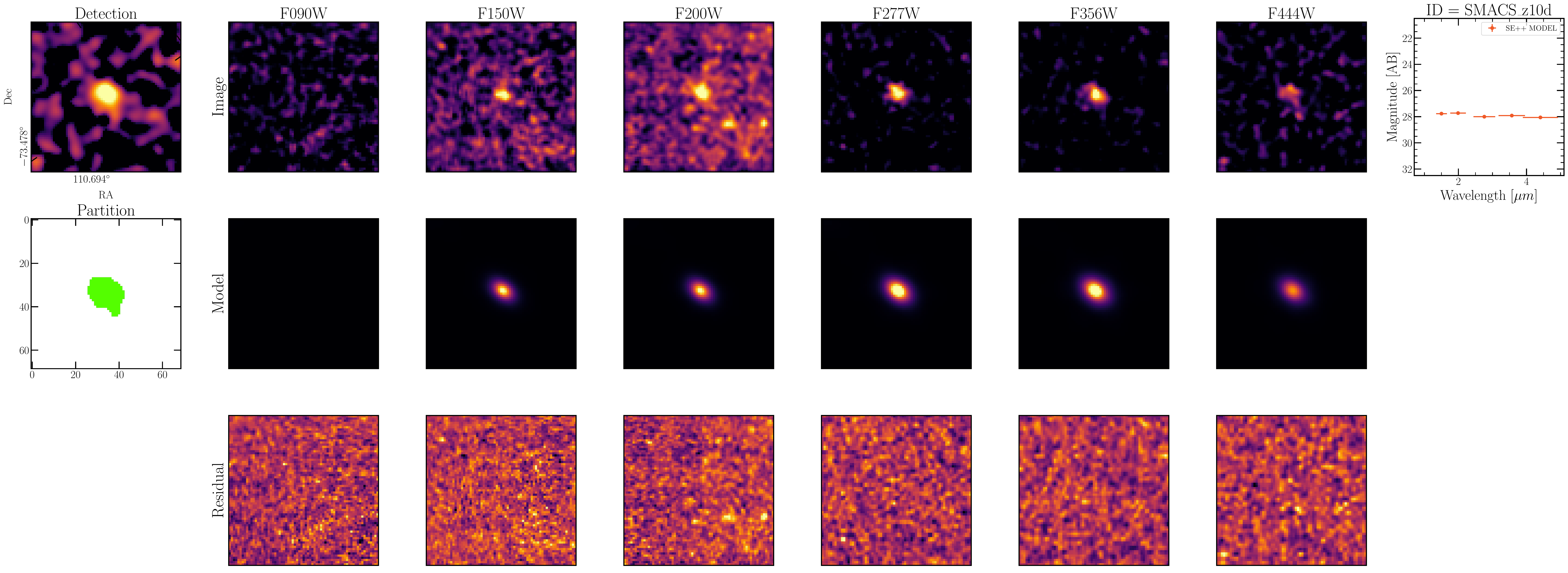}
    \includegraphics[width=0.49\textwidth]{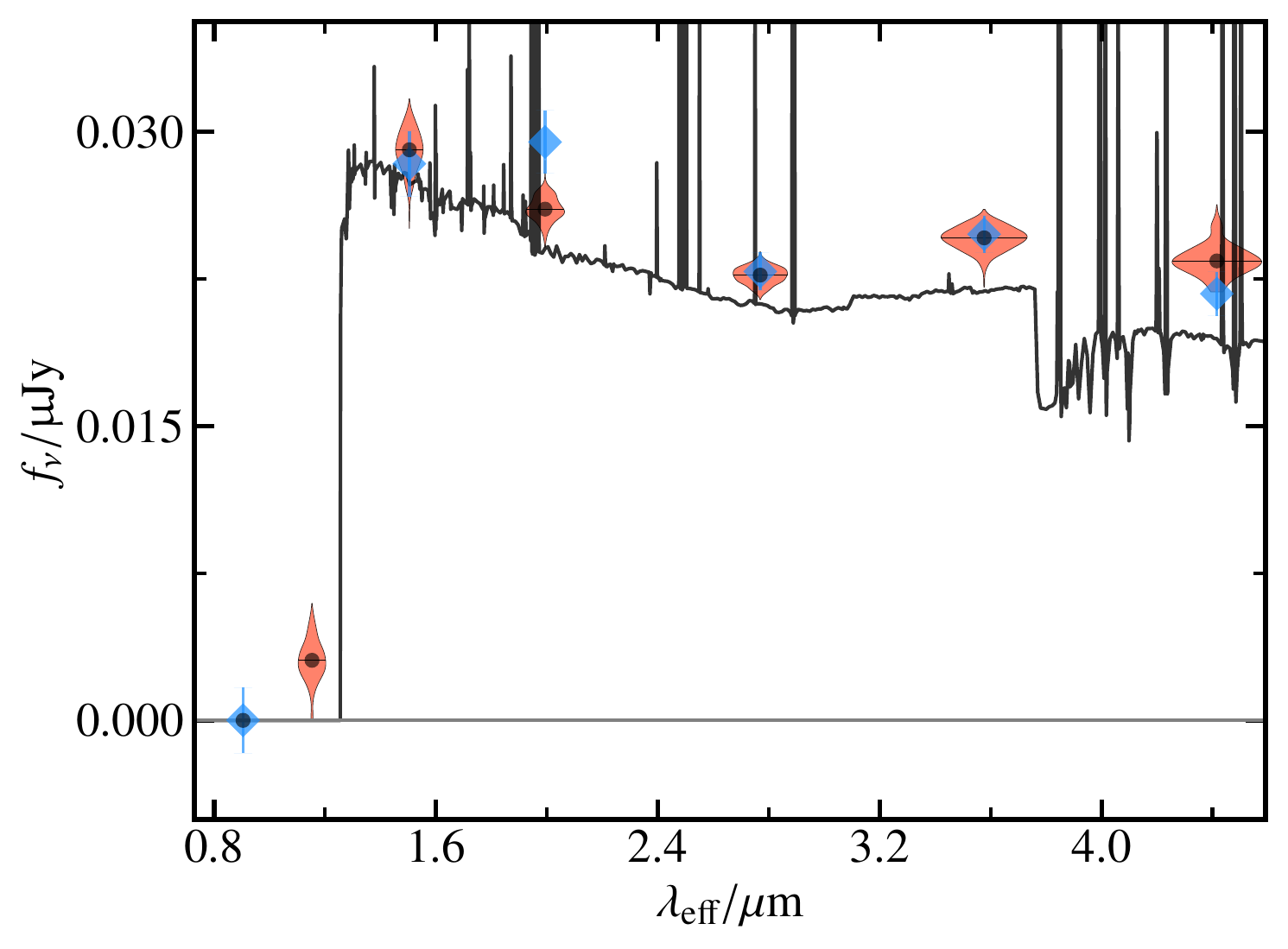}
    \includegraphics[width=0.49\textwidth]{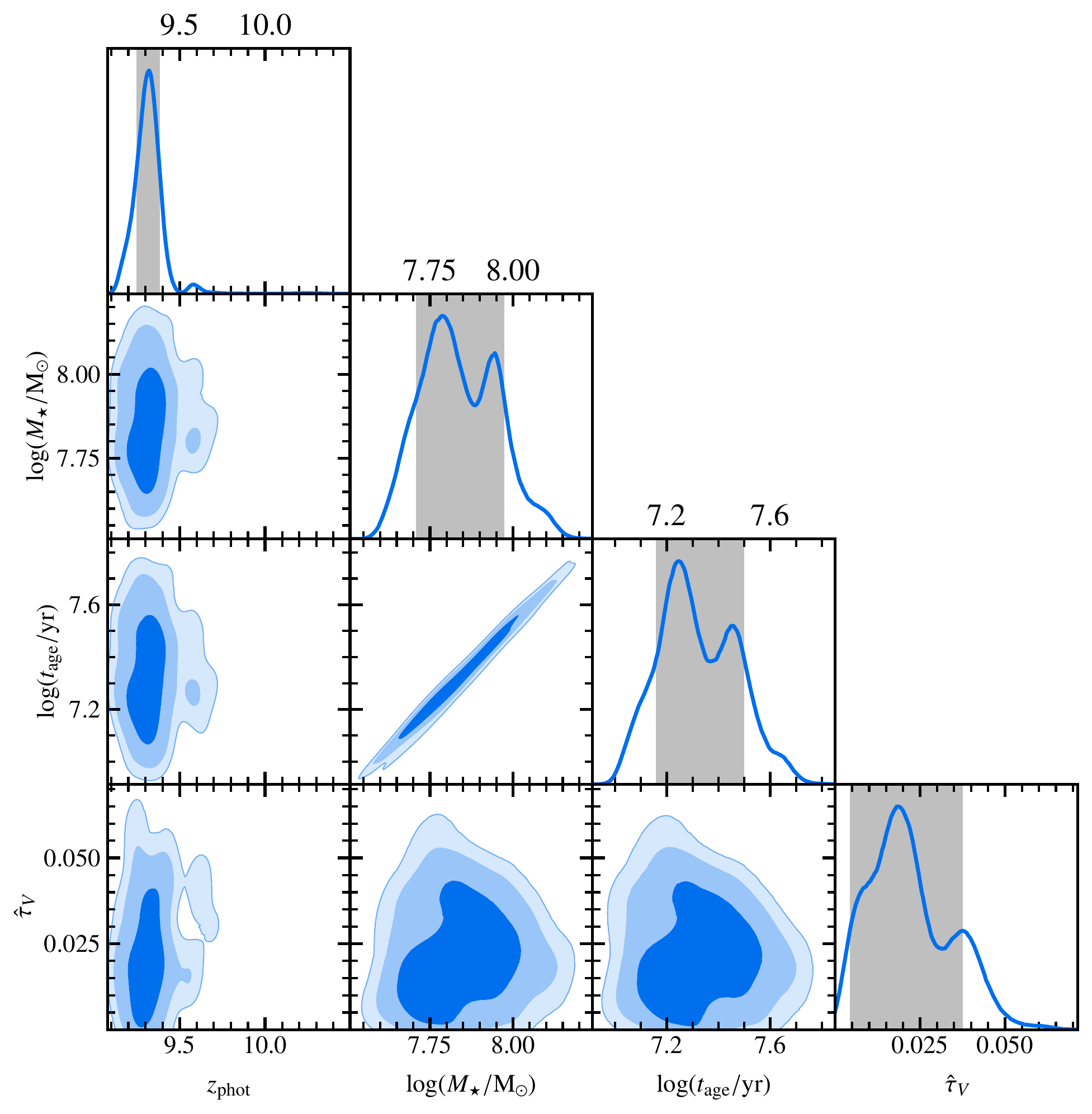}\\[1cm]
    \includegraphics[width=0.9\textwidth]{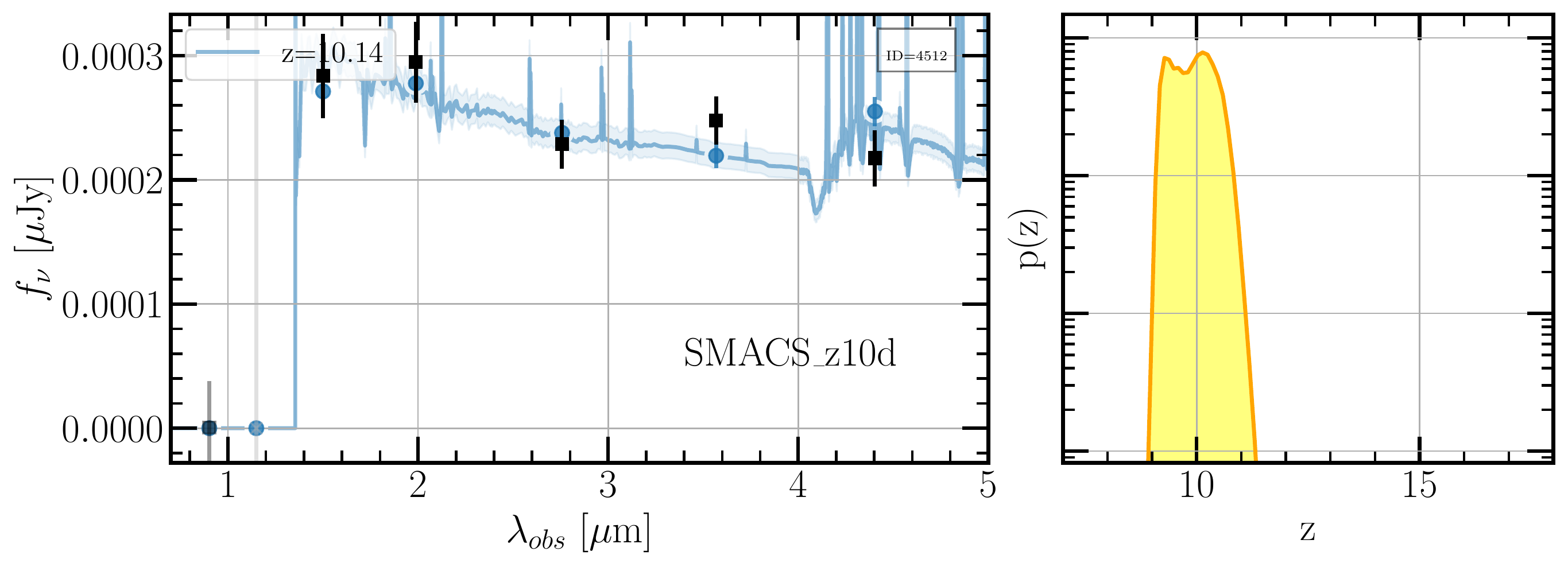}
    \caption{Photometric data and best-fit SED results for candidate SMACS\_z10d.}
    \label{fig:id-1457}
\end{figure*}

\begin{figure*}
    \centering
    \includegraphics[width=0.9\textwidth]{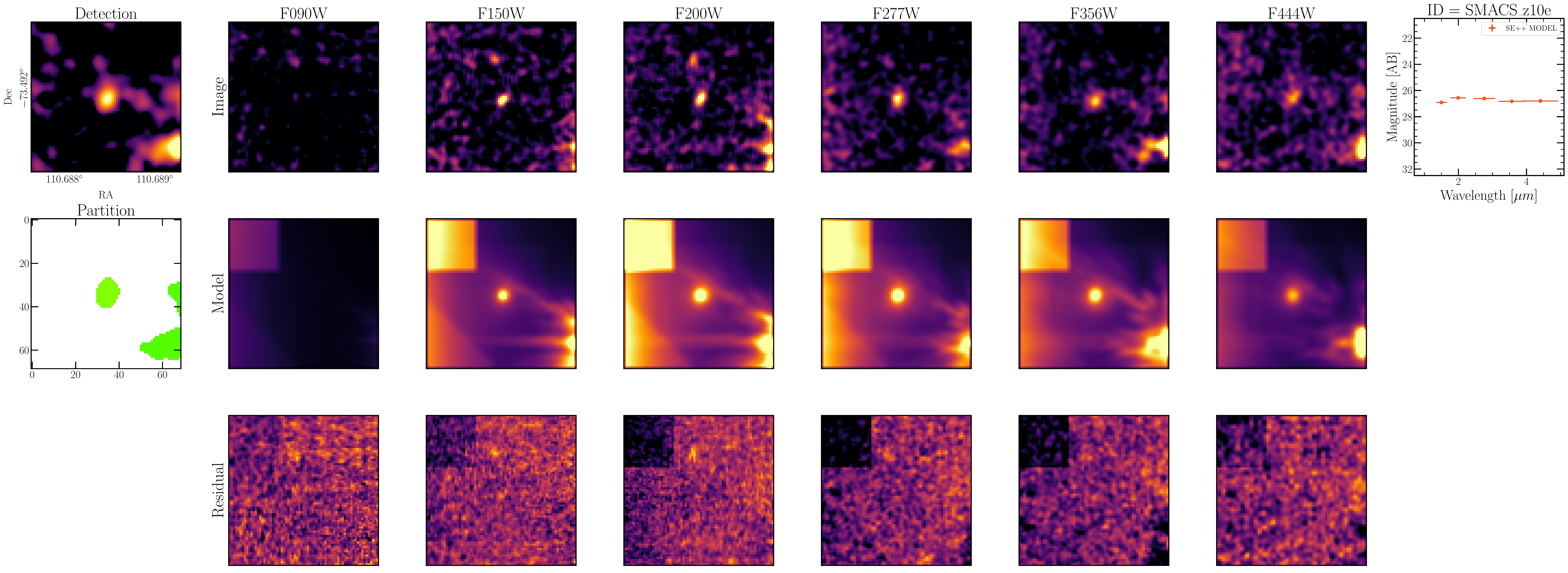}
    \includegraphics[width=0.49\textwidth]{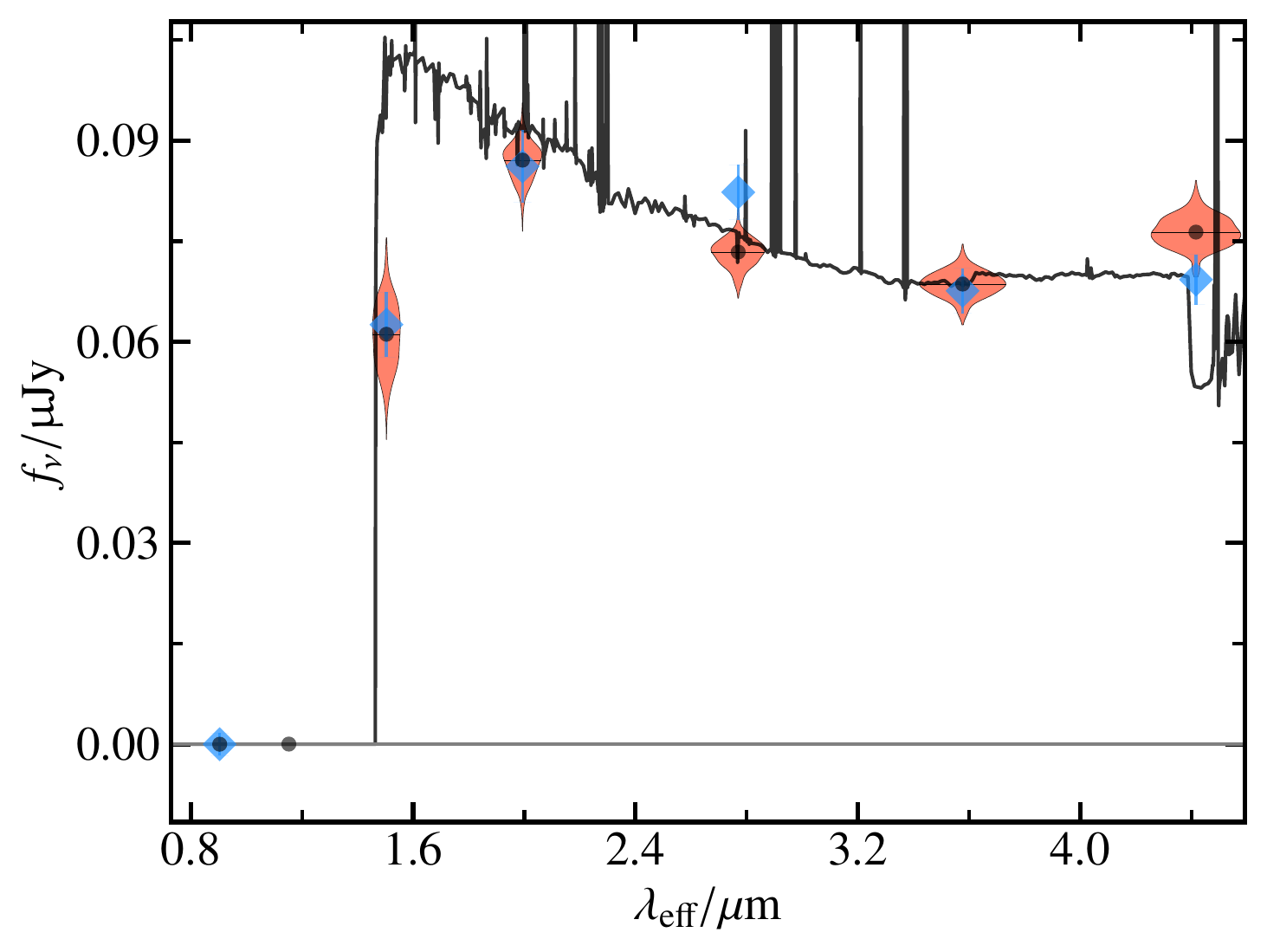}
    \includegraphics[width=0.49\textwidth]{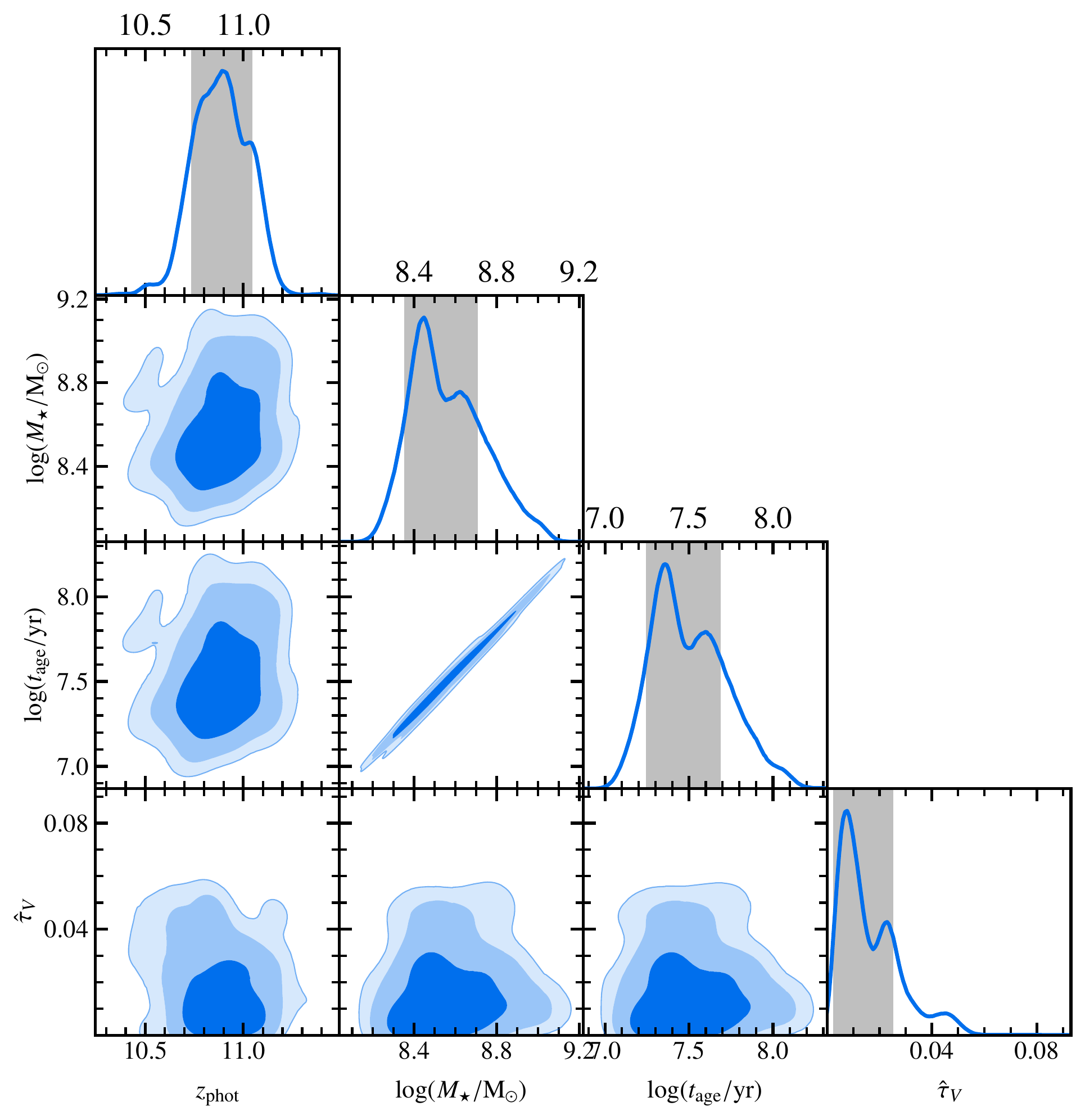}\\[1cm]
    \includegraphics[width=0.9\textwidth]{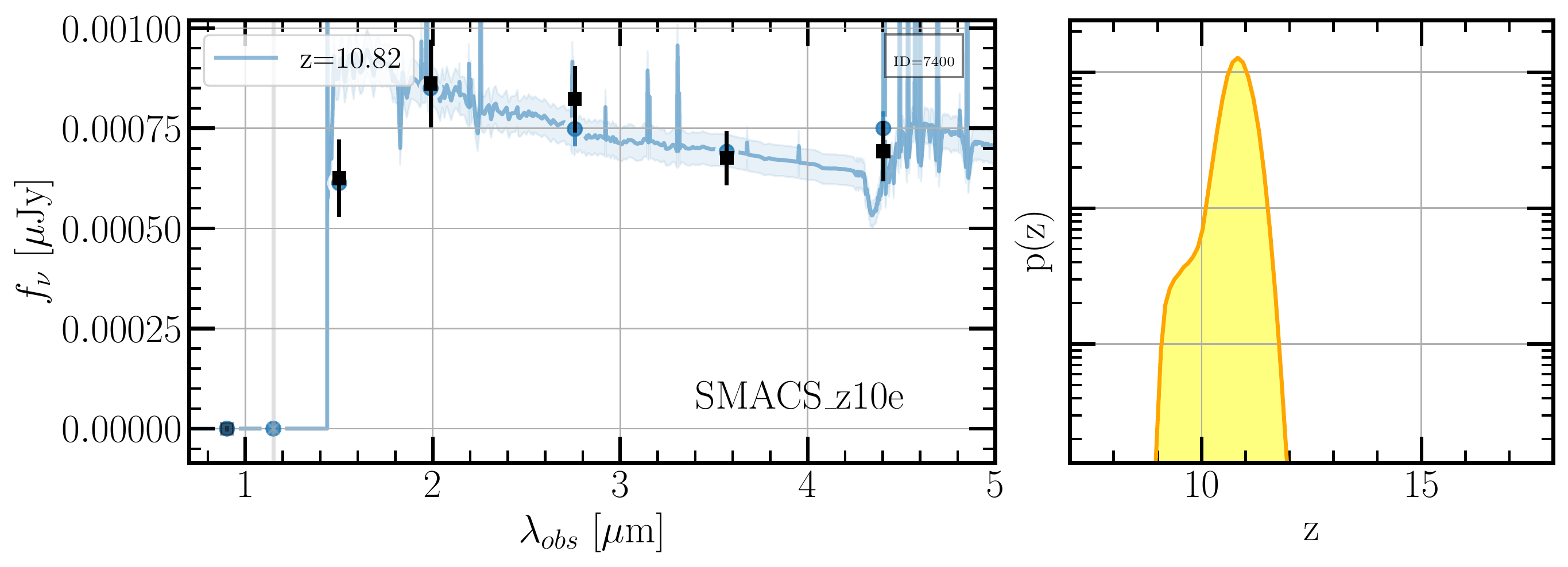}
    \caption{Photometric data and best-fit SED results for candidate SMACS\_z10e.}
    \label{fig:id-458}
\end{figure*}

\begin{figure*}
    \centering
    \includegraphics[width=0.9\textwidth]{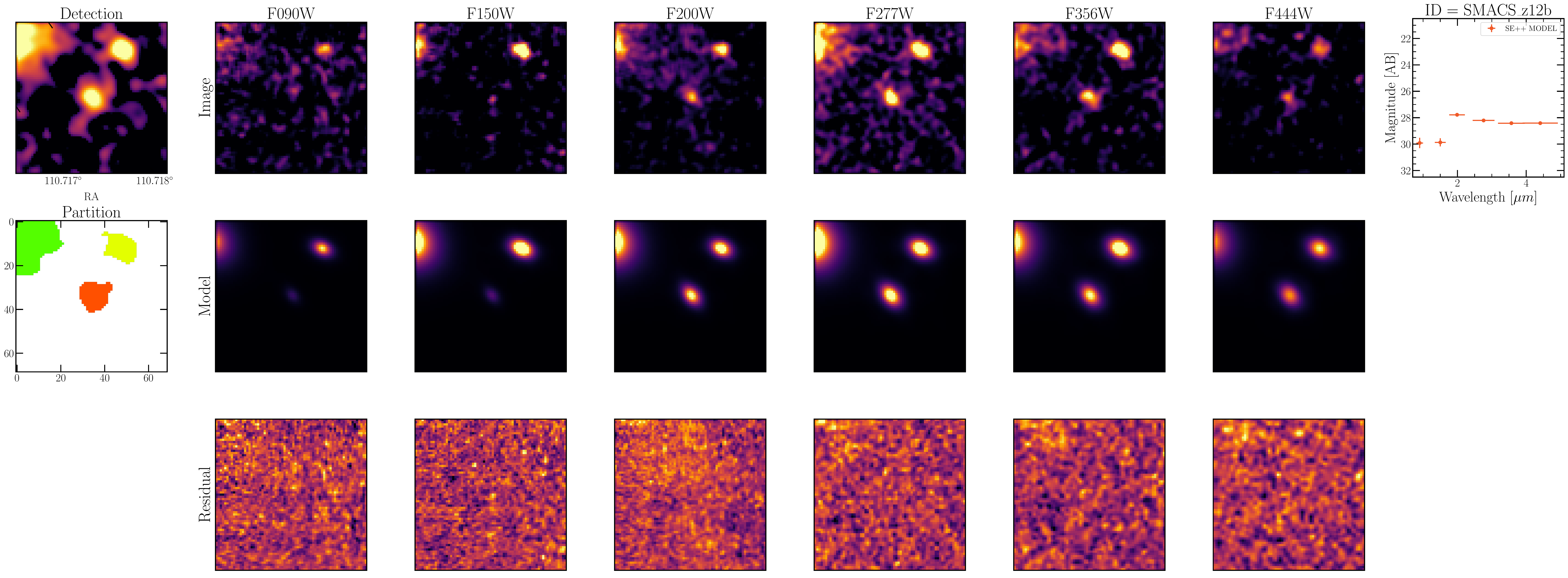}
    \includegraphics[width=0.49\textwidth]{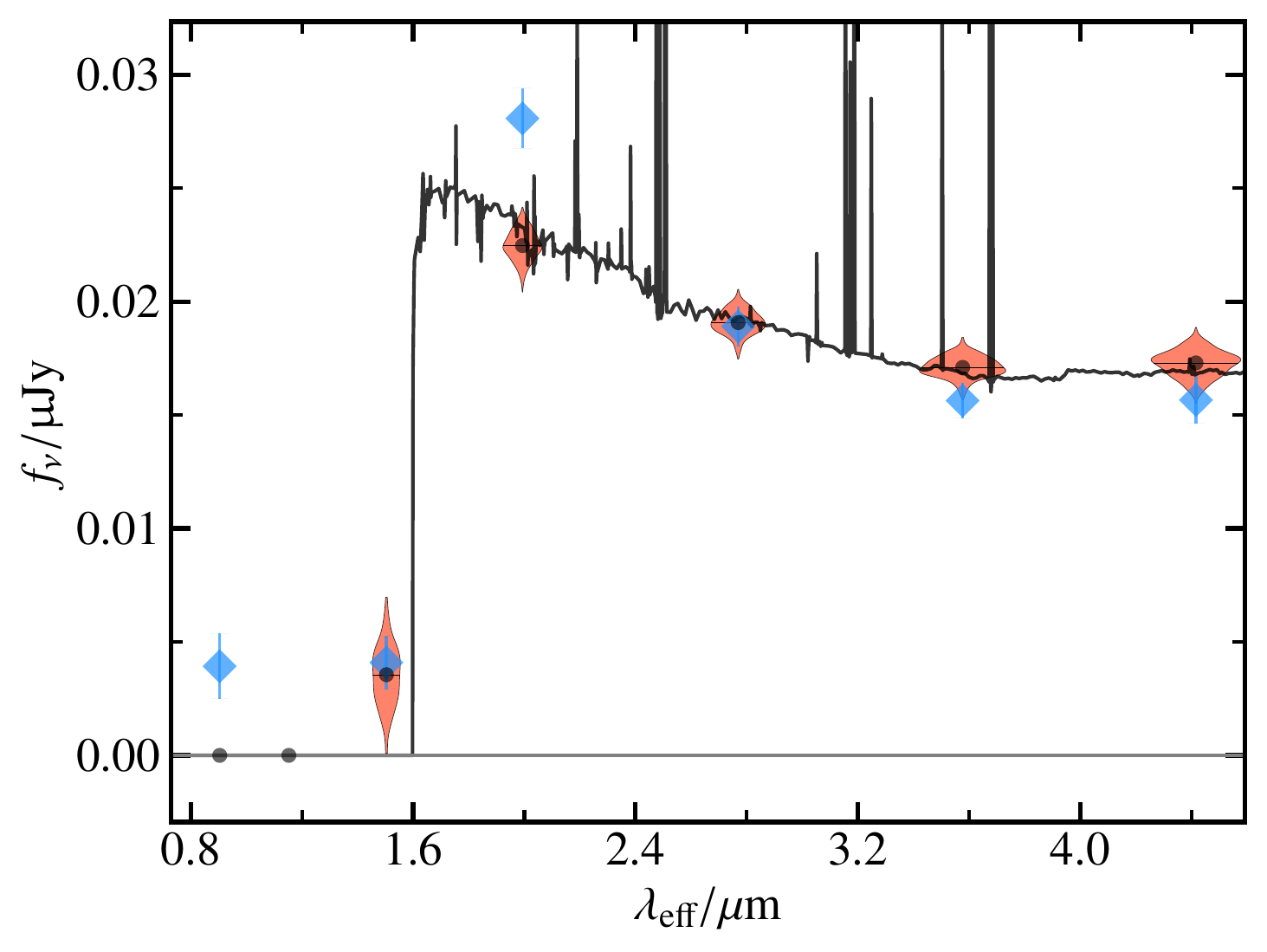}
    \includegraphics[width=0.49\textwidth]{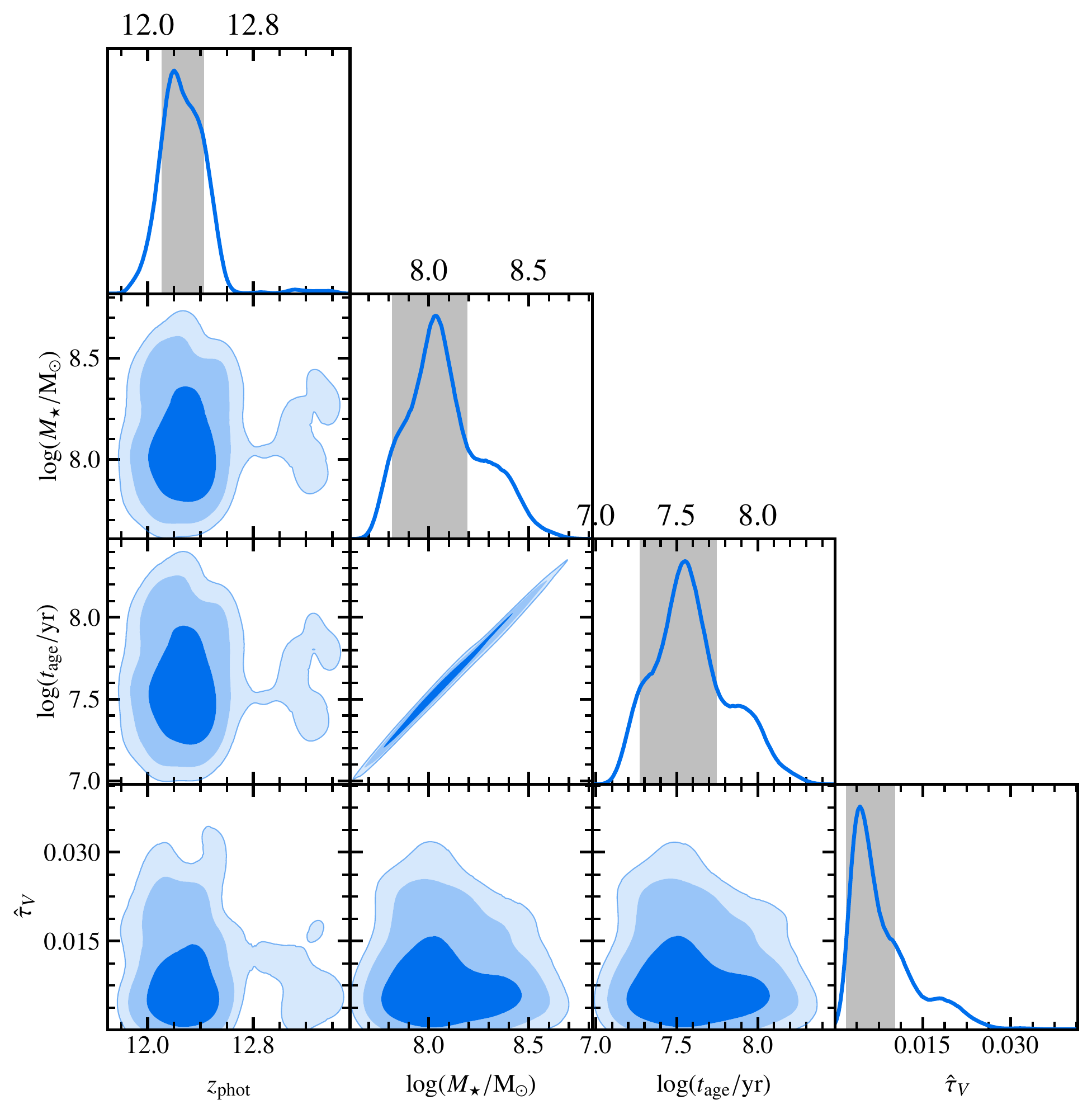}\\[1cm]
    \includegraphics[width=0.9\textwidth]{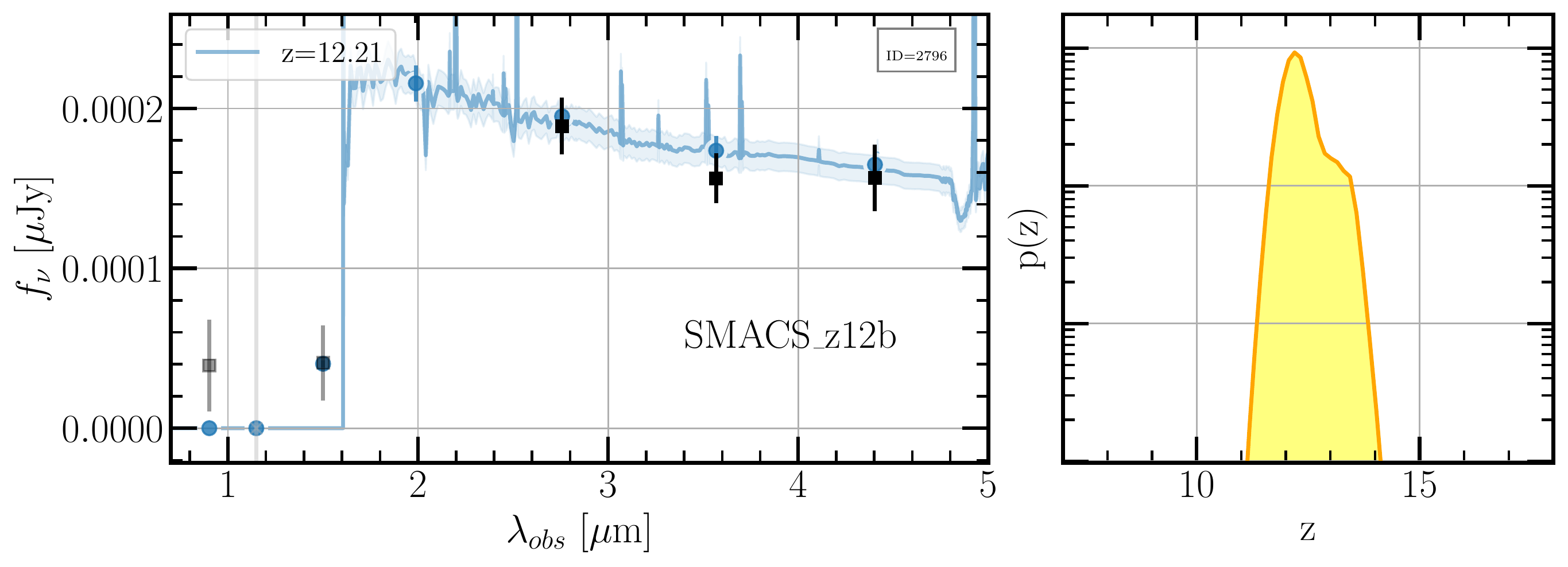}
    \caption{Photometric data and best-fit SED results for candidate SMACS\_z12b.}
    \label{fig:id-2720}
\end{figure*}

\begin{figure*}
    \centering
    \includegraphics[width=0.9\textwidth]{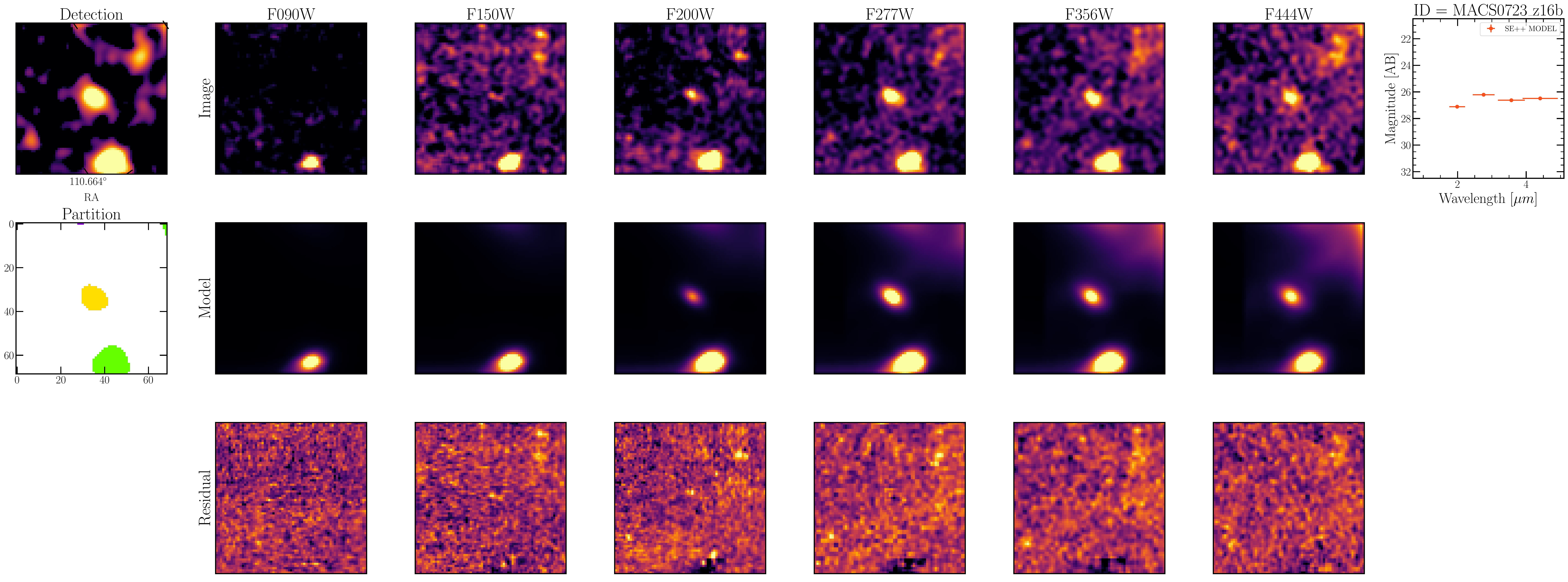}
    \includegraphics[width=0.49\textwidth]{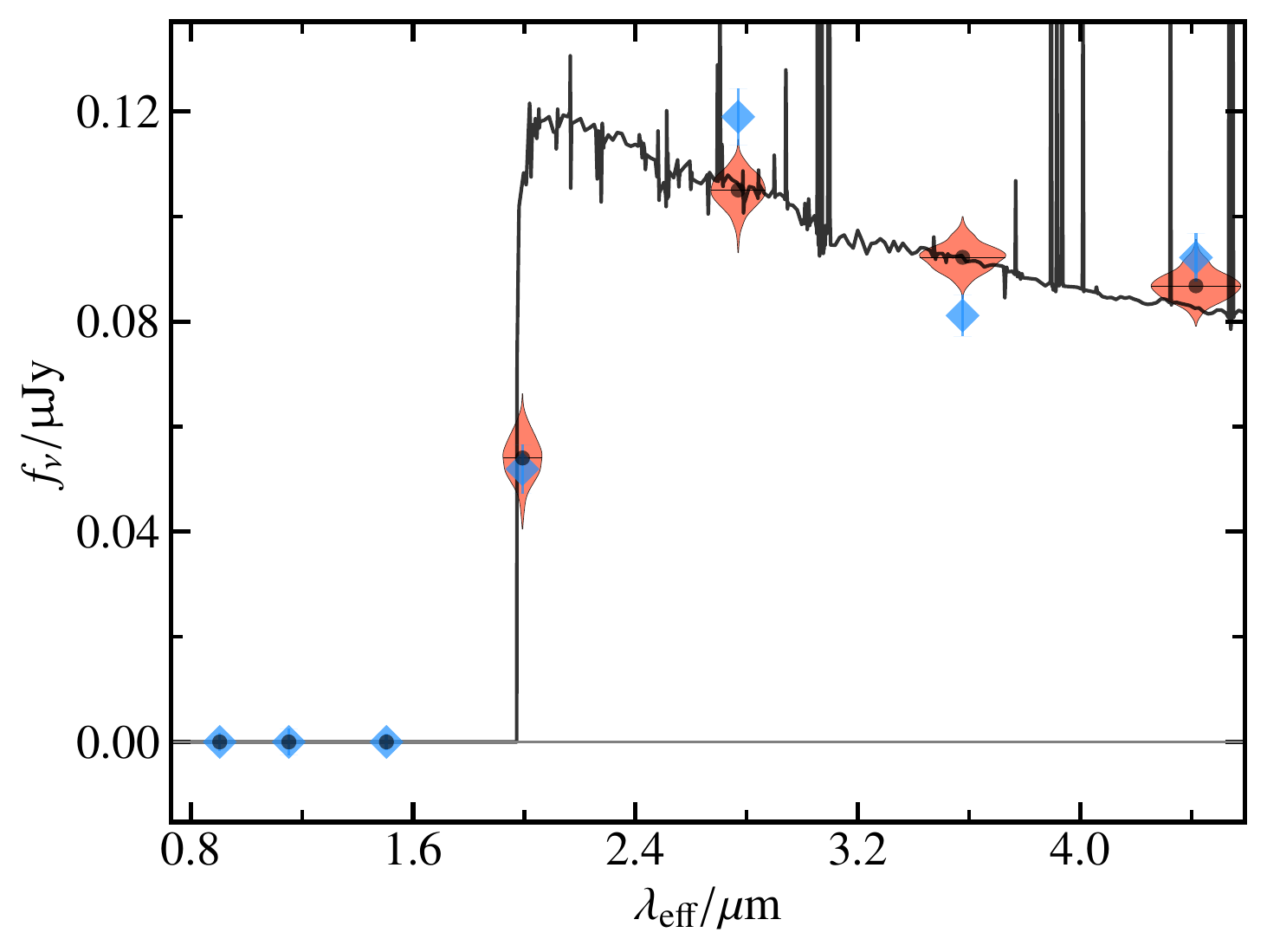}
    \includegraphics[width=0.49\textwidth]{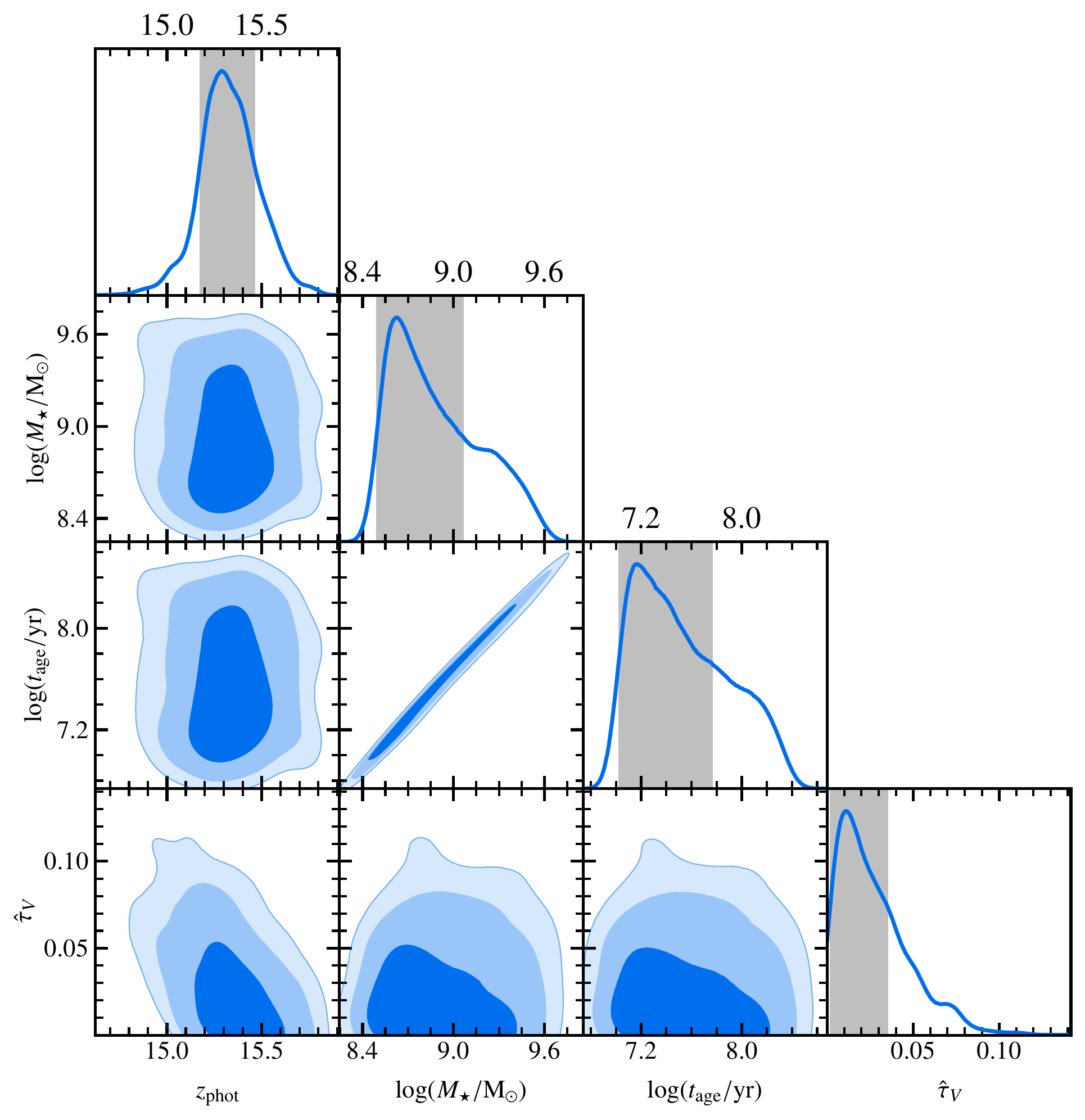}\\[1cm]
    \includegraphics[width=0.9\textwidth]{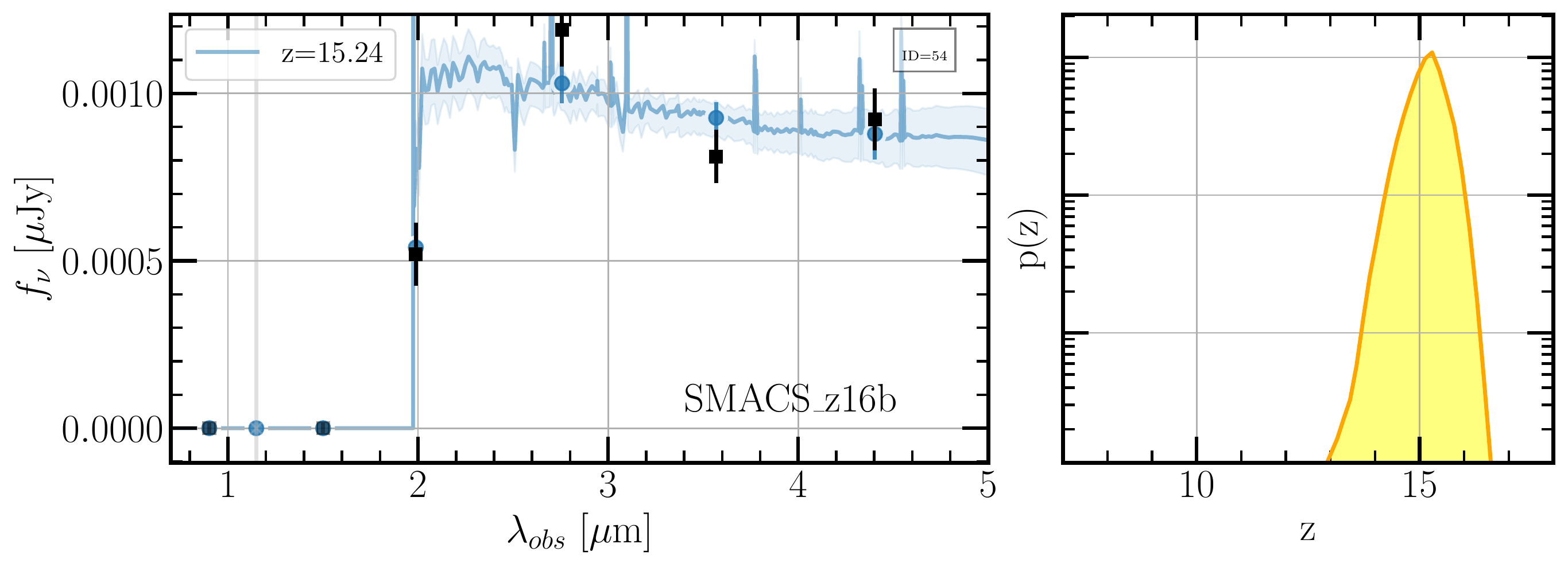}
    \caption{Photometric data and best-fit SED results for candidate SMACS\_z16b.}
    \label{fig:id-54}
\end{figure*}

\end{document}